\documentclass[preprint2]{pp7}
\usepackage[dvipsnames]{xcolor}
\usepackage{soul}

\input{pp7.h}

\usepackage[normalem]{ulem}

\definecolor{hlcolor}{rgb}{0.95, 1.00, 0.60}\sethlcolor{hlcolor}

\newcommand{\OIa}{[O\,{\scriptsize I}]\,$\lambda$6300}
\newcommand{\OIb}{[O\,{\scriptsize I}]\,$\lambda$5577}
\newcommand{\sii}{[S\,{\scriptsize II}]}
\newcommand{\SIIa}{[S\,{\scriptsize II}]\,$\lambda$4068}

\newcommand{\SIIc}{[S\,{\scriptsize II}]\,$\lambda$6731}
\newcommand{\oi}{[O\,{\scriptsize I}]}
\newcommand{\feii}{[Fe\,{\scriptsize II}]}
\newcommand{\neii}{[Ne\,{\scriptsize II}]}
\newcommand{\nii}{[N\,{\scriptsize II}]}
\newcommand{\hei}{He\,{\scriptsize I}}
\newcommand{\heilam}{He\,{\scriptsize I}\,$\lambda$10830}
\newcommand{\cii}{C\,{\scriptsize II}}
\newcommand{\ciilam}{C\,{\scriptsize II}\,$\lambda$1335}

\newcommand{\kms}{\,{\rm km/s}}
\newcommand{\ms}{\,{$M_\odot$}}

\begin{document}

\title{\textbf{\LARGE The Role of Disk Winds in the Evolution and Dispersal of Protoplanetary Disks}}

\author {\textbf{\large Ilaria Pascucci}}
\affil{\small\it University of Arizona}
\author {\textbf{\large Sylvie Cabrit}}
\affil{\small\it Observatoire de Paris}
\author {\textbf{\large Suzan Edwards}}
\affil{\small\it Smith College}
\author {\textbf{\large Uma Gorti}}
\affil{\small\it SETI Institute/NASA Ames Research Center}
\author {\textbf{\large Oliver Gressel}}
\affil{\small\it Leibniz Institute for Astrophysics Potsdam}
\author {\textbf{\large Takeru Suzuki}}
\affil{\small\it University of Tokyo}

\begin{abstract}
{\small  The assembly and architecture of planetary systems strongly depend on the physical processes governing the evolution and dispersal of protoplanetary disks. Since Protostars and Planets VI, new observations and theoretical insights favor disk winds as being one of those key processes. This chapter provides a comprehensive review of recent observations probing outflowing gas launched over a range of disk radii for a wide range of evolutionary stages, enabling an empirical understanding of how winds evolve. In parallel, we review theoretical advancements in both magnetohydrodynamic and photoevaporative disk wind models and identify predictions that can be confronted with observations. By linking theory and observations we critically assess the role of disk winds in the evolution and dispersal of protoplanetary disks. Finally, we explore the impact of disk winds on planet formation and evolution and highlight theoretical work, observations, and critical tests for future progress.     
 \\~\\~\\~}
\end{abstract}  


\section{\textbf{INTRODUCTION}}
Circumstellar disks of gas and dust are a direct consequence of star formation and harbor the birth sites of planets. Therefore, understanding how these protoplanetary disks evolve and disperse is crucial to understanding  the environment in which planets form, and how they are assembled and migrate. Since PPVI, theory and observations have challenged the notion that turbulence driven by magnetorotational instability \citep[MRI,][]{BalbusHawley1991ApJ...376..214B} is the main enabler of  accretion and disk evolution. Disk simulations with detailed microphysics find that MRI is suppressed in most of the planet-forming region ($\sim 1-30$\,au) and instead accretion is driven by radially extended magnetohydrodynamic (MHD) winds, i.e. outflowing gas
from the disk atmosphere \citep[e.g.,][for a recent review]{Lesur2021A&A...650A..35L}. At the same time, new observations have revealed a plethora of  wind diagnostics  that can be used to test theoretical MHD disk wind scenarios as well as thermal photoevaporative (PE) disk winds \citep[e.g.,][]{ErcolanoPascucci2017} which will not influence accretion but may play an important role in disk clearing.

Motivated by these recent advancements, here we juxtapose observations and theory to critically assess the role of disk winds in the evolution and dispersal of protoplanetary disks. 
There is growing evidence since PPVI that, even within 10\,au of the star,  wind velocities decrease with increasing disk radius, as expected in radially extended disk winds,  which will be a focus for the observational sections of this chapter. 

Our attention is on low-mass ($\sim 0.5-1$\,M$_\odot$) stars in low-mass star-forming regions. Hence, processes that may drive evolution and dispersal in higher-mass star-forming environments, e.g. external radiation driven PE winds, are not covered. Even in less complex environments, establishing evidence for disk winds is a challenging task 
due to the need for high spatial and spectral resolution, and interactions between the stellar magnetic field, the jet, the disk, and, at the earliest stages, an infalling envelope. Observationally, disk wind candidates are identified with gravitationally unbound/outflowing gas moving at moderate to low velocities ($< 30$\,\kms ) and, when sufficient spatial resolution is available, opening at wide angles from the disk.  
These two properties distinguish them from jets, a common mass ejection phenomenon characterized by collimated fast ($\ge 100$\,kms) outflowing gas \citep[e.g.,][]{Frank2014}. Jets and large scale, swept up bipolar molecular outflows have historically dominated the discussion of mass ejection from young stars \citep[e.g.,][]{Bally2016ARA&A..54..491B}, while direct evidence for radially extended disk winds has been scant. Although our focus here is on evidence for disk winds, we will address some observations of jets on  small scales, within a few hundred au, that are new since PPVI. 

Our chapter is structured as follows. First, we present observations of outflowing gas for systems in different evolutionary stages and discuss how flows evolve while accretion subsides and disks disperse (\S~\ref{sect:obs}). Next, we review the basic physics of MHD and PE winds  with focus on predictions that can be tested via observations (\S~\ref{sect:theory}). We interpret the observations in the context of theoretical models in \S~\ref{sect:link} 
and discuss the implications of disk winds for planet formation and migration  in \S~\ref{sect:implications}. We conclude in \S~\ref{sect:outlook} by critically assessing the role of winds in the evolution and dispersal of protoplanetary disks and by highlighting theoretical developments and observations for future progress.


\section{\textbf{\uppercase{Observations constraining disk wind models}}}\label{sect:obs}
We begin with an observational overview of mass ejection from low-mass stars and their disks.
With the goal of building a coherent picture, we cover all types of flows, from fast jets to winds launched from a large range of radial distances, at all stages of star and disk evolution (\S~\ref{sect:Class0I} and~\ref{sect:ClassII}).  An approximate age sequence for the three phases of star formation and envelope dispersal is assumed, where characteristic ages for Class 0 sources are on the order of $10^4$ yr, Class I $\sim 10^5$ yr and Class II $\sim 10^{5.5} - 10^7$ yr. We mostly rely on traditional boundaries between the different classes.  Class~0 are separated from  Class~I  based on the ratio of submillimeter to bolometric luminosity, a proxy for the envelope-to-stellar mass ratio which is larger than 1 for Class~0 sources \citep{1994ApJ...420..837A}. The infrared spectral index probes the presence, or absence, of a residual infalling envelope around the disk, hence it is used to distinguish Class~I from II \citep[e.g.,][]{1989ApJ...340..823W}. However, because
the spectral index depends on the viewing geometry, 
we also consider dynamical evidence of envelope infall or extended dust emission beyond the disk
as an additional discriminant between the two later stages  (hence, DG~Tau~B is classified here as Class~I, see \citealt{deValon2020}, while  HH~30 as Class~II, see \citealt{Louvet2018}). For the same reasons, flat-spectrum T~Tauri stars \citep[][]{Adams1987ApJ...312..788A,2016ApJS..224....5F} are also included in the Class~II category. 
Along with jets and wind properties, we also discuss  stellar accretion rates and disk radii as they  provide important constraints for disk wind models.  We  conclude with an empirical reconstruction of
how flows evolve in time while accretion weakens and disks dissipate (\S~\ref{sect:empiricalevolution}).

\subsection{\textbf{Embedded Class 0 and Class I stages}}\label{sect:Class0I}
Embedded protostars are known to drive powerful outflows on scales of 0.1-5\,pc, in the form of
collimated jets with velocities $\ge 100$\,\kms\ and wider colder large scale molecular outflows at $\simeq 10$\,\kms ,  interpreted as swept up material. These outflows are associated with embedded shock fronts, caused by jet variability or interaction with the ambient medium, and seen in a variety of tracers \citep[see e.g.,][]{Frank2014, Bally2016ARA&A..54..491B}. Here, we focus solely on jet and flow properties at the innermost scales, $\leq 500$\,au from the central source. These properties place
relevant constraints on ejection processes from both the inner ($< 1$\,au) and outer disk and represent new advancements  at high spatial {\it and} spectral resolution since PPVI. 

\subsubsection{\it{\textbf{Innermost regions of high-velocity jets}}}\label{sect:innermostjets}
Both Class 0 and Class I jets are detected in atomic and H$_2$ lines \cite[see][for reviews]{Frank2014,Bally2016ARA&A..54..491B}, but  Class~0 jets are also traced by pure rotational lines of CO, SiO, and SO, where advances in sub-mm interferometers have brought great progress in the last few years \citep[see][for a recent review]{Lee2020ARAA}. Concerning statistical trends, proper motions and precession models have yielded deprojected Class 0 jet speeds of $\simeq 100 \pm 50$\kms\ \citep{Lee2015ApJ...805..186L,Jhan2016ApJ...816...32J,Podio2016A&A...593L...4P,Yoshida2021ApJ...906..112Y}, slightly lower than in Class~I sources. 

Jets appear to be highly collimated.  The first resolved measurements of Class 0 SiO jet widths within $3-100$ au from the source \citep{Lee2020ARAA,Bjerkeli2019A&A...631A..64B} show a mean semi-opening angle $\simeq 4\degr$ and a typical jet width of  $\simeq 5$\,au.   Although the base of high-velocity Class~I jets was first probed 
 in atomic lines of \oi{}, \feii{}, and sometimes in H$_2$ \citep[see][for reviews]{Bally2007prpl.conf..215B,Frank2014}, resolving their transverse 
 structure within 1,000\,au of the source has remained difficult due to residual nebulosity from scattered light in narrow-band HST images (e.g., the recent tens of au widths inferred for HH34 and HH46, \citealt{Erkal2021ApJ...919...23E}), lack of optical reference for Adaptive Optics (AO) correction from the ground, and limited  
 resolution at mid- and far-infrared wavelengths. A notable exception is the optically visible, outbursting Class~I source SVS13 where AO-corrected Integral Field Spectroscopy resolved the  200 \kms\ \feii\ jet and measured a width of $3-6$\,au ($20 - 40$\,mas) within 150\,au of the source \citep{Hodapp2014ApJ...794..169H}, similar to the SiO Class 0 jets.

 Jet mass-fluxes could be also recently estimated for a dozen of Class~0 jets from CO millimeter data, assuming a standard CO ISM abundance ($\simeq 10^{-4}$ of H) and the following expression $\dot{M}_{\rm jet} = M v_{\perp}/l_{\perp}$ with $M$ being the optically thin emitting mass inside a projected length $l_{\perp}$ and $v_{\perp}$ the projected jet speed in the plane of the sky.  The CO jet mass-flux is  
found to correlate with the source bolometric luminosity \citep{Podio2021}, which is assumed to be dominated by accretion.  In sources with a dynamical estimate of the stellar mass, the CO jet mass-flux is found to be on average $\simeq 20\%$ of the mass-accretion rate onto the central protostar \citep{Lee2020ARAA}. A similar ejection to accretion ratio of $\simeq 10\%$ was inferred for atomic Class~0/I jets from optical and infrared lines of \oi{} and \feii{} resolved along the jet axis \citep[e.g., ][]{Hartigan1994ApJ...436..125H,Nisini2015ApJ...801..121N,Bally2016ARA&A..54..491B,Sperling2021A&A...650A.173S}.

Concerning jet chemistry, an important new result from a survey of 21 Class 0 sources is that the detection rate of SiO and SO jets increases with source luminosity, from $\le 33\%$ below $1 L_\odot$ to $\ge 80\%$ above \citep{Podio2021}. Since the jet mass-flux also correlates with source luminosity in Class 0 sources (see above), the implication is that SiO and SO are detected preferentially at high jet mass-fluxes which  is consistent with the non-detection of molecules other than H$_2$ in the lower mass-fluxes Class~I jets. 
 
Finally, two additional impactful results on the structure of Class~0 jets, particularly relevant to the issue of disk winds, were recently achieved thanks to the unprecedented angular resolution and polarization capabilities of ALMA. One of them is the detection of consistent rotation in the same sense as the disk down to a 10\,au scale in the bright SiO jet of HH~212 \citep{LeeHo2017} which  establishes a disk origin for the jet. The specific angular momentum suggests an origin from the innermost disk at $0.1-0.2$\,au, potentially inside the dust evaporation radius 
\citep[see][and \S~\ref{sect:SpatiallyResolvedWindDiagnostics}]{LeeHo2017,Tabone2017}. 
The second result is the first detection of SiO linear polarization by the Goldreich-Kylafis effect inside the HH~211 jet. The inferred magnetic field intensity (from the dispersion of polarization angles) and 
morphology are consistent with magnetic jet collimation \citep{LeeHwang2018}. This represents the first direct, and so far only, probe of magnetic field inside a jet from a low-mass protostar.

\begin{figure}[t]
   \centering
    \includegraphics[width=0.6\textwidth,angle=-90]{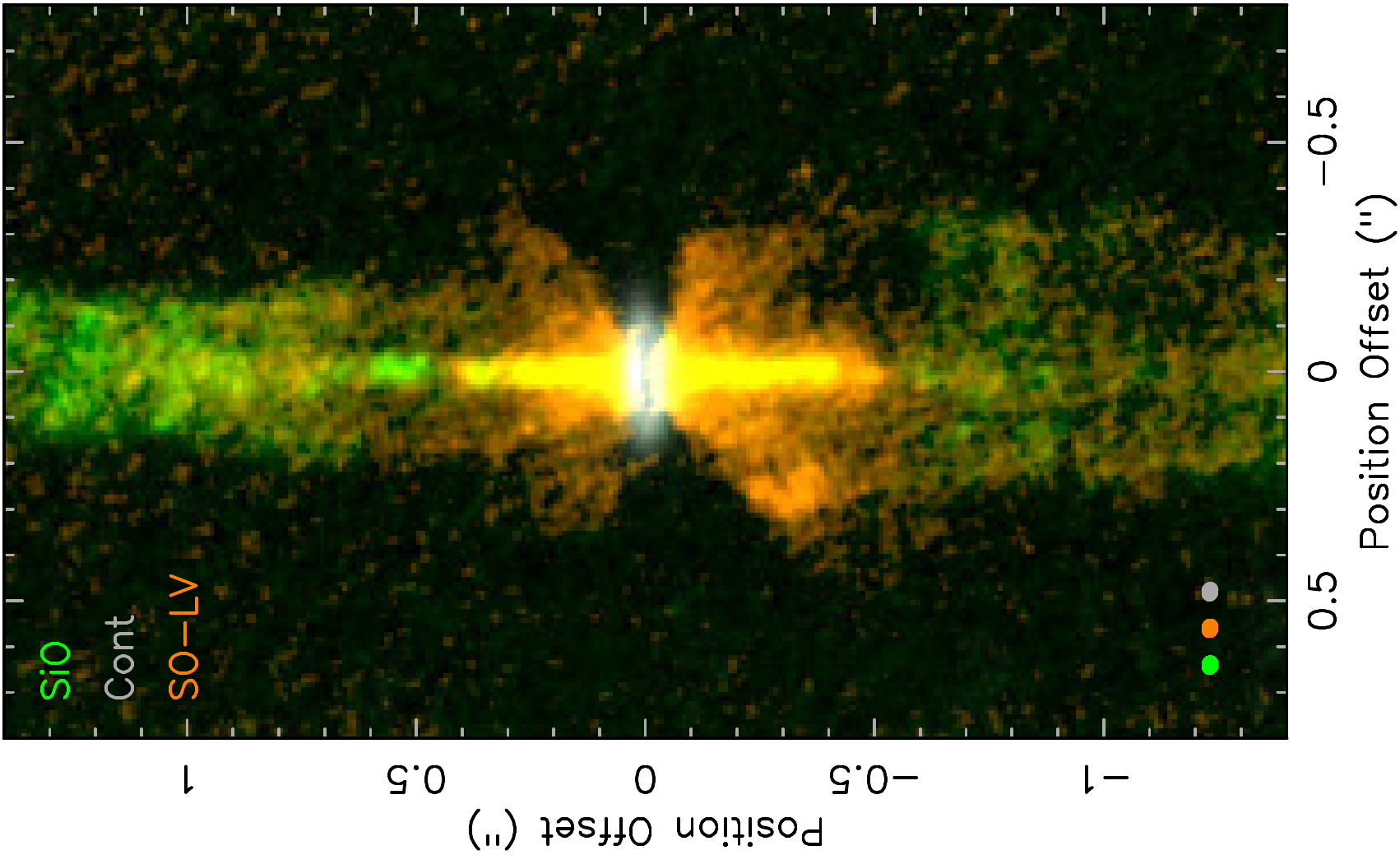}
    \caption{Morphology of nested outflow regimes in the Class~0 source HH~212 which is located at a distance of 400\,pc and has been observed with ALMA at 15\,au resolution  \citep{LeeTabone2021ApJ...907L..41L}. The SiO intensity map (green) traces the fast collimated jet while the SO-LV, within $\pm 3$\,\kms\ of the systemic velocity (orange), probes slower outflowing material surrounding the jet and has a wide-angle wind-like morphology. Maps are rotated by 22.5$^\circ$ clockwise to align the jet axis vertically. The dust disk emission is in grey.}
    \label{fig:hh212}
\end{figure}

\subsubsection{\it{\textbf{Slow molecular ``winds"}}} \label{sect:SlowMolecularWinds}
The first evidence for a slower and wider flow component surrounding the base of Class I jets was the ubiquitous detection in near-infrared spectra of a low-velocity component (LVC) blueshifted by $\simeq 5-20$ \kms\ in rovibrational H$_2$ emission \citep{Davis2001MNRAS.326..524D}. 
This component extends over a few hundred au along the jet \citep[e.g.,][]{GarciaLopez2010A&A...511A...5G} and more recent spectro-imaging suggests that it broadens out with a semi-opening angle $\simeq 10\degr-30\degr$ \citep{Davis2011A&A...528A...3D,GarciaLopez2013A&A...552L...2G}. An extended LVC is also sometimes detected in atomic lines of [OI] and [SII], e.g. in DG Tau B \citep{Podio2011A&A...527A..13P}.
However, detailed studies of its transverse structure in the optical or near-infrared meet the same challenges as for the base of jets in Class~I  sources (see \S~\ref{sect:innermostjets}). 
One exception is again SVS13, where AO-corrected observations resolved the slow H$_2$ into a series of expanding bright-rimmed ``bubbles", attributed to short-lived pulses of jet activity \citep{Hodapp2014ApJ...794..169H}.   In parallel, 
 CO ro-vibrational line profiles observed at high spectral resolution revealed broad blueshifted absorption features from -10 to -100\,km/s in 6 out of 18 ($\simeq$ 33\%) Class~I targets, with increasing columns at low velocities and gas temperature $\simeq$ 1200~K. Since the absorbed 4.6\,\micron{} continuum is presumably dominated by emission from  the inner disk, these features are suggestive of warm inner molecular disk winds with a broad range of flow speeds that produce detectable CO absorption at favorable viewing angles \cite[e.g.,][]{Herczeg2011}.  
 However,  flow structure and mass fluxes could not be derived due to lack of spatial information.

\begin{figure}[h]
  \includegraphics[width=0.45\textwidth]{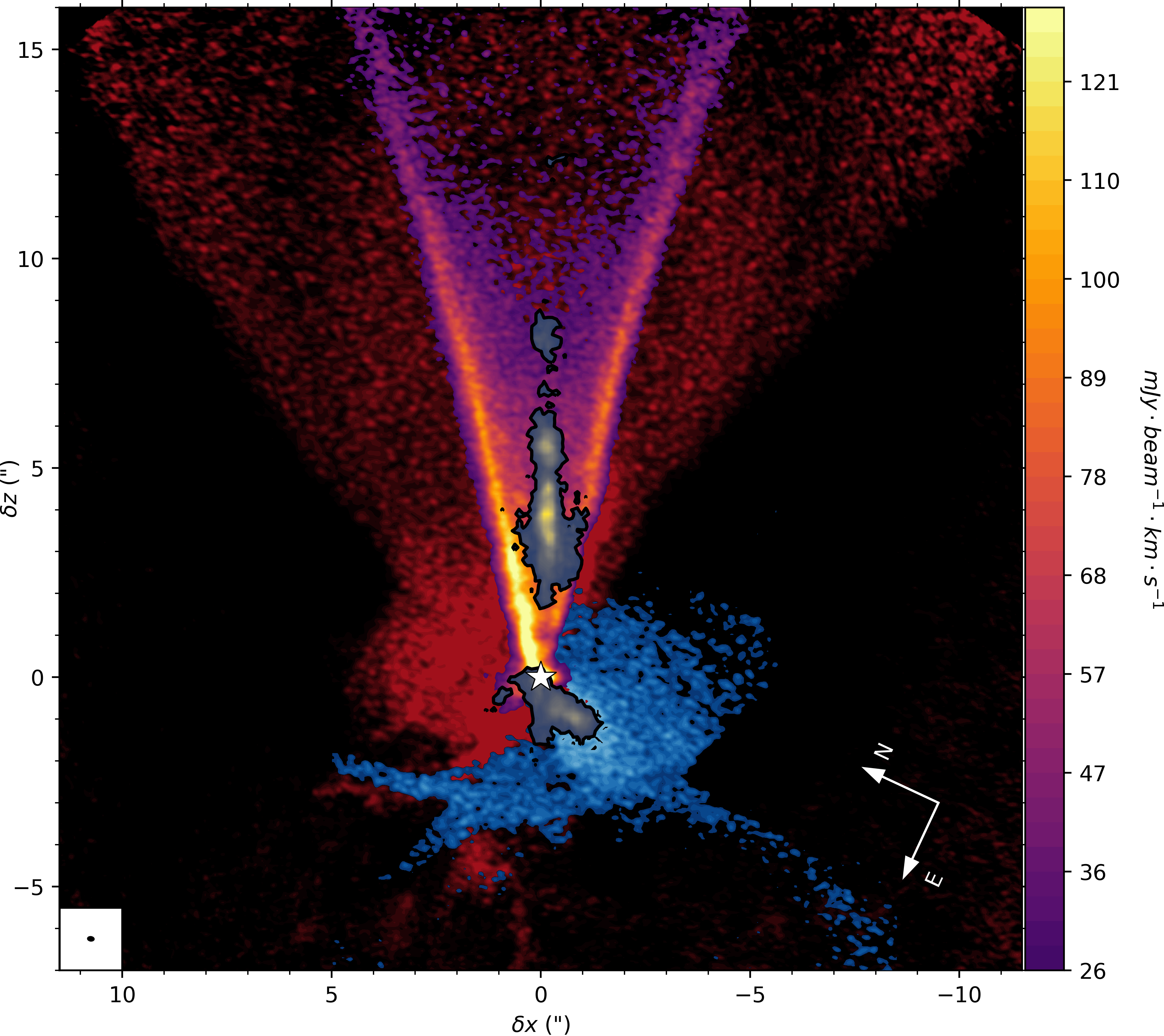}
  \caption{Morphology of nested outflow regimes in the Class~I source DG~Tau~B which is located at a distance of $\simeq 140$\,pc. The collimated jet emission (gray/yellow) is from the HST/WFPC2  \citep[F675W filter,][]{Stapelfeldt1997IAUS..182..355S}. Note that the jet emission is more vertically extended than shown, a flux cut has been applied to emphasize emission close to the star. The other structures are from various $^{12}$CO(2-1) kinematic components \citep{deValon2020}. The inner conical flow (purple to yellow shades) is detected at higher velocities ($V- V_{\rm sys}= +2.15$ to $+8$\,\kms ) than the wider angle and slower ($V- V_{\rm sys}= 1.19$\,\kms ) flow (in red).} \label{fig:dgtaub}
\end{figure}

A key new result provided by submm interferometers is the discovery  of small-scale ($\le$ 2000 au) molecular flows around a growing number of low-mass Class~I/0 sources that rotate in the same sense as the disk and are rooted within the disk at radii $\le 10-100$\,au. \cite[Class~I: CB26, TMC1A, DGTauB; Class~0:HH212, HH211, NGC1333-IRAS4C;][]{Launhardt2009,Bjerkeli2016,Zapata2015,deValon2020,Tabone2017,Lee2018ApJ...863...94L,Zhang2018}. This geometry hints at material directly ejected from the disk surface. In addition, in HH~212, the rotating flow is confined {\it inside} of the main swept-up outflow cavity walls delineated in CO and CS \citep{Tabone2017}, and inside the dense flattened rotating infalling envelope traced by HCO$^+$ \citep{LeeTabone2021ApJ...907L..41L}.  Therefore, in the following, we will refer to these small-scale rotating flows as molecular ``winds" to distinguish them from the classical bipolar molecular flows of swept-up ambient gas observed over much larger (parsec) scales around embedded sources \citep[e.g.,][]{Arce2007prpl.conf..245A,Frank2014}.

All small-scale rotating molecular winds display striking conical or parabolic shapes, with semi-opening angles 10\degr$-$40\degr and an inner low-brightness ``cavity" surrounding the jet beam.
Figures~\ref{fig:hh212} and \ref{fig:dgtaub} provide the illustrative cases of HH~212 (Class~0) and DG~Tau~B (Class~I) as observed by ALMA at a projected resolution of $\sim 15$\,au \citep{LeeTabone2021ApJ...907L..41L,deValon2020}. In the case of HH~212, a rotating wide-angle flow with semi-opening angle $\sim 40^\circ$ is detected in SO for velocities within a few \kms\ from systemic (orange). Toward DG~Tau~B, CO emission at a few \kms\ from systemic delineates a bright rotating cone of semi-opening angle $\simeq 15\degr$ (purple to yellow shades), anchored at $\le$ 40 au in the disk, surrounded by a broader and fainter  flow at lower velocities (red) that also shows clear signatures of rotation \citep{deValon2020}. 
The collimated jet emission, traced via ALMA SiO  for HH~212 and  with the HST/WFPC2 F675W filter for DG~Tau~B, is also shown in the figures. Interestingly, the optical LVC  of DG~Tau~B was found to have less refractory species (including Fe and Ni) than the HVC \citep{Podio2011A&A...527A..13P}, implying that the LVC arises from beyond the dust evaporation radius where refractories are locked into dust grains.
Considering all the spatially resolved flows to date, the distinction in morphology between co-existing jets and winds is striking, where the fast moving inner jets are highly collimated and precessing, with time variable knotty structure arising from moving bow shocks, while the surrounding slow moving winds are wide cones appearing quite smooth in structure and showing no signs of collimating, sometimes out to $\ge$\,1000\,au.

The wind kinematics can be recovered from transverse position-velocity (PV) cuts in a quasi model-independent fashion, assuming only axisymmetry (with slight biases introduced by inclination and velocity gradients, cf. de-
Valon et al. 2022, in press). 
Application to DG~Tau~B shows that the cone in Figure~\ref{fig:dgtaub} harbors a nested velocity structure, often called ``onion-like", with a decreasing flow speed (from 20 to 6\,\kms) and increasing specific angular momentum (from 40 to 100\,au\,\kms) at wider opening angles. 
Assuming constant opening angles down to the disk midplane 
gives geometric base radii ($r_{\rm geom}$) of $\sim 5-40$\,au for these nested cones, which, as discussed in \S~\ref{sect:SpatiallyResolvedWindDiagnostics}, set  upper limits to the MHD  wind-launching radii.

Published results for the six resolved rotating molecular winds from 
low-mass Class 0/I sources (3 in each stage, cf. list of objects above) shows that their properties are  similar across the  current sample, although the molecular tracers differ.  Deprojected vertical flow speeds span from 1 to $20\kms$ and lateral expansion speeds  are $\simeq 2-5$\,\kms . Measured specific angular momenta are in the range $20-300$\,au\,\kms\ and geometric radii range from $\sim 5$ to $\sim 100$\,au (see Table~\ref{tab:jetwindprop}).
In addition, the other two winds that are laterally resolved (from TMC1A and HH212) also show a nested velocity structure with progressively lower speeds and larger specific angular momenta on the outside, similar to DG~Tau~B,  hinting at the same physical process. 
In each case, the inner streamline r$_{\rm geom}$ lies well inside the gas disk radius, which is important to assess the flow origin (see \S~\ref{sect:SpatiallyResolvedWindDiagnostics}).

 Assuming that the flows do trace steady ejection from the disk, the estimated mass-flux would be very significant. In the Class 0 source HH~212, it would be $(1-10) \times 10^{-6} M_\odot$ yr$^{-1}$ (where the large range reflects the uncertainty in SO abundance), i.e. 1--10 times the jet mass-flux and 0.3--3 times the estimated mass-accretion rate onto the protostar \citep{Tabone2020A&A...640A..82T}. In the Class I source DG~Tau~B,  the mass-flux would be $\simeq 2 \times 10^{-7} M_\odot$ yr$^{-1}$, i.e. 40 times that in the fast axial atomic jet \citep{Podio2011A&A...527A..13P} and, assuming a standard ejection/accretion ratio in the jet of 10\%, $\simeq 4$ times the accretion rate onto the star  \citep{deValon2020}. Such massive disk winds would clearly play a major role in limiting the fraction of envelope mass that ends up on the star, in driving disk accretion if they are MHD-driven, and in dissipating the disk mass reservoir (\S~\ref{sect:MHD}, \ref{sect:EvolutionMassEjection}, and \ref{sect:SpatiallyResolvedWindDiagnostics}).

We note that the interpretation in terms of slow disk winds is still debated. An alternative interpretation is that these compact rotating flows might instead trace  the base of entrained ambient material swept up by a fast wide-angle X-wind.
We will discuss the X-wind scenario, and its main challenges in reproducing observations, in \S~\ref{sect:alternative}.

\subsection{\textbf{Class~II stage}}\label{sect:ClassII}
Class~II sources span a wide range of accretion and outflow properties, ranging from 
a handful of objects with accretion rates $\ge 10^{-8} M_\odot$ yr$^{-1}$  and spatially resolved microjets \citep[see][]{Ray2007} to, more commonly, accretion rates from $\sim 10^{-8}$ to $10^{-11} M_\odot$ yr$^{-1}$ \citep{Nisini2018}. As we will discuss in \S~\ref{sect:SpatiallyUnresolvedFlows}, outflows/winds from the majority of Class~II sources are inferred solely from the kinematics of spatially unresolved,  blueshifted forbidden emission including \oi , \sii , \nii, \feii , \neii\  as well as P Cygni profiles in strong permitted emission lines, such as \hei\ 10830\,\AA . 
The forbidden profiles often have two components, which were already early on interpreted as arising from a high velocity microjet (HVC: 50 to 300 km/s) and a low velocity disk wind (LVC: often with velocities less than 10 km/s), e.g., \citet{HEG1995}, hereafter HEG, and \citet{KwanTademaru1995}.  Although the association of the HVC with microjets  has been accepted for some time, firm evidence for a more radially extended disk wind is new since PPVI. The relation between the collimated jet and a possible extended disk wind, however, remains uncertain.

\subsubsection{\it{\textbf{Spatially resolved outflows}}} \label{sect:SpatiallyResolvedOutflows}
Resolved Class II microjets, with velocities on the order of a few hundred km/s and vertically extended up to a few hundred au from the star, are observed with high resolution spectro-imaging in optical and near-IR forbidden lines in a handful of high accretion rate sources, as summarized in \citet{Ray2007}. Although less extended, their structures and high velocities are reminiscent of Class I jets, with semi-opening angles $< 5^\circ$ and, in at least one source (DG~Tau~A), this tight collimation occurs as close as 10\,au to the star \citep{2014A&A...565A.110M}.
The formation and proper motion of knots of shocked gas in Class II microjets has been tracked with high angular resolution imaging, with new knots appearing every few years \citep{LopezMartin2003,Pyo2003ApJ590340P,Takami2020ApJ90124T},  attributed to episodic variations in the velocity of mass ejection  \citep{2018MNRAS.481.5532P}.   Small velocity shifts across jet widths hint at rotation but can also be attributed to jet wiggling and asymmetric shocks \citep[e.g.,][]{Erkal2021ApJ...919...23E}.

\begin{figure}[h]
  \includegraphics[width=\linewidth]{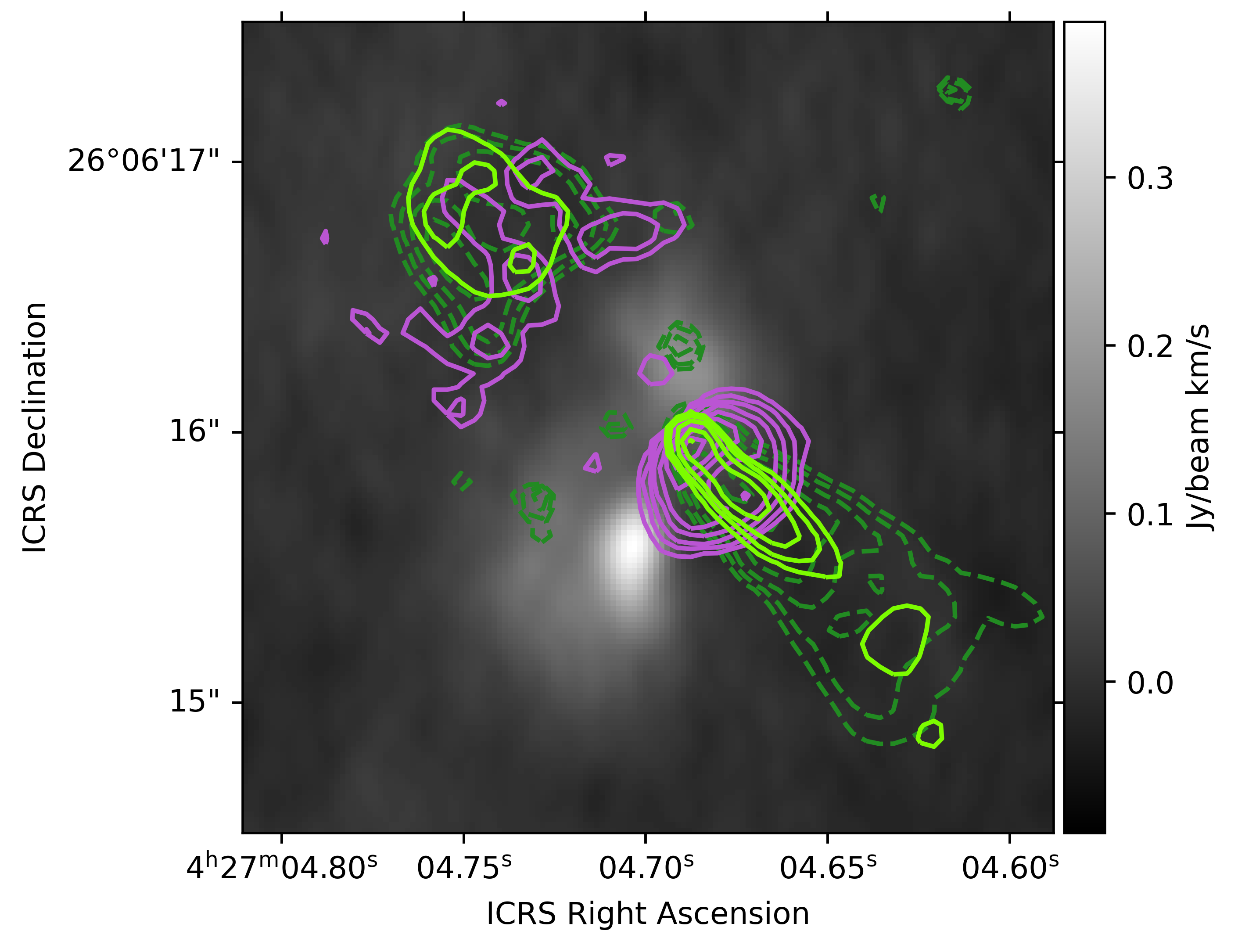}
  \caption{Morphology of nested outflow regimes in DG~Tau~A which is located at a distance of $\sim 121$\,pc. Contours of \feii{} emission at 1.64\,\micron{} show the narrow high-velocity jet ($V < -160$\,\kms{}  and $V > 120$\,\kms ; solid green) and wider slower flow ($-160 < V < 120$\,\kms ; dashed green). H$_2$ 1--0 S(1) emission at 2.12\,\micron{} (solid purple contours) traces an even wider cone \citep{AgraAmboage2014}. 
   The grey scale intensity map shows $^{12}$CO (2--1)  blueshifted emission within -0.7 to  -6.7 \kms\ of the systemic velocity, attributed to  disk emission (bright SE elongation) plus a slow disk wind outflowing at $\sim -3$\,km/s  \citep{Gudel2018}.  }
   \label{fig:dgtaua} \end{figure}

Since PPVI, it has become clearer that the fast, collimated, and knotty microjets are only the innermost region of flows that emerge from disk radii out to $\sim$ 100 au. The flows are characterized by a nested velocity structure, as first noted in optical forbidden lines of DG~Tau~A \citep{Lavalley-Fouquet2000,Bacciotti2000ApJ...537L..49B,Coffey2008} where the lower blueshifted velocities ($<$ 50\,km/s) are spatially wider with larger opening angles than the jet. The spatial morphology of the different velocity regimes is illustrated in  Fig.~\ref{fig:dgtaua} for DG Tau A, superposing emission from \feii, H$_2$ 1–0 S(1) and $^{12}$CO (2-1). The \feii\ emission shows the narrow jet ($V < -160$\,\kms{}  and $V > 120$\,\kms ; solid green) with a semi-opening angle of $\sim 4 ^\circ$ nestled inside of a wider slower ($-160 < v < 120$\,\kms ; dashed green) cone of outflowing gas with a semi-opening angle of $\sim 14 ^\circ$ \citep{AgraAmboage2014}. Interestingly, the faster jet is also more Fe-rich than the low-velocity ($-100 < v < 10$\,\kms) sheath surrounding it, suggesting that the latter arises beyond the dust sublimation radius \citep{AgraAmboage2011}.  Both \feii{} velocity components shown in Fig.~\ref{fig:dgtaua} lie interior to 
a slower and more iron-depleted flow of semi-opening angle $\sim 20 ^\circ$ seen only in \OIa\ and \SIIc\ \citep[][not shown]{2014A&A...565A.110M}. This flow is itself surrounded by an even 
wider cone of H$_2$ emission (purple contours) with deprojected semi-opening angle of $\sim 27^\circ$ mapped with near infrared AO IFU spectro-imaging at velocities of $\lesssim$-15\,km/s \citep{AgraAmboage2014}. 
The outermost flow component in Fig.~\ref{fig:dgtaua} is traced by low-velocity blueshifted CO emission in recent ALMA millimeter imaging (grayscale): in addition to the rotating disk contribution (asymmetric peak to the South),  an arc-like structure is present from SE to NW. This is interpreted as a slow wind at $\sim -3$\,\kms\ originating from close to the disk surface {at a disk radius $\approx 40-70$\,au}, and with a deprojected semi-opening angle of $\sim 25 ^\circ$ \citep{Gudel2018}, similar to that inferred from the H$_2$ near-infrared emission. 
Although not shown in Fig.~\ref{fig:dgtaua}, FUV fluorescent H$_2$ emission pumped by Ly$\alpha$ presents blueshifted emission with nearly identical  morphology and velocity to the H$_2$ 2.12\,\micron\  feature \citep{Schneider2013,AgraAmboage2014}.  Furthermore, 
there is  evidence for rotation across the spatially broad, lowest velocity  (from -75 to +50\,\kms ) atomic flow surrounding the jet, from transverse velocity shifts mapped in \oi\ and \sii\ lines \citep{Bacciotti2002}.

To date, there are a few other examples of ALMA millimeter CO imaging  interpreted as a slow, wide molecular flow which encompasses  a narrow jet and  sometimes also wider atomic and $ H_2$ outflows, extending  100s of au above the disk. In HH~30 the fast axial microjet seen in forbidden lines \citep{Bacciotti1999A&A...350..917B}, is enclosed within what appears as a hollow cavity defined by a cone-shaped CO molecular flow with semi-opening angle of $\sim 35 ^\circ$ and a  transverse expansion velocity of 5 \,\kms, suggesting a full deprojected flow speed $\sim 9$\,\kms\
\citep{Louvet2018}. The intersection of the CO cone with the  HH 30 disk midplane defines a geometric radius of 10\,au. Another example is HL~Tau with a slow CO flow seen with ALMA \citep[][Bacciotti, in prep.]{Garufi2020A&A...636A..65G} surrounding both the high velocity jet seen in forbidden lines and a wide angle H$_2$ cone \citep{Takami2007,Beck2008}. The more massive ($\sim 1.9$\,M$_\odot$) star HD~163296 also has an optical jet surrounded by both a very slow molecular outflow  ($\sim$30\,m/s) thought to originate beyond $\sim$300\,au \citep{Teague2019} and a much larger and faster ($20$\,\kms) molecular flow extending vertically out to $\sim$1000\,au  \citep{Klaassen2013,Booth2021ApJS..257...16B}.

In HH~30, mass loss rates in the fast optical jet and slow CO wind can now be compared. The jet rate, at $ 2 \times 10^{-9}$  $M_\odot$ yr$^{-1}$ \citep{Bacciotti1999A&A...350..917B}, is nearly 50 times lower than the $ 9 \times 10^{-8}$  $M_\odot$ yr$^{-1}$  of the disk wind \citep{Louvet2018}. Assuming a typical jet-to-accretion mass flux of 10\%, the wind mass loss rate is $\sim$5 times the accretion rate.
Also in HH~30, ALMA data revealed small rotation velocities ($\simeq 0.1-0.5$\,km/s)  in the CO molecular wind in the same sense as the disk. The rotation is traced up to vertical extents of 250\,au but  accounting for wiggling of the flow axis is necessary to avoid spurious rotation signatures \citep{Louvet2018}. 
As will be addressed in \S~\ref{sect:SpatiallyResolvedWindDiagnostics}, interpreting this outflowing, rotating molecular cone as an MHD wind is favored over swept up gas because of the high inferred  mass loss rate from radii $<$\,10\,au.

We anticipate that outer slow moving cones of cold molecular winds will be found more frequently with ALMA in the coming years. Even with the current data, the evidence is compelling that in spatially resolved Class II outflows mass ejection occurs over a wide range of disk radii, with an inner fast axial jet surrounded by a  wider, slower moving wind than can extend radially almost out to $\sim$ 100 au. These cone shaped molecular flows show no hints of recollimating into a jet, but maintain their opening angles up to 1,000\,au above the disk surface.

\subsubsection{\it{\textbf{Spatially Unresolved Flows}}} \label{sect:SpatiallyUnresolvedFlows}
 Spatially unresolved flows are widely traced by forbidden line profiles and surveys of nearby star-forming regions are based on the \OIa , the brightest of the forbidden emission lines detected in 80-90\% of the accreting T Tauri stars.

{\it{\textbf{HVC / jet.}}}  Among sources with \OIa\ detections,  30\% of the young ($\sim 1-3$\,Myr) and 10\% of the old ($\sim 5-10$\,Myr) T~Tauri stars show HVC emission (\citealt{McGinnis2018,Nisini2018} and Fang et al. in prep.). This HVC emission is preferentially detected among sources with higher stellar mass accretion rate and mass. The HVC can be linked to the spatially and spectrally resolved forbidden line images where it is associated with the fast collimated jet \citep[e.g.,][]{Bacciotti1999A&A...350..917B, AgraAmboage2011}.
    In agreement with past work, \citet{Nisini2018} find a strong correlation between the jet mass ejection rate (${\dot{M}_{\rm jet} }$) inferred from the HVC and the stellar mass accretion rate ($\dot{M}_{\rm acc}$) with an average ratio of 0.07, a spread of more than an order of magnitude, and  some examples of high accretion rate sources with no HVC.
  At least some of these outliers are likely sources where system inclination projects the HVC into the LVC velocity regime. 
  
A surprising new finding on the HVC/jet  kinematics  as $\dot{M}_{\rm acc}$ falls below $\sim 10^{-9}$  $M_\odot$ yr$^{-1}$ is reported by \citet{Banzatti2019}. 
Unlike the resolved jets in higher accretion rate sources which have outflow velocities of several 100\,\kms\ regardless of source luminosity, the centroid of the most blueshifted unresolved HVC component, after correction for inclination, correlates with the accretion luminosity. This suggests a continuum of HVC/jet terminal velocities from 300\,\kms{} down to 50\,\kms , decreasing systematically as the accretion luminosity falls by three orders of magnitude.

In addition to the forbidden line HVC, another indicator of outflowing gas with velocities up to several hundred km/s is blueshifted absorption in permitted lines, where the absorption occurs along the line of sight to the star. The blueshifts in strong lines such as  \heilam\ or \ciilam\ are inclination dependent, where deep and broad absorption, from  0 up to 400\kms , is seen only in sources with low inclinations \citep{Edwards2006,Xu2021ApJ...921..181X}. In contrast, at high inclinations  the absorption changes to shallow and narrow, with blueshifted centroids $< 50$\,\kms . The interpretation is that when viewed close to pole-on, the line of sight passes through a fast wind emerging from the star or its near environs, while when viewed closer to edge-on, the line of sight passes through a slower disk wind \citep{Kwan2007}.  Although mass loss rates from the atomic wind seen in  \heilam\ and \ciilam\ are difficult to quantify, the strength of these features correlates with the mass accretion rate \citep{Edwards2006}.
 
{\it{\textbf{LVC / wind.}}}
In contrast to the HVC, almost all stars with \OIa\ detections have LVC emission, covering the full range of T~Tauri accretion luminosities \citep{Natta2014,McGinnis2018,Nisini2018}.
 As with the HVC there is a good correlation of the LVC luminosity with $\dot{M}_{\rm acc}$, with the majority of non-detections falling among lower accretion rate sources. The LVC  component can be linked to spatially and spectrally resolved forbidden line images, corresponding to the slower, wider disk winds, e.g. DG~Tau~A (\S~\ref{sect:SpatiallyResolvedOutflows}) and the recent result from CT~Cha with VLT/MUSE (Haffert et al. in prep.).

\begin{figure*}[h]
  \includegraphics[width=\textwidth]{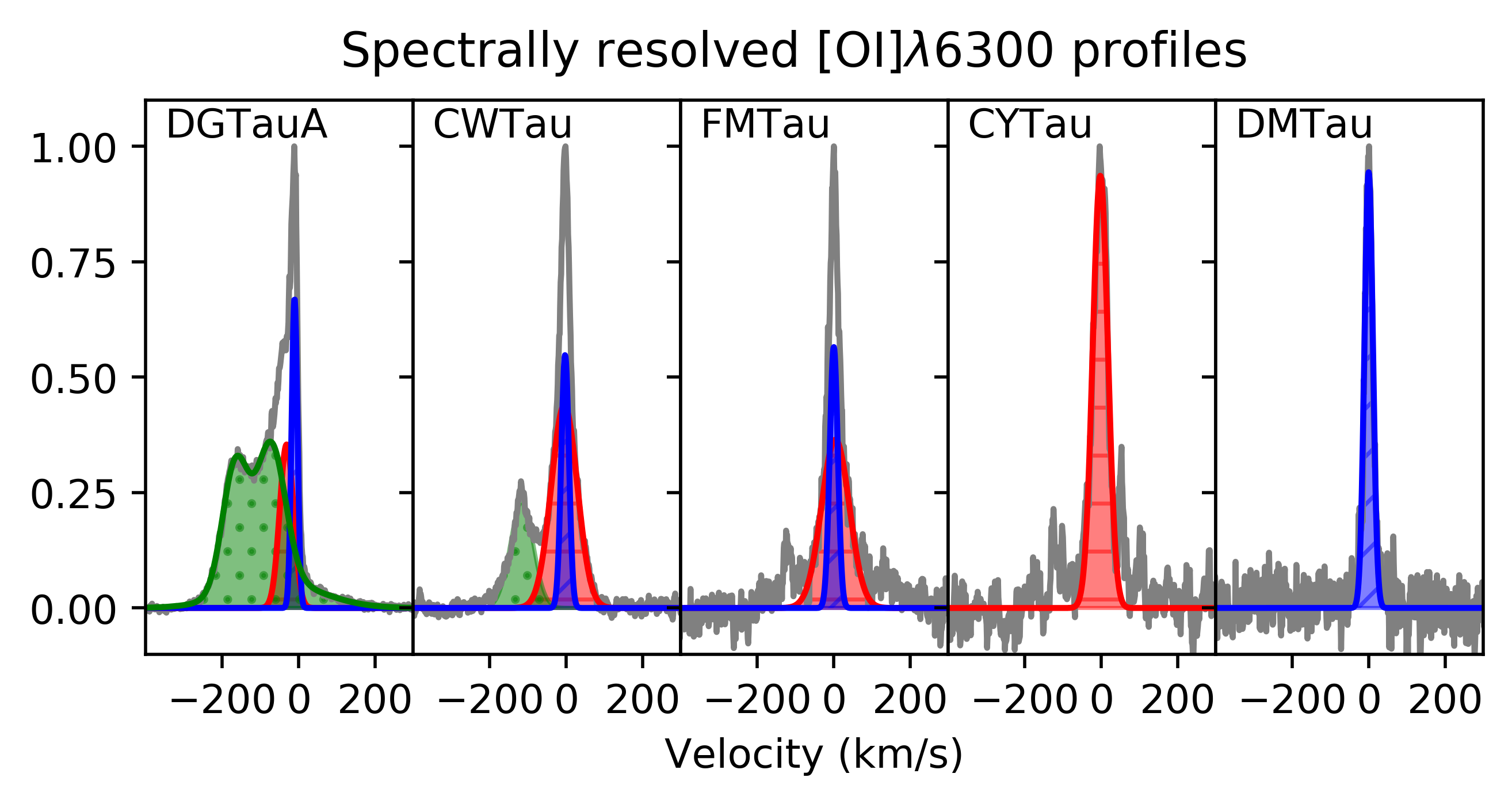}
  \caption{Sample \OIa{} profiles and kinematic decomposition into HVC (green and dots), LVC BC (red and horizontal lines), and LVC NC (blue and slanted lines). Spectra are from \citet{Banzatti2019} with kinematic decomposition following \citet{Simon2016}.} \label{fig:OIprofiles}
\end{figure*}

Since PPVI, high spectral resolution revealed that the \OIa{} LVC can either be decomposed via Gaussian fitting into two distinct kinematic components or is well described by a single Gaussian, see Fig.~\ref{fig:OIprofiles} for representative profiles. 
 We note that this decomposition is a tool to empirically describe line profiles and the identification of different kinematic components does not necessarily imply different physical mechanisms as further discussed below.
When two components are present, one is  broader and more blueshifted ($<$FWHM$>$\,$\sim 100$\,\kms\ and centroid velocity v$_{\rm c} \sim -10$\,\kms ) than the other one ($<$FWHM$>$\,$\sim 30$\,\kms\ and v$_{\rm c} \sim -3$\,\kms ), hence the naming BC and NC  \citep{Rigliaco2013,Simon2016,McGinnis2018,Fang2018,Banzatti2019}.  In those sources where \OIa{} shows both a BC and NC, a similar kinematic distinction is also seen at \SIIa{} and in the fainter \OIb{} line  \citep{Simon2016,Fang2018}. However, not all LVC show this two component structure, which is preferentially found among stars with higher accretion rates \citep{Banzatti2019}. In a survey of forbidden lines in 108 T Tauri stars in NGC 2264, \citet{McGinnis2018} found all but one had LVCs, where 27\% were BC+NC profiles, with the remaining single component profiles roughly evenly apportioned between those mimicking the properties of either the BC or the NC. 

The tendency for the BC and NC to have small blueshifts indicates they are formed in slow winds. Importantly, when the \OIa{} LVC does show a two component structure,  the BC and NC line widths,  equivalent widths, and centroid velocities correlate, clearly indicating a strong connection between what appear as two distinct kinematic features \citep{Banzatti2019}. Disk inclination, as inferred from resolved ALMA continuum images, appears to influence the observed centroid velocities of the BC, where the highest velocities (30 to 40 km/s) are found in systems with disk inclinations around $35^{\circ}$, consistent with the requisite semi-opening angle of early ideal MHD disk winds \citep[e.g.,][and \S~\ref{sect:MHD}]{BlandfordPayne1982}.  The much lower NC centroid velocities hint at a similar relation, as might be expected from the good correlation between BC and NC centroid velocities.

 The widths of both  BC and NC are also found to correlate with disk inclination. When the widths are normalized by the square root of the stellar mass, as expected if they are broadened by Keplerian rotation in the disk, then  the BC widths correspond to rotation in the inner disk ($<$ 0.5 au) and the NC widths to the outer disk, between 0.5 to 5 au \citep{McGinnis2018, Simon2016}. The inner disk wind traced by the BC is somewhat faster than the NC, reminiscent of the velocity-radius progression seen in resolved outflows. These disk winds are closely tied to accretion, as the  correlation between the \OIa{} LVC luminosity and the accretion luminosity and mass accretion rate extends to the individual NC and BC components as well \citep{Simon2016,Fang2018}. In contrast, there is only a weak correlation with the stellar luminosity and mass \citep[e.g.,][]{McGinnis2018,Nisini2018} and no correlation with X-ray luminosity  \citep{Rigliaco2013,McGinnis2018,Fang2018}. 

Mass loss estimates can be made for the BC and NC from the optically thin \OIa{} line, where its luminosity is combined with the measured centroid velocity, gas temperature, and an estimate for wind height. The temperature can be recovered from comparing \OIa{} and \SIIa{}, which have similar profiles \citep{Fang2018}, indicating they come from the same physical region. Unlike \OIa{}, which could be formed either via thermal excitation or by OH dissociation by UV photons, with very different implications for temperature, \SIIa{} can only be thermally excited \citep[e.g.,][]{Natta2014}. Thus, ratios of these two lines, plus \OIb{}, constrain the electron density and temperature of the thermally excited gas to $\sim 10^7 - 10^8$\,cm$^{-3}$ and 5,000$-$10,000\,K, respectively \citep{Fang2018}.  
Although the vertical heights are not constrained, 
assuming that they are the same as the radii  inferred from the FWHM of the BC ($\sim 0.15$\,au) and NC ($\sim 1.7$\,au), the mass loss rate in the BC is more than an order of magnitude higher than in the NC  \citep{Fang2018}. Importantly, unless the BC vertical extent is more than ten times the radius inferred from the FWHM, BC mass-loss rates are,  on average, of the same order as mass accretion rates, hence  significantly higher than the HVC/jet rates.

Evidence for a molecular component to the disk winds of the unresolved LVC sources, analogous to those seen in resolved outflows, comes from \cite{Gangi2020} who present the first high-resolution ($\Delta v <$7\,km/s) survey at optical {\it and} near-infrared wavelengths. Out of their 36 T~Tauri stars in Taurus they detect the \OIa\ line in all but one source and the H$_2$ $\nu=$1-0 S(1) at 2.12\,\micron{} in half of their sample. H$_2$ profiles are found to be less complex than the \OIa\ profiles,  and mostly single-peaked with only modest blueshifts ($< 20$\,\kms ), see also \cite{Bary2003}. Although the H$_2$ and \oi{} LVC-NC luminosities are not correlated, for sources where both lines are detected, their peak velocities are quite similar and their FWHMs correlate, pointing to a common origin. There is also a tendency for H$_2$ to be narrower than \oi{} LVC-NC, suggesting an origin at a larger disk radius.  Of the 17 sources with H$_2$ detection in \cite{Gangi2020}, seven have previous IFU high-resolution ($\sim$0.1$''$) spectro-imaging in this line, of which 5 show a wide angle morphology suggestive of a wind \citep[][and references therein]{BeckBary2019}. 
CO fundamental emission at 4.67\,\micron{} in T Tauri stars shares some properties in common to \OIa : the transition from the main isotopologue is detected in all accreting stars, profiles are predominantly, but only modestly, blueshifted, and often show a BC plus NC structure. For CO it can be shown that these two components are formed at different excitation temperatures, where the BC dominates the flux of $\nu$ = 2-1 (vibrationally hotter) and the NC of $\nu$ = 1-0 (vibrationally cooler), in accordance with the BC arising at closer stellocentric distances. At high inclinations CO profiles not only broaden but, unlike forbidden lines, begin to show central absorptions, also with small blueshifts, attributed to the line of sight grazing a low velocity wind over the disk surface. In all, the majority of T~Tauri CO profiles are interpreted as arising in a low velocity disk wind, although blueshifts above 5\,km/s are limited to a handful of sources  \citep{Bast2011A&A...527A.119B,Pontoppidan2011,Brown2013ApJ...770...94B,Banzatti2022}.
For one source, RU~Lup, this evidence is corroborated by similarly shaped spectro-astrometric signals in the LVC-NC optical forbidden lines \citep{Whelan2021ApJ...913...43W}. Interestingly, the peak velocities and astrometric offsets increase as the critical density of the transition drops (from CO fundamental, to \OIa{}, to \SIIc ) suggestive of material lifting from the disk.

The disk wind probed by the \OIa{} profiles  appears to evolve  as the inner dust disk is depleted, with  the latter traced by infrared spectral indices ($n_{3.6-8}$ and $n_{K-3.6}$, \citealt{McGinnis2018}; $n_{13-31}$, \citealt{Banzatti2019}; $n_{12-22}$, Fang et al. in prep.) which are often related to the accretion luminosity. The HVC and LVC with both BC + NC are only found in sources with optically thick inner disks.  As the disk is depleted, \cite{McGinnis2018} find that the proportion of single component profiles shifts from a higher percentage of BC to a higher percentage of NC, with the NC  persisting in optically thin disks with a dust cavity, the so-called transition disks \citep[][see also \S~\ref{sect:TransitionDisks}]{ErcolanoPascucci2017,Banzatti2019}. Additional evidence for an evolving disk wind comes from a recent high-resolution ($\Delta v \sim$10\,km/s) survey carried out with the VLT/VISIR2 spectrograph covering the  \neii{} line at 12.8\,\micron{} \citep{Pascucci2020}. This line shows an LVC similar to the optical forbidden lines in some sources \citep{Pascucci2011ApJ...736...13P,Alexander2014,Pascucci2020}. However, this new survey finds that the \neii{} LVC luminosity increases  as the dust inner disk is depleted, which is opposite to the \oi{} LVC. Moreover, the majority of full disks lack a \neii{} LVC, in spite of having an \oi\ LVC tracing an inner disk wind. These findings led \citet{Pascucci2020} to conclude that the $\sim$1\,keV hard X-ray photons needed to ionize Ne atoms (\S~\ref{sect:Freefree}) are somehow screened in full disks by inner MHD  winds that are mostly molecular, except for
a hot surface exposed to the star emitting in the \oi{} LVC. 

The nested velocity structure that is well documented in resolved outflows appears to persist when the spatial scale of the outflows has diminished and can only be traced spectroscopically.  This conclusion is based on attributing forbidden line HVC to a fast axial jet and the forbidden line widths of the (faster) broad and (slower) narrow components of the LVC to the radii at which the disk winds arise.  As in resolved jets and winds, the HVC is time variable in kinematic structure and luminosity, while the LVC is stable over decades \citep{Simon2016}. Several studies have examined whether the HVC and LVC have different excitation conditions, a question which is not yet resolved.  A multi-line analysis of forbidden lines covering a large range in excitation and ionization conditions in 7 high accretion rate sources found physical parameters, such as temperature, electron density, and ionization fraction, to vary smoothly with velocity across the forbidden lines, spanning both the HVC and LVC, suggesting similar physical conditions \citep{Giannini2019}. In contrast, a study of 48 T~Tauri stars covering a range of mass accretion rates suggests different excitation mechanisms for the HVC and LVC emission, where line ratios in the HVC indicate a shock origin in contrast to a thermal origin for the LVC \citep{Fang2018}. Furthermore, a recent spectro-astrometric study of one high accretion rate source, RU~Lup, shows different spatial extents and opposing velocity gradients in the HVC and LVC \citep{Whelan2021ApJ...913...43W}.  It thus appears more likely that the LVC and HVC  are dynamically separate flows. 

%
\begin{deluxetable}{l|c|c|c|c}
\scriptsize
\tablewidth{0pt}
\tablecaption{Relevant stellar, disk, and wind properties across evolutionary stage. \label{tab:jetwindprop}}
\renewcommand{\arraystretch}{.6}
\tablehead{
\colhead{Property} & \colhead{Class~0} & \colhead{Class~I} & \colhead{Class~II} & \colhead{Ref.}
}
\startdata
$\dot{M}_{\rm acc}$ (M$_\odot$/yr) & $\sim 10^{-5} - 10^{-6}$ & $\sim 10^{-6} - 10^{-9}$  & $\sim 10^{-6} - 10^{-12}$ & 1,2,3,4  \\
$R_{\rm gas}$ (au) & $\lesssim 50-300$ & $\lesssim 50 - 700$ &$\lesssim 20-600$ & 5,6,7,8\\
\hline
\sidehead{Jets (HVC)} \\
\hline
molecular tracers & CO, SiO, SO, H$_2$ & H$_2$ & H$_2$ & 1,9 \\
atomic tracers & \sii , \feii & H$\alpha$, \oi , \sii , \feii & H$\alpha$, \oi, \sii, \feii &  1,10\\
semi-opening angle ($^\circ$) & $\sim 1- 4$  & $\sim 1-4$& $\sim 1 - 5$ & 11,12,13,14,15  \\
$\dot{M}_{\rm jet}$  (M$_\odot$/yr) & $\sim 10^{-6} - 10^{-7}$ &  $\sim 10^{-6} - 10^{-8}$ & $\sim 10^{-7} - 10^{-10}$  & 1,16,17,18,19 \\
$\dot{M}_{\rm jet}$/$\dot{M}_{\rm acc}$ & $\sim 0.2$ &  $\sim 0.2-0.1$ & $\sim 0.1$ & 1,20,21,22 \\
\hline
\sidehead{Winds (LVC)} \\
\hline
molecular tracers & SO, SO$_2$  & H$_2$, CO  & H$_2$, CO & 1,23,24  \\
atomic tracers & unknown & \oi , \sii  & \oi , \sii & 25,26,27 \\
semi-opening angle ($^\circ$) & $\sim 20-45$  & $\sim 20-40$ & $\sim 25-35$ & 7,25,28,29,30,31 \\
$\dot{M}_{\rm wind}$ (M$_\odot$/yr) & $\sim 10^{-5} - 10^{-7}$   & $\sim 10^{-6} - 10^{-7}$ & $\sim 10^{-8} - 10^{-11}$ & 7,26,32 \\
$\dot{M}_{\rm wind}$/$\dot{M}_{\rm acc}$ & $\sim 1$ & $\sim 1$ & $\sim 0.1-1 $ & 7,26,32 \\
Specific AM\tablenotemark{*} (au km/s) & 20-100 & 40-250 & 20-300 & 7,28,29,31,33,34,35,36\\
$r_{\rm geom}$\tablenotemark{*} (au) & 20-120 & 7-40  & 5-10 &  7,28,29,31,33,34,35,37 \\
$r_{\rm MCW}$\tablenotemark{*} (au) & 0.3-40 & 0.8-32 & 0.5-3 & 7,28,29,31,33,34,35,36\\
\enddata
\tablecomments{{\bf Mass accretion rates} $-$ Class~0: eq.~2 in \citet{Lee2020ARAA} which assumes that the mechanical jet luminosity and the bolometric luminosity are generated by the release of potential energy at the protostellar radius. Class~I: near-infrared accretion-sensitive lines \citep[e.g.,][]{Fiorellino2021} and accretion-line luminosity relationships \citep[e.g.,][]{Alcala2014A&A...561A...2A}. Class~II: broadband flux-calibrated spectra that directly probe the excess emission from the accretion shock on the stellar surface (Manara et al. this volume). {\bf Gas disk radii} $-$ Class~0 and I:  modeling rotation curves from molecular lines (\citealt{Aso2015ApJ...812...27A} and \citealt{Najita2018}  for recent literature compilations and \citealt{deValon2020} for DG~Tau~B). Class~II: radius enclosing 90\% of the $^{12}$CO flux based on reconstructed images or Gaussian fits to interferometric data \citep{Sanchis2021A&A...649A..19S}. Note that the lower boundary of gas disk radii is set by the combination of spatial resolution and sensitivity adopted in the observations. See Manara et al. and Miotello et al. in this volume for more details on the Class~II disks. {\bf Jet mass loss rates} $-$ Class~0:  via molecular tracers, mainly high velocity CO millimeter emission \citep{Lee2020ARAA}. Class~I: via optical tracers (giving $\dot{M}_{\rm jet} \sim 10^{-8}- 10^{-7}$\,M$_\odot$/yr, \citealt{Hartigan1994ApJ...436..125H,Podio2011A&A...527A..13P}) and far-infrared \oi{} 63\,\micron{} emission (giving mostly $\dot{M}_{\rm jet} \sim 10^{-6}$\,M$_\odot$/yr, \citealt{Sperling2020}). Class~II: via optical tracers (HEG95). {\bf Wind semi-opening angle} $-$ Class~0: ALMA molecular imagery of HH~211 and 212. Class~I: spatially resolved $^{12}$CO rotating flows (CB26 and DG~Tau~B). Class~II: spatially resolved $^{12}$CO rotating flows (HH~30 and DG~Tau~A), \OIa{} wide-angle emission from CT~Cha, BC \OIa{} centroid velocities vs. disk inclination. {\bf Wind mass loss rates} $-$ Class~0 and I: same diagnostics and same sources for which the wind opening angle could be measured. Class~II: spatially resolved $^{12}$CO emission for HH~30 and LVC \OIa{} emission for many sources in $\sim 1-3$\,Myr-old star-forming regions (see \S~\ref{sect:SpatiallyUnresolvedFlows} for assumptions). The densities and temperatures probed by some of these diagnostics are discussed in \S~\ref{sect:thermal} and summarized in Fig.~\ref{fig:n-t-xe-lines}. }
\tablenotetext{*}{The specific Angular Momentum (AM), and the geometric ($r_{\rm geom}$) and cold MHD wind launching radii ($r_{\rm MCW}$) are  further discussed in  \S~\ref{sect:rotation_launch} and are only available for a few spatially resolved winds, see \S~\ref{sect:SlowMolecularWinds} for the Class~0/I and HH~30 and DG~Tau~A for the Class~II.}
\tablerefs{
(1) \citealt{Lee2020ARAA}; 
(2) \citealt{Fiorellino2021}; 
(3) \citealt{Alcala2017A&A...600A..20A};
(4) \citealt{Manara2017A&A...604A.127M};
(5) \citealt{Aso2015ApJ...812...27A};
(6) \citealt{Najita2018};
(7) \citealt{deValon2020};
(8) \citealt{Sanchis2021A&A...649A..19S};
(9) \citealt{Frank2014};
(10) \citealt{Bally2016ARA&A..54..491B};
(11) \citealt{Lee2017NatAs...1E.152L};
(12) \citealt{Lee2018NatCo...9.4636L};
(13) \citealt{Erkal2021ApJ...919...23E};
(14) \citealt{Erkal2021A&A...650A..46E};
(15) \citealt{Raga2001A&A...367..959R};
(16) \citealt{Hartigan1994ApJ...436..125H};
(17) \citealt{Sperling2020A&A...642A.216S};
(18) \citealt{HEG1995};
(19) \citealt{Podio2011A&A...527A..13P};
(20) \citealt{Ellerbroek2013A&A...551A...5E};
(21) \citealt{Sperling2021A&A...650A.173S};
(22) \citealt{Nisini2018};
(23) \citealt{Gudel2018};
(24) \citealt{Beck2008};
(25) Haffert et al. in prep.;
(26) \citealt{Fang2018};
(27) \citealt{Whelan2021ApJ...913...43W};
(28) \citealt{Lee2018ApJ...863...94L};
(29) \citealt{Tabone2017};
(30) \citealt{Padgett1999AJ....117.1490P};
(31) \citealt{Louvet2018};
(32) \citealt{Tabone2020A&A...640A..82T};
(33) \citealt{Bjerkeli2016};
(34) \citealt{Launhardt2009};
(35) \citealt{Zhang2018};
(36) \citealt{Pesenti2004};
(37) \citealt{2014A&A...565A.110M}
}
\normalsize
\end{deluxetable}

\subsection{\textbf{Empirical evolution}}\label{sect:empiricalevolution}
Spatially resolved outflows in Class 0, I, and II sources share many features in common, including both collimated jets and wider winds, implying that the accretion/ejection mechanisms are similar over a wide range of ages and accretion rates. 
We summarize the key outflow diagnostics and properties described in the previous two sections in Table~\ref{tab:jetwindprop}, including primary tracers, jet and wind opening angles, ejection rates, and  ejection to accretion ratios. We also include the range of mass accretion rates ($\dot{M}_{\rm acc}$) and gas disk radii  ($R_{\rm gas}$) as they are directly relevant to testing classic viscous disk models against wind-driven accretion models (\S~\ref{sect:EvolutionMassEjection} and \ref{sect:GasDiskRadius}). There is a trend for mass accretion rates to decline as the system evolves, both through Class 0 and Class I stages and also through the wide range of ages characterizing the  Class II sources, $\sim 10^{5.5} - 10^7$ yr. 
The Class II $\dot{M}_{\rm acc}$ encompass at the high end those sources with full dust disks with robust spatially resolved outflows to, at the low end, sources where the disk is dissipating, the accretion rate is dropping, and evidence for a jet/HVC disappears; yet a blueshifted LVC indicates a disk wind persists.

The behavior of $R_{\rm gas}$ with evolutionary stage is less clear. 
A tendency for the  maximum value of $R_{\rm gas}$ to increase going from Class I to Class II disks was reported by \citet{Najita2018} based on disk radii compiled from the literature. 
For the Class~II statistics in Table~\ref{tab:jetwindprop}, we relied solely on a recent ALMA study based on $^{12}$CO data, analyzed in a homogeneous way \citep{Sanchis2021A&A...649A..19S}, and expanded the sample of Class~0 and I sources. With this compilation, there is instead a suggestion that the Class 0 sources have maximum $R_{\rm gas}$ smaller than both Class I and II.  Thus at present, it is unclear if the maximum value of $R_{\rm gas}$ increases over time, as expected for viscous spreading,  given the wide range of $R_{\rm gas}$ at all stages, the differing diagnostics and approaches for the Class0/I disks, and a number of unresolved disks that are not included in these statistics.

When jets are resolved, their morphology is strikingly similar throughout the three evolutionary stages, including opening angles and knots of shocked gas advancing along the flow with three characteristic periods, a few years, a few 10 years, and a few hundred years \citep{Cabrit2002EAS.....3..147C,Frank2014,Lee2020ARAA}. Not only does the jet morphology persists over a wide range of ages and accretion rates, but the ratio $\dot{M}_{\rm jet}$/$\dot{M}_{\rm acc}$ $\sim$ 0.1 appears to be constant as the system evolves, with the jet ejection rate dropping in tandem with the stellar mass accretion rate (see also Fig.~7 in \citealt{Sperling2021A&A...650A.173S}).

What is new since PPVI is the growing evidence that a slow disk wind, not the jet, is the dominant source of mass outflow. Although the number of systems imaged with ALMA to date is limited, small scale, rotating cones of outflowing molecular gas are found to surround the narrow jet at all evolutionary stages. Their velocities are  $\le$ 10 km/s and their semi-opening angles of $\sim 20^\circ-45^\circ$ are significantly larger than those of jets and remain the same as the source evolves. Their base radii, when estimated from geometric projections, are of the order of tens of au (MHD launch radii are slightly smaller, see \S~\ref{sect:SpatiallyResolvedWindDiagnostics}). 
In a few cases, the CO cones appear to be entirely one-sided, although the disk is symmetric and the atomic high-velocity jet is bipolar, e.g.  DG~Tau~B (Class~I) and HH30 (Class~II).
When both jet and molecular wind ejection rates can be estimated, $\dot{M}_{\rm wind}$ are 50x $\dot{M}_{\rm jet}$ and $\dot{M}_{\rm wind}$/$\dot{M}_{\rm acc}$ are on the order of unity. If this turns out to be typical, the conclusion is that there is significantly more mass ejected in the slow winds than in the jets and wind mass loss rates may be sufficient to drive disk evolution.

The behavior of disk winds in spatially unresolved outflows, which constitute the majority of Class II sources, is traced by  LVC forbidden line emission. As described in \S~\ref{sect:ClassII}, mass loss rates can be estimated from these lines but strongly depend on the assumed gas temperature and wind geometry, especially the vertical extension of the emission \citep{Natta2014,Fang2018}. Even with these caveats, the mass loss rates for the LVC BC are higher than for the LVC NC  and also  higher than for the HVC \citep{Fang2018}.  Pinning down the wind geometry in forbidden line emission is necessary to reduce the uncertainty  of $\dot{M}_{\rm wind}$/$\dot{M}_{\rm acc}$ ($\sim 0.1-1$) estimated from optical tracers in the Class~II stage. In any case, if Class~II winds are mostly molecular \citep[e.g.,][]{Pascucci2020}, these optically inferred values will be lower limits. 

Evidence for a transformation in the character of jets and winds comes during the later Class II phase when the mass accretion rate falls below $\sim 10^{-9}$\,M$_\odot$/yr and infrared spectral indices indicate disk dissipation. The general trend, although with a lot of scatter,  is that first the terminal velocity in the \OIa\ HVC drops, then the HVC becomes undetectable but LVC BC and NC emission from the disk wind is still present \citep{Simon2016,Banzatti2019}. As the inner disk continues to thin out, the BC disappears but the NC persists, sometimes no longer showing a clear blueshift at \oi\ \citep{ErcolanoPascucci2017,McGinnis2018}. However, in the lowest accretors a narrow and blueshifted \neii{} feature appears, with Ne atoms presumably ionized by hard X-rays  no longer screened by a dense inner disk wind, and the luminosity of this feature increases as the inner disk is dissipated, in contrast to the \oi{} LVC-NC \citep{Pascucci2020}. This latter phase includes transition disks with large inner dust holes such as T~Cha and V4046~Sgr.

\section{\textbf{\uppercase{The theory of disk winds}}}\label{sect:theory}

This section provides a thorough overview of MHD disk winds as well as PE winds.
To bridge the gap between theorists and observers, we adopt a pedagogical approach and focus on the salient physics and assumptions that these models are built upon. We discuss model predictions that can be tested by observations in \S~\ref{sect:link}. 

The most significant theoretical insights since PPVI  have been achieved by advanced computational methods.
It has now become possible to couple gas dynamics with magnetic fields, thermal energy transport, and chemistry in global simulations of the entire disk thus moving toward a more complete view of disk evolution.
Physical conditions in the outflowing gas, such as the density, temperature, and ionization fraction are central to the development of a comprehensive theory. Along with an understanding of the chemical state and abundances of tracer species, they are essential to interpreting and making inferences from the observations summarized in \S~\ref{sect:obs}.  

We first begin with a basic overview of gas heating and ionization, especially as it pertains to observed line emission (\S~\ref{sect:thermal}). Next, we discuss MHD winds along with their launching and mass loading mechanisms, describe recent global disk simulations, and briefly summarize different wind types in the inner ($\leq$\,0.5\,au) disk, see \S~\ref{sect:MHD}. This is followed by a summary of our current understanding of how disks evolve, covering a description of the evolution of the magnetic field (\S~\ref{sect:EvolutionPoloidalFlux}),  PE winds and disk dispersal at the final epoch (\S~\ref{sect:photoevaporation}), and the expected evolution of mass ejection and accretion (\S~\ref{sect:EvolutionMassEjection}). 

\subsection{Thermal and ionization state of outflowing gas}\label{sect:thermal}
The local flux of stellar high energy photons determines the chemical state in  wind regions and, to some extent, the thermal energy balance. 
The flux is generally characterized as Far UltraViolet (FUV; $6\,$eV$<h\nu<13.6\,$eV),  Extreme Ultraviolet (EUV; $13.6\,\rm{eV} <h\nu<100\,\rm{eV}$), and X-ray ($h\nu\gtrsim 100\,$eV). Terms  such as XEUV and XEFUV thus refer to a combination of flux bands, more specifically X-ray + EUV for the former and X-ray + EUV + FUV for the latter. Of these bands, hydrogen-ionizing EUV photons penetrate the least into the wind (or disk) material, up to column densities $\lesssim 10^{19-20}\,$cm$^{-2}$ before they are absorbed, and heat gas to temperatures $\sim 8,000-12,000\,$K. For young, accreting, and chromospherically active stars, X-ray photons 
range in energy from $0.1-10\,$keV. They have an absorption cross-section $\sigma_X \sim 10^{-22} (E/1\,{\rm keV})^{-2.6}\,$cm$^{-2}$ \citep{wilms2000} and their penetration depth and heating ability via the Auger effect 
on metals depends on photon energy. Ionization determines heating; ejected primary photoelectrons have abundant kinetic energy which can lead to further (secondary) ionization and eventually convert the excess kinetic energy to gas thermal energy via collisions.  Soft X-rays at the low end of the spectrum can ionize at levels of $\sim$ 10\% and heat gas to $\gtrsim 10,000\,$K; however they are readily absorbed in atomic gas at column densities of $\sim 10^{20}\,$cm$^{-2}$. Harder 1 keV X-rays penetrate deeper to $\sim 10^{22}\,$cm$^{-2}$ and their ability to heat gas to high temperatures depends on the chemical state of the gas \citep{maloney1996}. Atomic gas can attain a few thousand K, but molecular gas cools efficiently due to higher available degrees of freedom and leads to lower gas temperatures. 

At column densities $\gtrsim 10^{21}\,$cm$^{-2}$, FUV photons can dominate heating although FUV heated gas seldom exceeds $\sim 5,000\,$K. FUV photons can ionize gas up to levels of ion abundances $\sim 3 \times 10^{-4}$  relative to H nuclei, 
depending on metallicity, but heating is not as strongly tied to ionization as in the case of EUV and X-rays.  FUV heating can occur by several gas photoprocesses as well as photoelectric ejection of electrons from small dust grains, including polycyclic aromatic hydrocarbons (PAHs). Due to the latter, FUV heating is strongly coupled to the dust content in winds; high amounts of small dust can heat gas more efficiently but also severely attenuate the FUV flux and decrease its penetration depth into the wind. Depending on the dust absorption cross-section per H nucleus in the wind, determined by the dust/gas mass ratio, dust size distribution and material properties, FUV photons can penetrate up to column densities of $\sim 10^{24}\,$cm$^{-2}$, at which point gas opacity gains in significance. Lyman-$\alpha$ excited H in the $n=2$ level, C, Fe, Si, S, Mg atoms, the common molecules H$_2$, CO, OH and H$_2$O, and many other abundant gaseous species have photoionization  and photodissociation thresholds in the FUV energy range ($6-13.6\,$eV) and excess energy from the photo-processes are a source of heating even in the absence of dust. FUV pumping, where H$_2$ (or other species) absorbs a photon and  collisionally de-excites in high density gas, can be an additional significant source of heating \citep{TielensHoll1985}. In low density gas, pumping may lead to radiative de-excitation \citep[see][for \oi\ emission produced by UV pumping]{Nemer2020}.

While order-of-magnitude estimates were introduced above, the gas temperature is a key thermodynamic variable that needs to be accurately determined when solving for pressure gradients and wind flow properties, or when interpreting line emission. The gas temperature is determined by a complex balance between various heating processes, and cooling mechanisms like collisions with dust grains and line emission from ions, atoms and molecules (whose abundances in turn depend on chemical processes). At the surface, where winds are launched, grain photoelectric heating by FUV photons, formation of molecular hydrogen whereby some of the binding energy is converted to heat, and excess electron energy from photo-dissociation  (by FUV) and from photo-ionization (FUV, EUV and X-rays) all heat the gas. Gas in magnetized jets and winds can also be heated by dissipative non-ideal MHD processes: Ohmic heating due to electron collisions with neutrals and ions, and ambipolar diffusion heating due to ion-neutral drag,
in particular, can be dominant 
 in both atomic and molecular MHD disk winds, at altitudes where photo-heating is inefficient \citep[e.g.][]{Garcia2001A&A...377..589G,Panoglou2012,Wang2019}.
Adiabatic terms --- expansion cooling in winds and compression heating in jet collimation regions --- can become significant if dynamical timescales are short relative to heating and cooling timescales. 

\begin{figure}[h]
    \center\includegraphics[width=1.1\linewidth]{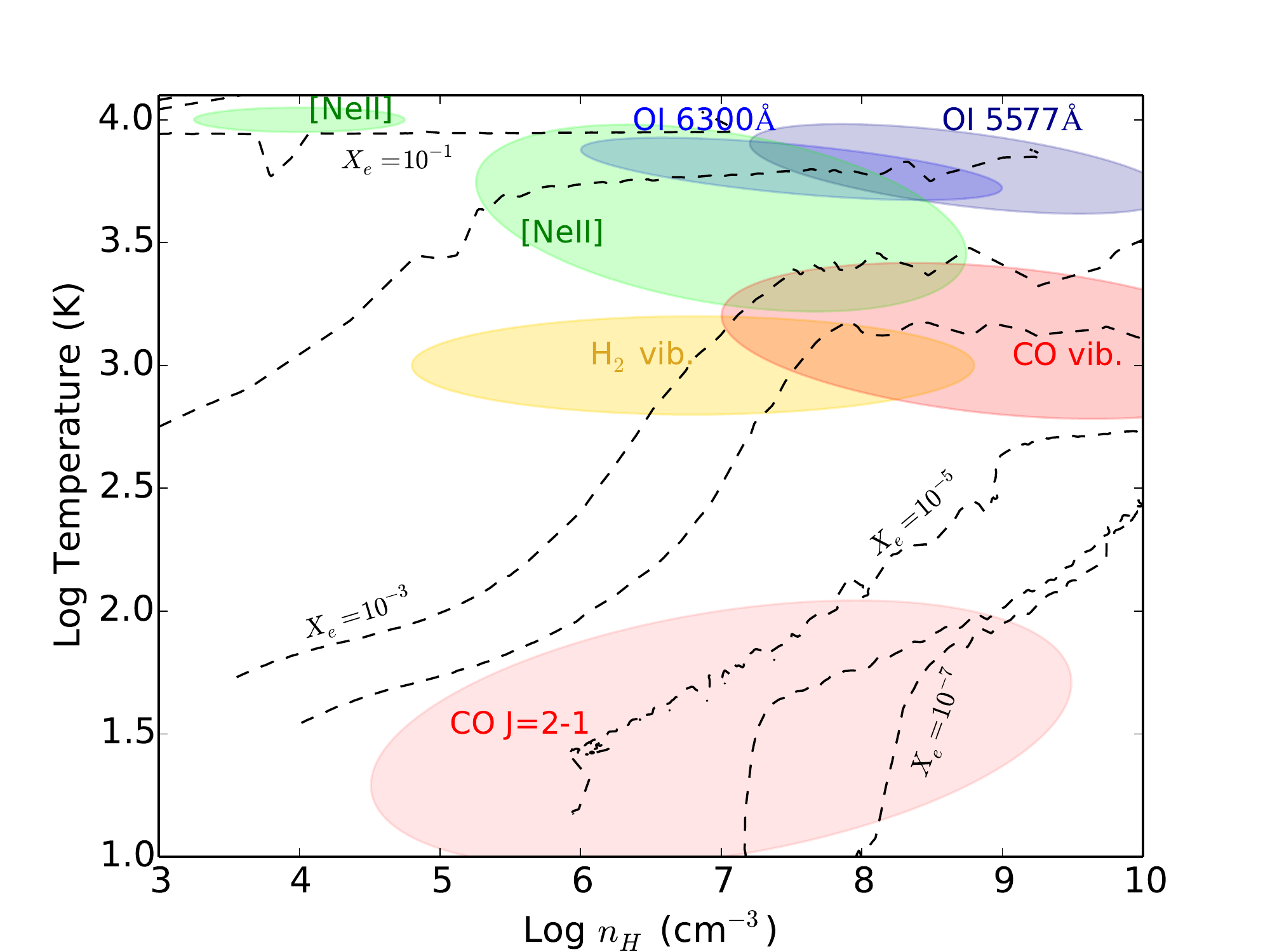}
    \caption{Regions in number density-temperature space where emission lines originate as calculated from a library of disk atmosphere models where heating is mainly due to high energy photons (magnetic heating is not included). Dashed lines illustrate as an example the electron abundance contours for one such model; this will in general vary depending on stellar and disk parameters. Optical forbidden lines (\oi{} lines are shown) need both hot and dense gas while \neii\ can originate both from low density but hot and fully ionized gas (by EUV) and also from partially ionized, cooler (X-ray ionization) regions. H$_2$ vibrational emission is restricted to a very narrow range of temperature, $\sim 3000$K; at lower temperatures the $2.12\mu$m line is not excited while at higher temperatures H$_2$ gets collisionally dissociated. CO vibrational emission can originate in gas where H$_2$ is absent. CO rotational lines probe a range of densities and typically arise in cooler regions. } \label{fig:n-t-xe-lines}
\end{figure}

The chemical state of the gas--- whether ionized, atomic or molecular---affects line cooling/emission and sets the gas temperature. Apart from the requirement that densities and temperatures be high enough for collisional excitation,  photoionization/photodissociation constraints need to be met for an emission line to be observed. Fig.~\ref{fig:n-t-xe-lines} delineates the emission regions of commonly observed wind diagnostic lines discussed in \S~\ref{sect:obs} on a H nuclei number density $-$ temperature plot. The ellipses indicate approximate locations drawn from a large library of disk atmospheric thermochemical calculations \citep{Gorti2009,Gorti2015}, and the electron abundance (often similar to the ionization fraction) is indicated for one such model. Note that the electron abundance $X(e^-)$ contours can vary depending on the high energy radiation spectrum and other disk and stellar parameters.  Magnetic dissipative heating is not considered in these models and  heating is primarily radiative therefore low gas densities typically result in greater penetration of high-energy photons and temperatures tend to be high. The forbidden lines of [OI] need both hot and dense gas \citep[see, e.g.,][]{Fang2018} for excitation. [NeII] can originate both from fully ionized gas (EUV-ionization) and from partially ionized gas (X-ray ionization), see \cite{Pascucci2014} and \S~\ref{sect:Freefree} for an approach on how to distinguish between the two ionization mechanisms.  
There is a notable paucity of emission line diagnostics in  low temperature and low density regions.
This is because atomic gas in the few 100 -- few 1,000\,K temperature range can be quite unstable due the absence of strong coolants (except for the low energy \oi{} and [\cii ] far-infrared fine structure lines)
and even a marginal  increase in heating can result in $\sim 5,000$\,K gas where optical forbidden lines do most of the cooling.  Lyman alpha and 2-photon cooling from H\,{\scriptsize I} prevent the gas from becoming hotter than $\sim 10,000$\,K \citep[see e.g.,][]{Draine2011}. 

Molecular cooling can result in gas temperatures of $\sim 1,000-3,000\,$K provided adequate heating mechanisms operate at the high densities ($n_H \gtrsim 10^7$ cm$^{-3}$), and more importantly, provided molecules are retained. Advection and non-equilibrium chemistry can maintain significant fractions of H$_2$ in dense, irradiated winds \citep{Panoglou2012}, while gas-phase chemical reactions (by H$^-$ for example) can form trace amounts of H$_2$ in lower density, atomic winds where dust may not be efficiently entrained \citep[e.g.,][]{Tabone2020A&A...636A..60T}. Even low abundances of H$_2$ open formation pathways to molecules such as CO which can survive in the wind \citep{Glassgold2004, Tabone2020A&A...636A..60T}. Both CO and H$_2$ can get collisionally dissociated at high temperatures, which sets an upper limit of $\sim 3,000\,$K on gas temperatures in predominantly molecular winds.  Cooler molecular gas ($\lesssim 500\,$K) is readily observed using pure rotational lines of CO and its isotopologues;  the CO rotational ladder extends from sub-millimeter to infrared wavelengths and can probe a wide range of conditions.

\subsection{\textbf{From idealized to realistic MHD disk winds}}\label{sect:MHD}
The term `MHD disk wind' originally refers to flows driven primarily by magneto-centrifugal processes  \cite[e.g.,][]{BlandfordPayne1982, PelletierPudritz1992, Ferreira1997}. In the simple ``bead on a rigid wire'' analogy \citep{1971MNRAS.152..323H}, material forced to corotate with the poloidal field is flung out when the field inclination angle from the polar axis $\theta > 30\degr$ (see the effective gravitational + centrifugal potential surfaces in Fig.~1 of  \citealt{BlandfordPayne1982}). In reality, the field is not strictly rigid but wound up by disk rotation; a vertical gradient of the toroidal $B_\phi$ component arises, which exerts both a braking torque inside the disk, leading to accretion, and an accelerating torque at the disk surface, leading to ejection \citep[see e.g.,][]{Wardle1993,FerreiraPelletier1995A&A...295..807F,
Ferreira1997}.
The transfer of angular momentum from the disk to the wind, mediated by $B_\phi$, is thus what ultimately powers the MHD disk wind. 
The toroidal $B_\phi$ component also produces a gradual wind recollimation \citep{BlandfordPayne1982}
with a jet-like density enhancement towards the polar axis on large scales \citep{Shu1995ApJ...455L.155S, Cabrit1999A&A...343L..61C}. 

Early global models solving for the disk vertical structure were very idealized (self-similar and steady), but allowed to explore a wide parameter space and establish key aspects of steady wind physics \citep[e.g.,][]{Ferreira1997}. The role of thermal effects was already investigated by \citet{CasseFerreira2000A&A...361.1178C} who showed that surface heating dramatically enhances the vertical mass flux and lowers the flow speed. 
Thus, warm winds are both denser and slower than cold winds, see \S~\ref{sect:massloading} for more details and recent results. 
A limitation of these early models was, however, the assumption of a strong magnetic field in the disk midplane, with the ratio of thermal to magnetic pressure $\beta(z) = P(z)/(B^2/8\pi)$ obeying $\beta_{\rm midpl} \equiv \beta(z\!=\!0) \simeq 1$ and the midplane $B$  taken as the global vertical field $B_z$ threading the disk (excluding any turbulent component).
This induces midplane accretion at a sizable fraction of the sonic speed, and results in disk surface densities much smaller than the  minimum mass solar nebula (MMSN) for typical accretion rates \citep{Combet2008A&A...479..481C}. Although it provides an attractive explanation for inner holes in accreting transition disks (\S~\ref{sect:TransitionDisks}), such a property appears inconsistent with planet formation and with disk lifetimes at larger radii if $\beta_{\rm midpl} \simeq 1$
 were to hold across the whole disk.

Since PPVI, a new generation of more realistic ``low magnetization" models has emerged, considering denser disks comparable to the MMSN with $\beta_{\rm midpl} \gg 1$. They reveal that an MHD wind is still ejected from the magnetically dominated upper layers where $\beta \lesssim 1$. Hence MHD disk winds are now established as a robust outcome of an organized poloidal field in the disk \citep[e.g.,][]{2019MNRAS.490.3112J,Lesur2021A&A...650A..35L}.
We will review below the wind launching (\S~\ref{sect:launching}) and mass-loading (\S~\ref{sect:massloading}) processes, with special emphasis on two families of models that have received particular attention since PPVI. The first type considers the ``dead-zone" (typically around 1-50 au), where ionization is so low that MRI turbulence is quenched by non-ideal effects, and only a surface layer remains sufficiently ionized by stellar FUV irradiation to drive MHD ejection. Wind properties thus depend on thermo-chemistry as well as non-ideal diffusion coefficients. The second type considers well-ionized disk regions where ideal-MHD applies and MRI operates in the disk interior. Turbulent pressure forces then also contribute to wind launching, generate fluctuating ejection, and aid mass-loading. Both types predict slow disk winds, 
but with different ratios of ejection to accretion rates (\S~\ref{sect:massloading}).

A fundamental aspect of the physics of MHD disk winds
is that the disk accretion rate they induce 
is mostly determined by the magnetic field strength threading the disk. This occurs because, ultimately, the accretion rate is regulated by the braking torque exerted on the disk  by the wound up magnetic field lines. This fact can be expressed as a simple scaling, common to both early ``high magnetization" and new "low magnetization" models \citep[e.g.][]{FerreiraPelletier1995A&A...295..807F,Wardle2007,Bai2016,Hasegawa2017}
    \begin{eqnarray}
        \dot{M}_{\rm acc} (R) \simeq \frac{2R}{\Omega} (B_z B_\phi) 
        \label{eq:bfield-macc}
    \end{eqnarray}
where $\dot{M}_{\rm acc}(R)$ is the mass accretion rate at the radial distance $R$, which differs from the value onto the star, $\dot{M}_{\rm acc}$, used elsewhere in the chapter, and  $B_\phi$ is the toroidal field at the wind base. From  this equation we can obtain an order of magnitude value for the mean vertical magnetic field $B_z$ required to induce disk accretion solely via an MHD disk wind  
    \begin{equation}
    \begin{split}
        B_z \simeq 100 mG \left(\frac{10B_z}{B_\phi}\right)^{1/2}
        \left(\frac{\dot{M}_{\rm acc}(R)}{10^{-7} M_\odot {\rm yr}^{-1}}\right)^{1/2} \\
        \left(\frac{R}{\rm 1 au}\right)^{-5/4} 
        \left(\frac{M_\star}{1 M_\odot}\right)^{1/4}.
        \label{eq:bfield-macc-two}
    \end{split}    
    \end{equation}
this expression highlights that new ``low magnetization" models in the dead zone have a dominant toroidal component at the wind base (typically $B_\phi \sim 10 B_z$), hence they require a  $B_z$ that is $\sim 3$ times lower than early ``high magnetization" models  where $B_\phi \sim B_z$  \citep[see e.g., eq.~5 in][]{Garcia2001A&A...377..589G}. 
A more precise empirical scaling for disk winds ejected from the dead-zone can be found in eq.~18 of \citet{Lesur2021A&A...650A..35L}. 
Direct measurements of the magnetic field strength in protoplanetary disks by Zeeman effect 
remain extremely challenging even with ALMA and only  upper limits in one disk have been reported so far \citep[][]{Vlemmings2019A&A...624L...7V} with values compatible with the above scaling. Nevertheless, non-ideal MHD simulations of protostellar collapse that include ambipolar diffusion suggest that values of order 100 mG can be easily advected in the disk at the Class 0 stage \citep{Masson2016A&A...587A..32M}. 
Paleomagnetic measurements in solar system meteorites indicate organized magnetic fields of $\sim 500$\,mG ($\pm 200$\,mG) at $\sim 1-3$\,au until $\sim 2$\,Myr from the formation of the first solids \citep{2014Sci...346.1089F,2021SciA....7.5967W} suggesting that, at least in the case of the Solar System, sufficient magnetic flux was retained at later stages. As shown in eq.~\ref{eq:bfield-macc}, the accretion rate, and therefore the wind ejection rate, scales with the field strength. Hence, the long-term radial transport (inward by advection, and outward by diffusion) of the entrained $B_z$ is critical to how disks evolve (see \S~\ref{sect:EvolutionPoloidalFlux}).

\subsubsection{\it{\textbf{Launching Mechanisms}}}
\label{sect:launching}
The detailed dynamics of the disk atmosphere leading to wind ejection
depend on the level of MRI-induced turbulence which, in turn, depends on the ionization state of the gas, itself determined by the local disk temperature and surface density. In the inner disk, $r\lesssim 0.3$\,au, thermal ionization of alkali metals is expected to make the disk fully MRI active \citep{Desch2015} while, beyond, X-ray and FUV ionization sufficiently ionize only the surface $N\sim 10^{22-23}$ cm$^{-2}$ layer (see \S \ref{sect:thermal}). 
In typical disks, the surface density declines with radial distance and this can result in MRI-active outer disks because column densities are low enough for
ambient radiation fields and cosmic rays
to enhance the ionization fraction at the midplane. Importantly, in the planet-forming region, within $\sim 1-30$\,au, the disk is expected to be only weakly turbulent due to insufficient ionization and non-ideal MHD effects such as Ohmic resistivity and ambipolar diffusion \citep[see e.g.,][and Lesur et al. chapter for more details]{Turner2014}.

While disk accretion in the MRI-active regions has been traditionally attributed to radial outward transport of angular momentum (e.g., Hawley et al. 1995, Sorathia et al. 2010), recent work has shown that fluctuating winds can also develop in these regions. The local 3D ideal-MHD simulations by \cite{SuzukiInutsuka2009} were the first to find both turbulence-induced radial transport and structured disk winds caused by the breakup of linear MRI ``channel modes'', which are restricted to regions where ideal MHD is a good approximation (i.e., near the star and in sufficiently ionized layers). As noted in \citet{Lesur2013A&A...550A..61L}  and
in the PPVI chapter by \citet{Turner2014}, 
there is a continuity of behaviors between magnetocentrifugal winds \citep[MCW alas e.g.,][]{2012MNRAS.423.1318O} and  stratified MRI channel modes which are near-exact non-linear solutions of the MHD equations in the limit of weak magnetic fields \citep{2010MNRAS.406..848L}. 
The more recent global, ideal-MHD, and net-vertical-field accretion disk simulations of \cite{2018ApJ...857...34Z} 
 identify a turbulent accretion layer at high altitudes ($z \simeq R$) causing an inward pinch of field lines which is responsible for launching these fluctuating winds. This effect should also be relevant in the surface layer of protoplanetary disks and has been recently confirmed in full 3D global MHD simulations by  \citet{2021A&A...647A.192J}. 
 The latter authors also find that, when fluctuations are averaged over several orbital timescales, the wind obeys steady characteristics. 
 
It thus appears that the main launching mechanism of steady MCWs  is 
largely agnostic to the basic laminar or turbulent state of the gas and only depends on the presence of suitably inclined magnetic field lines \citep[cf.,][]{Lesur2013A&A...550A..61L}. Therefore, in principle,  MCWs can be launched from all disk radii.  
For the centrifugal acceleration to work, magnetic field lines that have their footpoints anchored at some height in the disk can be somewhat compared to rigid wires. 
The specific angular momentum carried by the wind is increased from the disk specific angular momentum by a factor $\lambda \simeq (R_A/R_0)^2 > 1$ (see \citealt{BlandfordPayne1982}) where $R_0$ is the footpoint radius and $R_A$ is the Alfv{\'e}n point radius (where the flow poloidal speed reaches the propagation speed of Alf{\'e}n waves in the poloidal plane). This is the same result as if the wind were forced to strictly corotate with a rigid wire anchored at $R_0$ up to radius $R_A$. Hence the ratio $(R_A/R_0) > 1$ acts like a mechanical ``lever arm", and $\lambda$ is called the ``magnetic lever arm parameter". 
For a significant magnetic  lever arm parameter  ($\lambda \ge 2$), magnetic \emph{tension} forces drive the acceleration \citep[as, e.g., found in][]{Gressel2020}, causing super-rotation with respect to the local Keplerian speed, and one speaks of a \emph{true} MCW.
In the opposite case of small $\lambda < 2$,  the flow is primarily driven by vertical gradients of the magnetic pressure (dominated by the strongly wound-up azimuthal component) and one speaks of a magnetic pressure--driven MHD wind 
\citep[see][where the term ``magneto-thermal wind" was used]{Bai2017}. 
This type of wind, sometimes referred to as a magnetic ``tower'', can even remain sub-Keplerian in its rotation. While this distinction at first glance may appear as a dichotomy, it really is a continuum with respect to the extent of the magnetic lever arm 
\citep{2019MNRAS.490.3112J,Lesur2021A&A...650A..35L}. We therefore subsume both these limiting cases (i.e., tension-force versus pressure dominated) under the common MCW class. 

Beyond $R_A$, magnetic forces gradually become negligible. If the contribution of thermal and turbulent pressure gradients to the total energy budget can be neglected  (``cold wind" regime), the asymptotic flow speed is given by the Bernouilli invariant as \citep{BlandfordPayne1982}
\begin{equation}
    V_p^\infty \simeq \sqrt{2 \lambda - 3} \quad V_{\rm Kep} (R_0).
    \label{eq:vp-lambda}
\end{equation}
While early high magnetization models could only reach small values of $\lambda$ with strong additional heating \citep{CasseFerreira2000A&A...361.1178C}, new models of winds from weakly magnetized disks have intrinsically small values of $\lambda \simeq 1.5-5$ even in the `cold' regime, due to their strongly wound-up field \citep[e.g.,][]{2021A&A...647A.192J,Lesur2021A&A...650A..35L}. 
They are therefore slower than early cold wind models from strongly magnetized disks.

As we will discuss in the next subsection, the extent of the magnetic lever arm  is intricately coupled to the loading of material onto the field lines,  
so as to remove a certain amount of angular momentum from the disk (eq.~\ref{eq:lambda_xi}).
This obviously has important implications for the overall mass budget of the disk's accretion/ejection flow and potentially its dust content.  

\subsubsection{{\it{\textbf{Mass loading mechanisms and wind mass-flux}}}}
\label{sect:massloading}
The overall mass loss rate is primarily determined by the density, $\rho$ at a ``wind base'', which is located  around the slow magnetosonic point \citep{Bai2016}. The mass flux (per logarithmic radius) onto the wind is then given by
\begin{equation}
  \frac{1}{2\pi} \frac{{\rm d} \dot{M}_{\rm wind}}{{\rm d}\log R} \approx \left. 2 R^2 \rho\, c_{\rm s} \frac{B_z}{B_R}\right|_{\rm slow}\,,
\end{equation}
where $B_z$ is the vertical field,  $B_R$ is the local radial component of the field (which defines the field-line inclination angle), and $c_s$ the sound speed \citep{1995MNRAS.275..244L,2011ppcd.book..283K}. As discussed in \citet{Gressel2020}, the bending of field-lines in non-ideal MHD simulations is at least partly due  to ambipolar diffusion. As such,  mass loading  depends  not only on the thermal (and turbulent) pressure gradient near the footpoint, but also on the disk ionization. 

Because only charged species are subject to the Lorentz force, sufficient ionization is required at and above the wind base to drive MCWs. 
Although the ionization requirement for MCWs is in principle the same as that for driving MRI turbulence, the latter is active only in  layers where $\beta(z) \gtrsim 1$. 
 In contrast, MCWs are present where the gas dynamics is dominated by the coherent magnetic field (i.e., $\beta(z)<1$), such that the two phenomena may, in a sense, appear mutually exclusive
 at a given altitude in the disk 
 --- but see \S\ref{sect:launching} for the possibility of an MRI-turbulent layer at high altitude with $\beta(z)<1$ when $(B_\phi/B_z) \le 1$.

At typical disk radii that harbor dead zones in their interior, the altitude of the wind base is determined by FUV ionization, in a region where the density is sufficiently low  to ensure $\beta <1$. 2D global simulations covering from $1$\,au to $\sim 20$\,au show that the wind base is located at $\gtrsim 5 H$, where $H$ is the local pressure scale height \citep{Bai2017,Bethune2017,Gressel2020}. Since the density at the wind base is much lower than the density at the midplane, the magnetic pressure of the global field dominates over the gas pressure ($\beta < 1$), and consequently laminar MCWs are driven while MRI is quenched. The wind mass-loading is thus fully controlled by the thermal pressure gradient, without any local turbulent fluctuations.

In MRI-active regions where fluctuating winds can develop (\S~\ref{sect:launching}) mass loading occurs differently. The break-up of MRI 'channel' modes happens deeper in the disk where the gas pressure dominates over the magnetic pressure ($\beta > 1$).  Upflows lift gas from the lower turbulent region to the upper B-field dominated laminar region. This uplift adds to the thermal 
pressure gradient  discussed above  and results in much higher mass being loaded into the wind than without the help of MRI-turbulence. The uplift of material also results in overall higher gas density than in hydrostatic disks.  For instance, in the ideal-MHD limit, the density distribution deviates from the hydrostatic profile already above $2H$ and it is two orders of magnitude higher at $4H$ \citep{SuzukiInutsuka2009,SuzukiInutsuka2014}. 
However, we should note that the specific density increase and launching height may be affected by the vertical upper boundary condition which is not sufficiently large in these early simulations. Recent global ideal-MHD simulations covering the entire polar angle show that the disk wind from the MRI-active region arises from higher altitudes ($z \ge 10 H \sim R$) and therefore is not very massive \citep{2021A&A...647A.192J}. Furthermore, it only removes a small fraction, about  $10\%$, of the angular momentum flux, the rest being extracted by turbulent stresses.

In disk regions where the mass accretion is induced solely by the removal of angular momentum via an MCW, 
the accretion rate and the wind mass loss rate have the following relation
\citep{PelletierPudritz1992,Bai2016}:
\begin{equation}
 \begin{split}
   \dot{M}_{\rm acc}(R) \frac{d}{dR}(R^2\Omega)   \simeq \frac{d \dot{M}_{\rm wind}(R)}{dR}\Omega(R_A^2  - R^2)
   \label{eq:Macc_Mwind_relation}
\end{split}   
\end{equation}
where $\Omega$ can be usually approximated by the Keplerian rotation frequency at radius $R$. 
 The left term, $\dot{J}_{\rm acc}(R)$, is the rate at which angular momentum is extracted to allow accretion across a disk annulus at  $\dot{M}_{\rm acc} (R)$, counted positive inward. The right term, $\dot{J}_{wind}(R)$, is the rate at which the MHD disk wind removes excess angular momentum from the underlying disk annulus. The wind magnetic torque thus imposes a minimum accretion rate through the underlying disk. 

If all of the MHD disk wind is ejected from a small region at the footpoint radius $R_0$, the derivative terms in eq.~\ref{eq:Macc_Mwind_relation} may be replaced by local values to obtain an order-of-magnitude scaling  \citep{PelletierPudritz1992}
\begin{equation}
    \frac{\dot{M}_{\rm wind}(R_0)}{\dot{M}_{\rm acc}(R_0)} \sim \frac{R_{0}^2}{R_{\rm A}^2} \approx \lambda^{-1}\,.
    \label{eq:pp92}
\end{equation}
When the disk wind is radially extended, however, 
the derivatives in eq.~\ref{eq:Macc_Mwind_relation} may no longer be ignored, and the ejection to accretion ratio can be substantially larger than predicted by the above scaling.  
Mass conservation imposes that the integrated  wind mass-flux ejected between radii $r_{\rm in}$ and $r_{\rm out}$ is given by
\begin{equation}
  \begin{split}
\dot{M}_{\rm wind} = \dot{M}_{\rm acc}(r_{\rm out}) - \dot{M}_{\rm acc}(r_{\rm in})  \\
       = \dot{M}_{\rm acc}(r_{\rm in}) \left[ \left(\frac{r_{\rm out}}{r_{\rm in}}\right)^\xi-1 \right],
      \label{eq:Mwind_Macc_rin}
  \end{split}
\end{equation}
where $\xi$ is the ``mass ejection efficiency" of the radially extended disk wind, defined as \citep{Ferreira1997} 
\begin{equation}
    \xi  \equiv \frac{d \log \dot{M}_{\rm acc}(R)}{d \log R} 
       = \frac{1}{\dot{M}_{\rm acc}(R)} \frac{d \dot{M}_{\rm wind}(R)}{d\log R}.
    \label{eq:xi_relation}
\end{equation}
Inserting eq.~\ref{eq:xi_relation} into eq.~\ref{eq:Macc_Mwind_relation} and taking $\Omega \simeq \Omega_{\rm Kep}$ yields a fundamental relation between the magnetic lever arm parameter $\lambda$ and the ejection efficiency $\xi$ when disk accretion is driven by an MHD disk wind \citep{Ferreira1997}:  
\begin{equation}
    f_J (R) \equiv \frac{\dot{J}_{wind}(R)}{\dot{J}_{\rm acc} (R)} = 2 \xi(R) [\lambda(R)-1] \simeq 1.
    \label{eq:lambda_xi}
\end{equation}
    Since the MHD wind is driven by angular momentum extraction from the disk, this ratio cannot exceed one. Hence, a more heavily mass-loaded MHD disk wind (larger $\xi$) will necessarily have a smaller magnetic lever arm parameter $\lambda$, i.e. a smaller angular momentum per unit mass, and vice-versa. 
    This is indeed verified in all idealized self-similar solutions with a dominant wind torque, at both high and low magnetization \citep[e.g.,][]{CasseFerreira2000A&A...361.1178C,Lesur2021A&A...650A..35L}. 
  
Assuming that $f_J$ is constant over radius and inserting eq.~\ref{eq:lambda_xi} in eq.~\ref{eq:Mwind_Macc_rin} 
gives the predicted ejection to accretion ratio for a radially extended disk wind: 
\begin{equation}
  \begin{split}
\frac {\dot{M}_{\rm wind}}{\dot{M}_{\rm acc}(r_{\rm in})} = 
\left[ \left(\frac{r_{\rm out}}{r_{\rm in}}\right)^{f_J/[2(\lambda-1)]} -1 \right].
      \label{eq:Mwind_Macc_lambda}
  \end{split}
\end{equation}
where $f_J$ is the fraction of angular momentum flux removed by the wind.
 Unlike in the case of a radially very narrow wind (eq.~\ref{eq:pp92} ), the expression above shows that the integrated wind mass loss rate can easily exceed the accretion rate at the inner radius. As an example, if 
accretion through the disk dead-zone is 100\% wind-driven 
(i.e., $f_J = 1$) with
$\lambda \simeq 2$ 
and $r_{\rm out}/r_{\rm in} \geq 10$ (as suggested by recent non-ideal MHD simulations) 
    one expects an integrated wind mass-flux of more than twice the mass-accretion rate at the inner launch radius, instead of a half using the approximate local scaling in eq.~\ref{eq:pp92}. Inside the MRI-active region (0.05--0.5 au), instead, recent global simulations suggest $f_J \simeq 0.1$ and $\lambda \simeq 5$ \citep{2021A&A...647A.192J}. Here, eq.~\ref{eq:Mwind_Macc_lambda} predicts  a disk wind mass-flux in this region of only 3\% of the accretion rate. We will test these predictions against specific observations in \S~\ref{sect:SpatiallyResolvedWindDiagnostics}.  

\begin{figure}[t]
    \centering
     \includegraphics[width=\columnwidth]{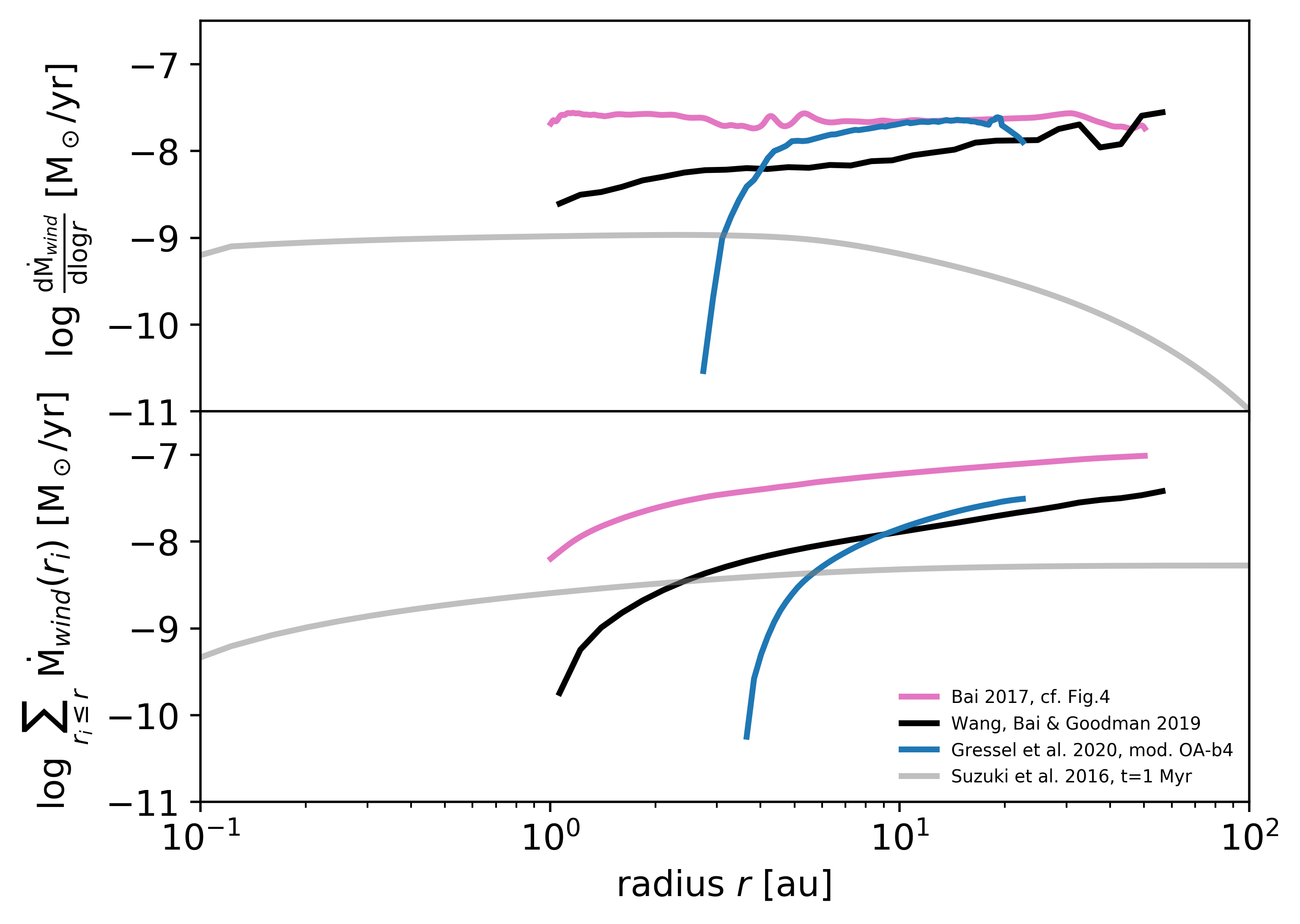}
    \caption{Compilation of  wind mass-loss rate vs. radial distance from the star from recent global MHD disk wind  simulations.}
    \label{fig:r2-Sigma-dot}
\end{figure}

\subsubsection{\it{\textbf{Global Non-ideal Disk Wind Simulations}}}\label{sect:GlobalDiskWindSimulations}
The most important development since the PPVI review  by \citet{Turner2014} is the advent of reasonably realistic global non-ideal MHD simulations in terms of the considered microphysical effects and thermodynamics. Local shearing box simulations of winds have been recognized as having fundamental limitations, primarily their inability to account for the net rate of magnetic field transport and to capture the dependence of the wind characteristics on  global geometry. Hence, they
have been abandoned since PPVI and their summary in Section 4.2 of \cite{Turner2014} remains comprehensive. 

Besides efforts to combine local and global approaches \citep{2017MNRAS.471.1488N}, new models include
one-dimensional self-similar \citep{Lesur2021A&A...650A..35L}, as well as two- and three-dimensional semi-global, and global simulations. Most of these properly include multiple non-ideal MHD effects, more precisely, ~i)~Ohmic resistivity, ~ii)~ambipolar diffusion \citep[both considered in ][]{Gressel2015}, and ~iii)~Hall drift \cite[e.g.,][]{BaiStone2017,Bethune2017}, even though the assumptions on the underlying ionization model can vary rather significantly. While earlier  semi-analytical calculations \citep[e.g.,][]{Wardle1993,Ferreira1997} found the launching of winds to depend rather sensitively on assumptions, recent global non-ideal MHD simulations demonstrate that wind solutions are a robust outcome of the assumed initial conditions even when the disk midplane is not coupled to the ionized component. 

We highlight the recent advancement to include disk thermodynamics in global non-ideal MHD simulations: from an \textit{ad hoc} prescribed thermal structure \cite[e.g.,][] {Bai2017,Bethune2017} to the addition of stellar irradiation and diffuse re-radiation \citep{Gressel2017}, thermochemical 
and ambipolar heating \cite[e.g.,][]{Wang2019,Gressel2020} self-consistently.  Because the wind base is typically found to coincide with the FUV-ionized upper disk layers, the thermochemistry is built around chemical processes \citep{2011ApJ...735....8P} driven by FUV irradiation from the central star (but see \citealt{Rodenkirch2020} for an Ohmic-resistivity-only, i.e. lacking ambipolar diffusion, model with stellar EUV and X-rays instead of FUV). These simulations consider weak fields ($\beta_{\rm midpl} \simeq 10^4$), roughly consistent with meteoritic constraints on the solar nebula magnetic field \citep{2014Sci...346.1089F},  and the added external heating and thermal energy are found to contribute to wind energetics, 
 unlike in the ``cold" winds. Further, the addition of self-consistent chemistry allows identification of potential tracer species and simulations of synthetic observations for detecting and characterizing the outflows  \citep{Gressel2020}, a predictive approach previously limited to idealized self-similar models \citep[e.g.,][]{Yvart2016}.

The inclusion of thermal effects, hence radial thermal pressure gradients, also results in wind opening angles that are closer to the 30$^\circ$ minimum of the early cold \citet{BlandfordPayne1982} model. For instance,  typical values of the field-line inclination angle at the wind base (at $r=10\,$au) are, e.g., $\simeq 46\degr$ for the AD model `fid0` of \citet{Bai2017} and $\simeq 34\degr$ for the `OA-b4` model of \citet{Gressel2020}.
Importantly, beyond the launching radius, the wrapping of field lines by wind inertia inevitably produces a gradual magnetic  collimation (see e.g.,  Fig.~11 in \citealt{Wang2019}, and Fig.~2 in \citealt{Gressel2020}). Hence, when comparing models with data the scale probed by observations needs to be taken into account (\S~\ref{sect:SpatiallyResolvedWindDiagnostics}). 

A somewhat puzzling feature of global simulations \emph{without} an enforced symmetry (i.e., with regards to reflection upon the disk midplane) is the emergence of one-sided outflows -- or at least asymmetric wind strengths on the top and bottom half of the simulated disk. While these are predominantly seen when including the Hall effect \citep{Bethune2017,Bai2017}, asymmetric winds have also been found in simulations including Ohmic and ambipolar diffusion alone \citep{Gressel2020}.
These asymmetries are related to the sporadic/stochastic transport of azimuthal flux into the highly diffusive midplane and ongoing work is testing the role of  initial and/or boundary conditions on breaking the wind symmetry.
At least observationally, there is \emph{some} evidence for asymmetric winds (\S~\ref{sect:obs} and \ref{sect:SpatiallyResolvedWindDiagnostics}).

To illustrate just one central quantity that is obtained as an outcome of such comprehensive simulations, Fig.~\ref{fig:r2-Sigma-dot} showcases radial profiles of the mass loss rates obtained from 2D-axisymmetric MCW global simulations from \citet{Bai2017}, \citet{Wang2019}, and \citet{Gressel2020}. Compared to the latter, the former two models generally display smaller lever arms, hence a higher contribution from magnetic pressure rather than the tension force in driving the wind (see discussion in \S~\ref{sect:launching}). 
Some of the differences between the results of \citet{Bai2017} and \citet{Gressel2020}
may be due to differences in mass loading onto the field-line at the wind base, which is determined by thermal heating processes and can vary significantly depending on assumptions (see \S~~\ref{sect:massloading}).
We caution the reader that the figure is meant to represent a \emph{compilation} rather than a true comparison, as the individual models in fact cover a range of mass accretion rates. Yet, despite the considerable number of assumptions going into the various MHD models, they overall agree fairly well with respect to their inferred mass-loss rates. 
For further comparison, we also show a curve that includes the effect of fluctuating winds in the inner MRI-active region in the form of a 1D long-term model calculation by \citet{Suzuki2016}. The considerable spread in the obtained levels of ejection of material illustrates that the mass loss is sensitive to the model inputs and assumptions, particularly the magnetic field strength. We note that the evolution of the disk mass and surface density is also very sensitive to the assumed evolution of the poloidal magnetic flux (see \S~\ref{sect:EvolutionPoloidalFlux}).

\subsubsection{\it{\textbf{Jets and Winds from the Inner Disk Region}}} \label{sect:innerdisk}

Jets  (HVC) and inner winds traced by the LVC BC are thought to originate within disk radii 
$\lesssim$
0.5\,au (see \S~\ref{sect:Class0I}).  This is a complex region that will have multiple contributions to outflowing gas, including the star, the star-disk interaction region, and the disk itself
\citep{Ferreira2006} in a regime where the disk is likely MRI-active \citep[e.g.,][]{Armitage2011,Flock2017}.

The innermost component of the ejection is a wind from the central star. In addition to an ordinary stellar wind driven by coronal gas pressure,  accretion energy coupled with a strong magnetic field and surface convection can drive powerful winds radially from  polar regions along open field lines  
\citep{Matt2005ApJ...632L.135M,Matt2008,Cranmer2008}.
The velocity of an accretion-powered stellar wind is typically comparable to or exceeds the escape velocity from the central star, $\gtrsim$ a few hundred \kms , and is likely the source of the deep and broad blueshifted absorption at \hei{} and \cii\ in sources with small disk inclinations (\S~\ref{sect:SpatiallyUnresolvedFlows}). These accretion powered stellar winds are invoked to spin down the accreting central star \citep[e.g.,][]{Matt2005,2014EPJWC..6404001P}.

In the star-disk interaction region, close to the co-rotation radius $\sim 0.1$\,au, some stellar field lines remain closed and deliver disk material to the star while others are opened up and may launch a wind. Different configurations can lead to ejection and we refer the reader to \citet{Ferreira2006} for a summary of the main possibilities. 

In the ``X-wind'' scenario, the disk 
hijacks opened stellar field lines in 
a narrow region near the co-rotation radius ($\sim 0.1$\,au), generating a steady MCW that 
is assumed to carry away sufficient angular momentum to spin down the accreting star \citep{Shu1994}. The fanning out of the field lines from this anchor point 
creates an MHD wind with the dual character of a dense axial jet surrounded by a wide angle wind, both moving at a fairly uniform velocity  $\sim 150$\,\kms\  \citep[e.g.,][]{Shang1998ApJ...493L..91S,Shang2007prpl.conf..261S}. As will be addressed in \S~\ref{sect:alternative}, the  wide angle component of the X-wind is invoked as one possible means of sweeping up ambient gas into 
slow molecular outflows \citep[e.g.,][]{Lopez2019}.

MHD simulations of the star-disk interaction favor a different ejection geometry, dominated by an unsteady ``conical wind'' launched at $\simeq 45\degr$ \citep[e.g.,][]{Romanova2009,Zanni2013A&A...550A..99Z}. 
Magnetic field lines anchored at the stellar surface and the disk are stretched by the differential rotation between the two footpoints, winding  up the toroidal magnetic field and accelerating mass outward  to $\gtrsim 100$\,\kms\ from $\approx 0.1$\,au \citep{Liffman2020}.
If the disk harbors a net vertical field aligned with the stellar dipole, its reconnection with closed stellar field lines near the magnetopause will also drive powerful unsteady ejections at intermediate latitudes \citep{Hirose1997,Ferreira2000MNRAS.312..387F}.

MCW disk winds launched from the co-rotation radius outward are also likely contributors to outflows within 0.5\,au  (\S~\ref{sect:launching}). Such a centrifugal disk wind will result in a range of wind speeds, reflecting the underlying range of launch radii  and creating a nested ``onion-like" kinematic structure \citep[e.g.,][]{Cabrit1999A&A...343L..61C}. The innermost part of such a disk wind is thought to be the origin of the slow, narrow blueshifted absorption at \hei{} and \cii\  seen in systems at high inclination, where the line of sight to the star passes through the disk wind. Also the \oi{} LVC BC line width is interpreted as arising from a wind in the inner disk (\S~\ref{sect:SpatiallyUnresolvedFlows}). 

Additionally, within several stellar radii, the possibility for failed winds may also occur. 
In this instance, some of the vertical upflows do not reach escape velocity and instead fall back and accrete onto the star \citep[see the 3D global simulations by][]{Takasao2018}.  When the strength of the stellar magnetic field is less than 1\,kG, a magnetosphere does not develop well around the star  \citep{Takasao2018,Takasao2019} and the failed  winds accrete onto mid-latitude regions with high velocity due to stellar gravity \& magnetic braking, at nearly the Keplerian velocity at the surface, $\approx 100-200$ km/s. A characteristic feature of this funnel--wall accretion is that the accretion column covers a wide belt around the latitude $15^\circ-40^\circ$. This is in contrast to magnetospheric accretion that occurs in localized magnetic flux tubes connecting the star and the disk \citep[e.g.,][for a review on accretion onto young stars]{Hartmann2016ARA&A..54..135H}. No observations have been associated with MRI failed winds to date, but they may be a contributor to some of the complex variability of emission lines formed in near-stellar regions. 

It is likely that more than one of these outflow processes expected within 0.5 au, several of which will be variable due to changing accretion rates onto the star, together form and collimate the jet, and  give rise to its variability. We do not pursue the formation of jets further, as the major focus for this chapter is new evidence supporting radially extended, slower disk winds that surround the jet, addressed in \S~\ref{sect:link}.

\subsubsection{\it{\textbf{Evolution of the Poloidal Magnetic Flux}}}\label{sect:EvolutionPoloidalFlux}

As alluded to above, protoplanetary disks are believed to inherit a certain amount of magnetic flux aligned with the rotation axis from the pre-stellar core phase. The typical `hourglass' morphology of a collapsing, rotating object can be explained by the fundamental effect of `magnetic braking'. It is furthermore believed that ambipolar diffusion \citep[e.g.,][]{Masson2016A&A...587A..32M,2021MNRAS.502.4911X} as well as the Hall effect \citep[e.g.,][]{2012MNRAS.422..261B,2015ApJ...810L..26T,2020MNRAS.492.3375Z} play a key role in mediating the amount of magnetic flux retained in forming the disk. An alternative explanation involves turbulent reorganisation of the flux \citep[e.g][]{2013MNRAS.432.3320S,2017ApJ...846....7K}. Once a rotationally supported disk has formed, essentially the same agents continue to drive its secular evolution. 

The standard model for this inward ``dragging'' of the flux \citep{1994MNRAS.267..235L}, applying to fully turbulent accretion disks, has recently attracted new interest  \citep{2008ApJ...677.1221R,2014ApJ...785..127O,2014ApJ...797..132T,2014MNRAS.441..852G}. From the study of \citeauthor{2014ApJ...785..127O}, the prediction for the maximum attainable field strength is about $100\,{\rm mG}$ at 1\,au, and about $1\,{\rm mG}$ at 10\,au from the central star for an external field B$_\infty$ of 10\,$\mu$G outside a disk of radius 100\,au -- which agrees broadly with what is needed for obtaining  mass accretion rates of $10^{-7}$\,M$_\odot$/yr from the MCW, or via net-vertical-field MRI.
None of these approaches include the evolution of the poloidal flux by means of an $\alpha$~effect dynamo -- despite ample evidence \citep[e.g.,][]{2011ApJ...735..122F,2015ApJ...810...59G,2020MNRAS.494.4854D} from local and global MRI simulations. An 
important idea, first sketched-out in \citet{2008ApJ...677.1221R}, pertains to the dragging-in of flux in the magnetically dominated surface layers, and this is indeed observed in ideal-MHD global disk simulations \citep{SuzukiInutsuka2014,2018ApJ...857...34Z,2021A&A...647A.192J}.

In the case of non-ideal MHD (arguably more realistic for significant parts of the disk, between $\sim 1-30$\,au), the outward transport of flux by means of turbulent eddies has to be replaced by the microphysical coefficients (Ohmic diffusion and ambipolar diffusion acting in the ``dead zone''). The Hall-effect can, moreover, lead to a pronounced pinching of field lines in the disk midplane -- a consequence of the escalating feedback of Hall rotation and background shear when the poloidal field is 'aligned' with disk rotation  \citep{BaiStone2017}. The resulting field configuration leads to fast inward transport partly counter-acted by outward drift through ambipolar diffusion. For the opposite case of `anti-aligned` flux, the combined Hall \& ambipolar diffusion effects render the effective transport a factor of about two more efficient, and comparable to the Hall-free case. The magnitude of the flux transport generally scales with the overall magnetization, and is moreover found to be sensitive to the assumed microphysical coefficients. Simulations with realistic ionization models \citep{Bai2017,Bethune2017,Gressel2020,Lesur2021A&A...650A..35L} agree on the result that the vertical flux diffuses radially out of the disk on a secular timescale (i.e., a fraction of an au per century), but there is considerable scatter amongst the various results in terms of the precise diffusion rate measured.

If accretion torques from MCWs indeed dominate viscous redistribution of angular momentum and mass, the timescale of disk evolution and dispersal becomes primarily steered by the amount of magnetic flux threading the disk. In this case, the question of how rapidly the disk loses the flux (or whether it is able to retain substantial parts of it) becomes a crucial one. At the time of writing, the long-term impact of the flux evolution from global simulations remains an open question. Also,  there are no simplified evolutionary models  that self-consistently include the flux evolution. Existing models of \citet{Armitage2013}, \citet{Bai2016}, \citet{Suzuki2016}, and \citet{Tabone2021MNRAS.tmp.3115T} include only the limiting cases for the evolution of the total magnetic flux: i) constant total magnetic flux (i.e., a fixed radial distribution of $B_z$ in time from the balance between inward drag due to accretion and outward drift/diffusion due to non-ideal MHD effects and turbulence); and ii) decreasing flux in accordance with surface density.
 In the constant flux scenario, disk accretion can possibly become ``super-charged'' in the presence of significant entrained flux, eventually overcoming any residual level of viscous disk spreading by disk turbulence. This may  lead to a two-timescale behaviour for disk dispersal \citep[see discussion in ][]{Armitage2013,Bai2016} and rapid dispersal \citep[][see  Manara et al. in this volume]{Tabone2021MNRAS.tmp.3115T}. In contrast, efficient loss of magnetic flux (e.g., from  slippage of field lines mediated by ambipolar diffusion) would imply a stalled disk evolution, leading to massive old disks. 
One possible example is TW~Hya where \citet{Vlemmings2019A&A...624L...7V} found  small vertical field strengths $< 0.8$\,mG (1$\sigma$) at $\sim 40$\,au that the authors interpret as  faster  diffusion over advection if the protostellar envelope had an initial B$_\infty \ge$ 1 \,mG.
 There is, in principle, no reason, why the entire disk should adhere to either of these extreme cases during its lifetime. For example, flux could be advected inwards in the inner MRI-active regions and diffuse outwards in the weakly ionized regions. This opens the possibility to construct evolutionary pathways \citep[e.g.,][]{Bai2016,2021A&A...647A.192J} that, for instance, lead to the formation of inner disk cavities with efficient accretion due to an MCW at high disk magnetization $\beta_{\rm midpl} \simeq 1$ \citep[][and discussion in \S~\ref{sect:TransitionDisks}]{Combet2008A&A...479..481C, WangGoodman2017transitions}.

\subsection{\textbf{Photoevaporative winds and disk dispersal}}\label{sect:photoevaporation}
While MHD winds may presumably drive stellar accretion via mass loss at \textit{all} epochs, thermal winds have been proposed to disperse protoplanetary disks. These so-called photoevaporative winds --- which are purely thermally driven and play no role in angular momentum transport --- were first invoked as a disk clearing mechanism to explain relatively short ($\sim$ few Myr) disk lifetimes \citep{Hollenbach1994}. At a basic level, these are conceptually similar to the Parker solar wind.  Gas at the surface is heated by photons and ``evaporates" in a wind if the rotating gas has sufficient thermal energy to escape the stellar gravitational field---typically beyond a radius $r_{crit} \sim 0.1-0.2 G M_*/c_s^2$ where the thermal speed $c_s$ is $\sim 11$\,\kms\ for $10^4\,$K fully ionized gas, $\sim 4-7$\,\kms\  for $3,000-8,000$\,K neutral gas, and $\sim 3$\,\kms\ for  $2,500$\,K fully molecular gas. The thermal speed and the density $\rho_{base}$ to which photons penetrate, which depend on the column density (\S \ref{sect:thermal}), determine the resulting mass loss rate $\dot{\Sigma}(r) = \rho(r_s) c_s $ where $\rho(r_s)$ is the wind density at the sonic point. This has also been approximated by \citet{Gorti2009} as $\dot{\Sigma}(r) \sim f\,c_s \rho_{base}$, with the flow velocity at the wind base typically some fraction $f$ of the thermal speed, of the order unity. 
Note that the flow speed reaches $c_s$ at the sonic radius with the density $\rho(r_s)$ here determined by the flow characteristics from the base to $r_s$, and there is a need for an accurate computation of both the thermochemistry and hydrodynamics.
The driving agent (EUV, FUV or X-rays) plays a key role in determining how the spatial distribution of mass in the disk evolves with time. The critical radius within which gas remains bound is $\lesssim 1-2$ au for EUV-driven winds, $\lesssim 2-4$ au for X-ray driven winds, and $\lesssim 3-12$ au for FUV-driven winds. Gas temperatures resulting from EUV and X-ray heating can be more easily approximated to allow full hydrodynamical simulations \citep{Owen2012, Rodenkirch2020}; in contrast, the determination of the temperature for FUV-irradiated gas is considerably more complex and impacted by both dust evolution and chemistry \citep{Gorti2015,WangGoodman2017}. 

The most significant new developments in photoevaporation theory, relevant both for internal and external photoevaporation, are related to coupling hydrodynamics with chemistry \cite[e.g.,][]{WangGoodman2017,Nakatani2018a,Nakatani2018b,Haworth2018} and to coupling dust dynamical evolution with FUV photoevaporation and disk dispersal  \cite[e.g.,][]{Gorti2015,Facchini2016,Carrera2017,Sellek2020a}. Prior to the PPVI review by \citet{Alexander2014}, models considered either hydrodynamics or chemistry in detail but not both simultaneously; modest size chemical networks are now being incorporated into hydrodynamical models. \citet{WangGoodman2017} conducted hydrodynamical simulations with a ray tracing method for radiative transfer and were the first to include FUV heating and gas line cooling calculated consistently with chemistry. They found that mass loss-rates are sensitive to the incident radiation field (especially the EUV and Lyman-Werner FUV bands), thermochemistry and hydrodynamics.  An additional complication is introduced by dust \citep{Gorti2015,WangGoodman2017}, which simultaneously evolves in disks en route to planet formation. The decrease in abundance of small grains increases penetration depth of FUV photons and hence the mass loss rate, but also decreases photoelectric heating to lower the temperature and mass loss rate. Photoevaporation rates are tied to dust disk evolution, but \citet{Gorti2015} find that  these competing effects offset each other to a large extent.

Related to dust evolution is the effect of varying metallicity. Using radiation hydrodynamical models,  \citet{Nakatani2018} find that FUV photevaporation rates dominate X-ray rates, and that FUV mass loss increases with decreasing metallicity due to reduced opacity. While X-rays do not drive strong flows in their models, they act to enhance mass loss rates by affecting the ionization degree in the gas which increases the FUV heating rate. \citet{Ercolano2018} and \citet{Wolfer2019} consider the effects of gas phase depletion of carbon on X-ray photoevaporation (carbon contributes to the opacity for the soft X-rays in their models), and find that heating is increased to drive greater mass loss, see \S~\ref{sect:TransitionDisks} for an application to transition disks.

There appears to be an emerging consensus from recent hydrodynamical models that include EUV, FUV and X-ray photoevaporation that FUV photons determine the base height from where flows are launched and that EUV and X-ray photons aid in the acceleration of the flow \citep{WangGoodman2017, Nakatani2018,Komaki2021}. 
Integrated mass loss rates range from a few $10^{-9}$ to $10^{-8}$ \ms\ yr$^{-1}$. 
While \citet{Nakatani2018a} find FUV mass loss rates comparable to results from  XEFUV driven photoevaporation without hydrodynamics \citep{Gorti2009}, \citet{WangGoodman2017} conclude that the static models  over-estimate mass loss rates by a factor of a few. \citet{WangGoodman2017} attribute the discrepancy 
to the difficulty in  sustaining the thermal flow to the sonic radius where gas escapes, which was tacitly assumed by \citet{Gorti2009}. Previous XEUV models with radiation hydrodynamics and simplified temperature profiles \citep{ErcolanoOwen2010MNRAS.406.1553E, Owen2012} were found to overestimate mass loss rates as well, new results indicating a minor role for X-rays and EUV in driving photoevaporation---these differences are attributed to the simplified temperature approximation used in the XEUV models and neglect of important cooling terms \citep [molecular cooling and adiabatic expansion in atomic gas, see ][]{WangGoodman2017}. 
Improved models by \citet{Picogna2019}, however, find results similar to their earlier work \citep{Owen2012}, and that adiabatic expansion cooling can be neglected as  advection timescales always exceed recombination timescales in their models.

\begin{figure}[t]
    \centering
       \includegraphics[width=\columnwidth]{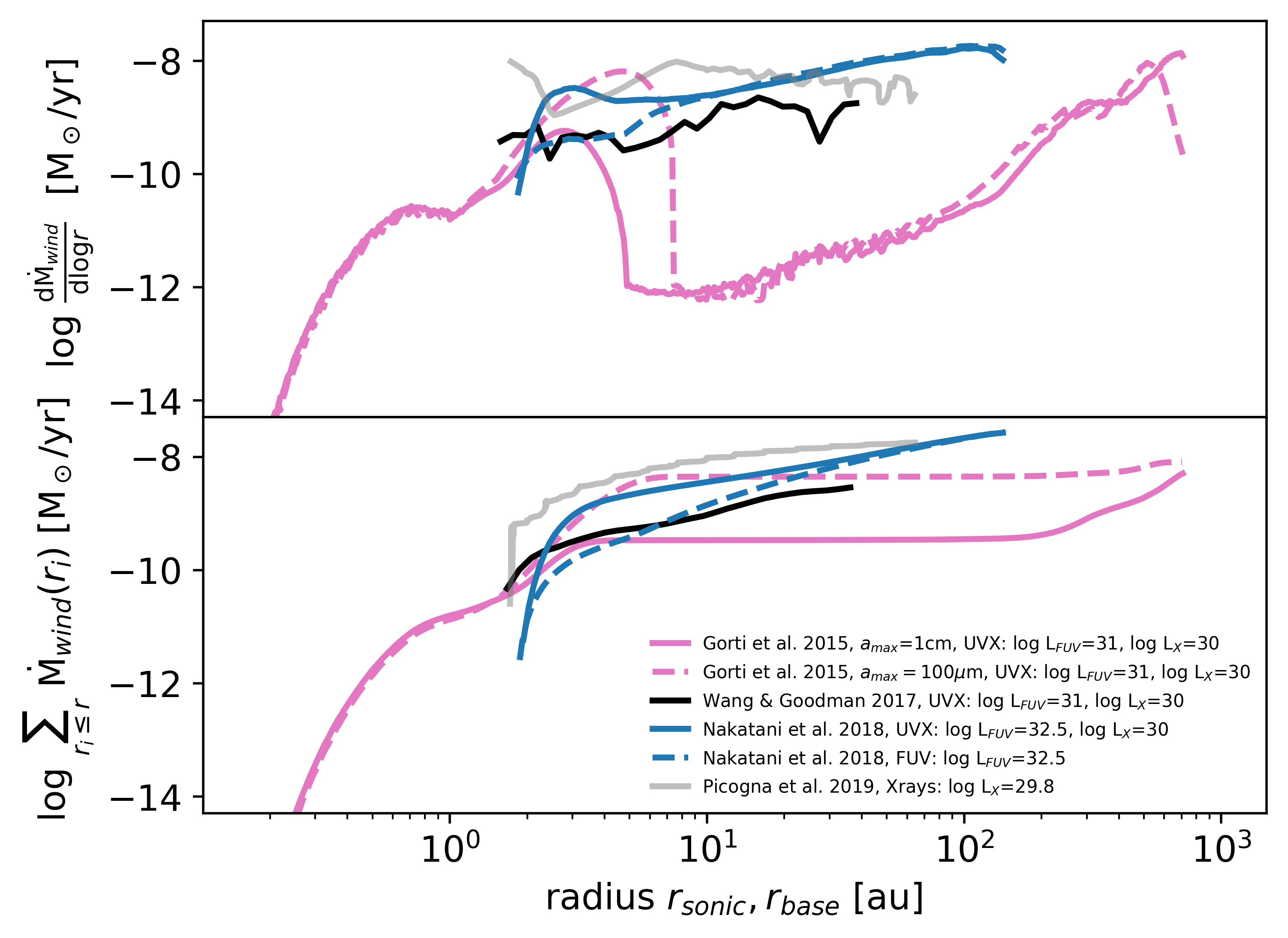}
    \caption{Compilation of wind mass-loss rates vs. radial distance from the star from several recent 
    photoevaporative models.}
    \label{fig:PE-r2-Sigma-dot}
\end{figure}

Fig.~\ref{fig:PE-r2-Sigma-dot} shows a compilation of mass loss rates from pure PE models, analogous to the MHD results presented in Fig.~\ref{fig:r2-Sigma-dot}. 
In contrast to the MHD disk winds, PE winds rely solely on the thermal pressure gradient and mass loss rates are consequently lower. Three representative classes of models are shown with the radial profile in the upper panel and the cumulative mass loss rate in the lower panel. 

The reference model (with a fixed grain size distribution), and a model with an enhanced population of small grains (including PAHs) after 1 Myr of evolution are shown from \citet{Gorti2015}. They use
an analytical prescription for the mass loss rate and include EUV, FUV and X-rays with full thermochemistry; here $L_{FUV}=10^{31}$ erg s$^{-1}$ and $L_X=L_{EUV}=10^{30}$  erg s$^{-1}$.  A similar model from \citet{WangGoodman2017}, but advanced in that it includes full hydrodynamics along with some chemistry, is also shown. The mass loss rate in these models is calculated at the base of the flow ($r_{base}$) which is indicated along the $x-$axis. The XEUV model result from \citet{Picogna2019} has nearly the same X-ray and EUV luminosities but does not include FUV heating. \citet{Picogna2019} treat full radiation hydrodynamics with simplified ionization and thermochemistry, and the mass loss rate is computed at the sonic radius ($r_{sonic}$). The models from \citet{Nakatani2018b} feature full radiation hydrodynamics, X-ray, EUV and FUV heating with thermochemistry and indicate that FUV heating mainly drives mass loss (gray solid and dotted lines are similar). However, note that the FUV luminosity for the gray curves is unrealistically high ($L_{FUV}\sim 3\times 10^{32}$ erg s$^{-1}$) for low-mass stars \citep{Yang2012ApJ...744..121Y}. 

It is evident from Fig.~\ref{fig:PE-r2-Sigma-dot} that model assumptions can affect the mass loss rates by orders of magnitude,
and the radial profile can differ significantly. 
Some of these differences could be due to the adopted abundances of small grains for FUV photoevaporation, chemical coolants included, and choice of UV and X-ray spectra that all affect the gas temperature and hence mass loss rate.
Other differences are that the X-ray mass loss rate peaks at $r\lesssim$ 10au \citep{Picogna2019}, while the models including FUV have mass loss rates that increase with radius; the \citet{Gorti2015} models are evolution models that follow the disk to dispersal with a mass loss rate that depends on time (see Fig.~\ref{fig:t-Mdot} and related discussion in \S \ref{sect:EvolutionMassEjection}), and employ a larger radial grid with most of the mass loss occurring beyond $\sim$ 100\ au. 

FUV heating additionally depends on the dust disk properties (compare dashed and solid magenta lines in Fig.~\ref{fig:PE-r2-Sigma-dot}) but may gain in significance over MHD processes if the disk is sufficiently extended. At large enough radii, the gas temperature, which roughly scales as $r^{-0.5}$,  begins to exceed the escape temperature needed for thermal pressure to overcome gravity, which instead scales as $r^{-1}$, thus promoting PE wind launching.

\begin{figure}[t]
    \center\includegraphics[width=0.9\linewidth]{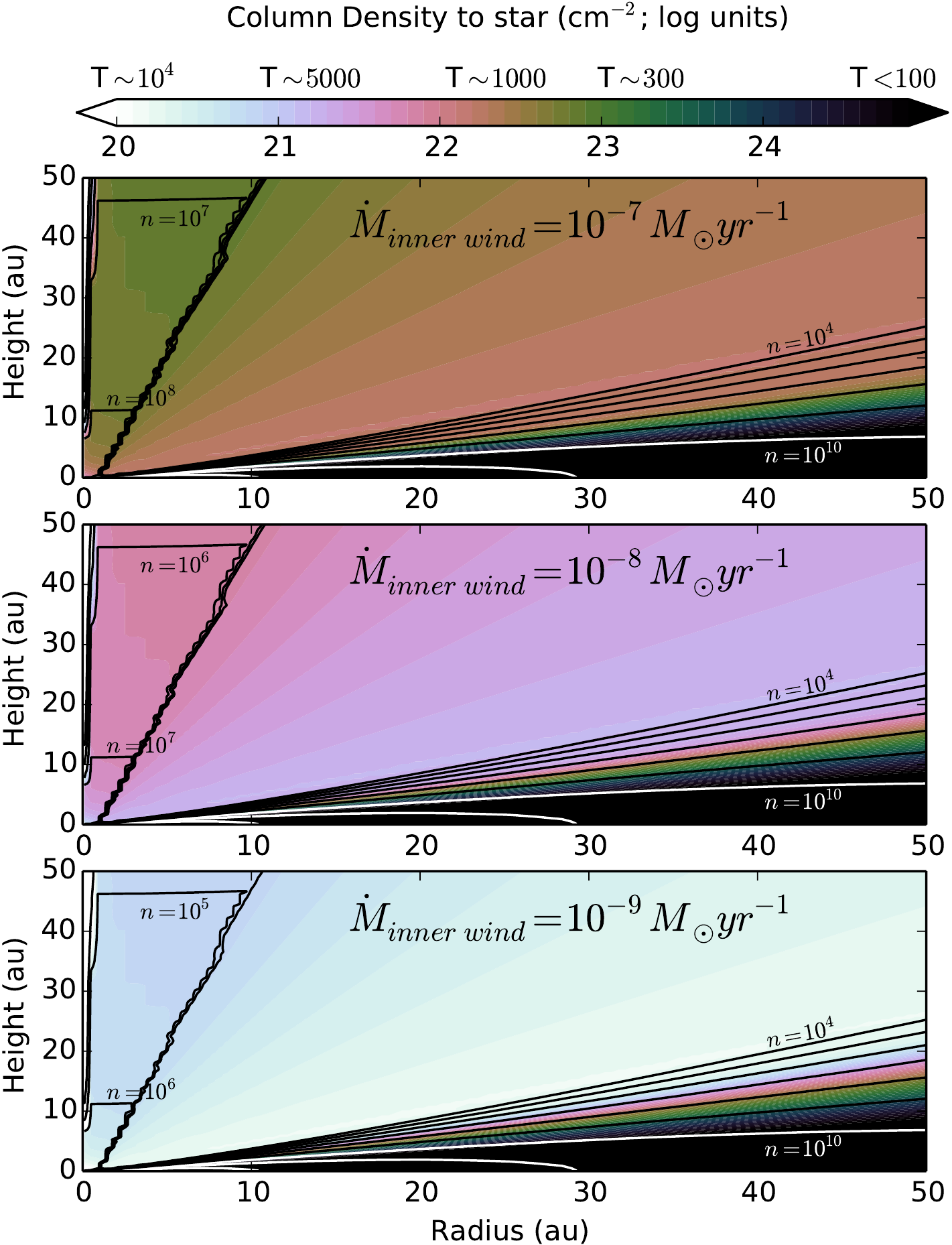}
    \caption{A toy model is used to estimate the local column density to the star, shown above for a $0.01$M$_{\odot}$ disk with an inner ($0.05-1$\,au) disk wind at 70\,km/s, and for three different wind mass loss rates. White/black contours indicate decades in the gas number density in the disk and inner wind. At early stages of evolution when $\dot{M}_{inner \, wind}$ is high, XEFUV photons may not be able to penetrate the dense wind and heat the outer disk surface, while this becomes possible at later stages when mass loss decreases in the inner wind. The colorbar shows the column density in log units and indicates the corresponding approximate gas temperatures that can be attained: columns $\lesssim 10^{19-20}$\,cm$^{-2}$,  T$\sim 10^4\,$K for EUV; columns $\sim 10^{20-21}$\,cm$^{-2}$, T$\sim 1000-10^4\,$K for X-rays and columns $\lesssim 10^{21-23}$\,cm$^{-2}$, T$\sim 300-5000\,$K  for FUV photons. } \label{fig:colden} 
\end{figure}

The launching of purely thermal flows is moreover contingent on the ability of the XEFUV photons to penetrate the jet/winds that are observed from the inner disk and to heat the gas at the surface beyond $r_{crit}$. EUV and soft X-ray photons can heat gas to $10^4$K but even then $r_{crit} \gtrsim 1$ au which implies that photoevaporative flows can only be launched when the column density through the inner MHD winds (within 1 au) is low enough. Fig.~\ref{fig:colden} shows an illustrative model providing the column density to the star where the disk density structure and wind velocity are kept constant, while the density in the wind declines as $\sim 1/r^2$ for  different $\dot{M}_{inner wind}$. 
FUV photons, with their lower gas absorption cross-sections, may be able to penetrate the inner wind even when  mass loss rates are high; however, dust is less evolved at early stages and introduces additional complications \citep[see][]{Gorti2015}. Dusty MHD winds with small grains could absorb FUV photons \citep[for e.g., see][]{Panoglou2012} and thwart photoevaporation beyond $r_{crit}$. 
It is therefore likely that photoevaporative winds are not very effective in early Class 0 and Class I stages of disk evolution, and become significant only at later stages. 

\subsection{\textbf{Evolution of Mass Ejection and Accretion}} \label{sect:EvolutionMassEjection} 
\begin{figure}[h]
\includegraphics[width=\columnwidth]{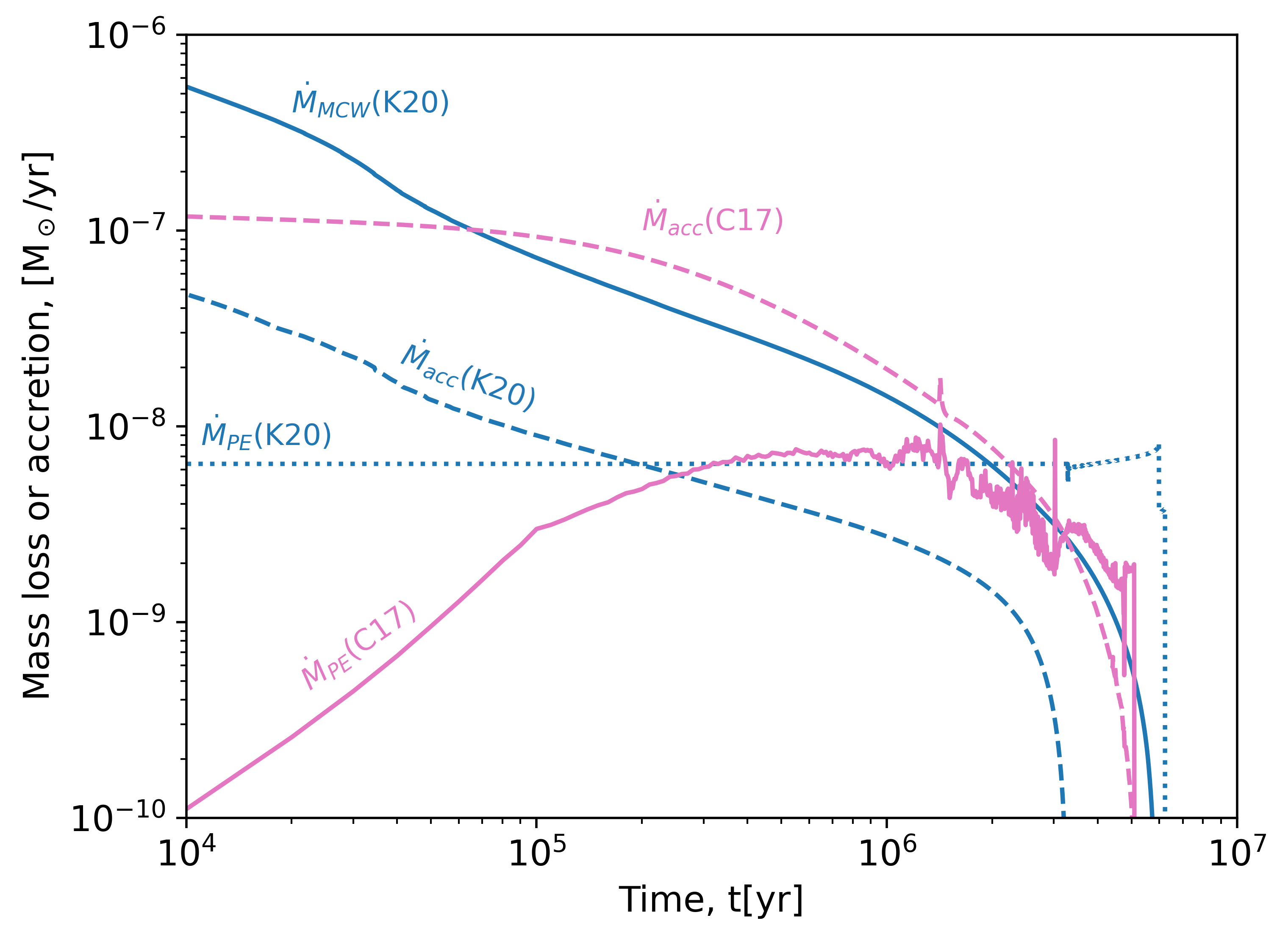}
\caption{
Comparison of mass loss (solid lines) and accretion rates (dashed lines) from the 1+1D time evolution models of \citet{Kunitomo2020}, MCW in blue, and \citet{Carrera2017}, PE in pink. The photoevaporative mass loss rate assumed in \citet{Kunitomo2020} is indicated with a dotted blue line: this rate includes X-ray and EUV photoevaporation but the latter is about 40 times lower.  The horizontal axis shows the time elapsed from the beginning of the simulation and does not correspond to the Class0/I stage. Note how the wind mass loss and mass accretion rates decrease in tandem with time in the MCW model.    
}
\label{fig:t-Mdot}
\end{figure}

The relation between disk mass accretion rates and wind ejection rates has been followed in some 1D evolutionary models. In Fig.~\ref{fig:t-Mdot} we compare predictions for two such models, one where disk accretion is driven largely by the removal of angular momentum through MCWs with a small contribution from turbulent viscosity, \citep{Kunitomo2020} and another with a disk evolving viscously \citep{Carrera2017}. Both models assume a similar initial disk mass (0.118\,M$_{\odot}$ in \citealt{Kunitomo2020} and 0.1\,M$_{\odot}$ in \citealt{Carrera2017}) and obtain a similar dispersal timescale $\approx 5-6$ Myr but the initial conditions and input physics  are quite different.
 
 The curves in Fig.~\ref{fig:t-Mdot} for the \citet{Kunitomo2020} model include a small turbulent viscosity, $\alpha = 8\times 10^{-5}$ (their Fig.~2) and show the evolution of disk mass accretion rate, and mass ejection from MCW and from PE, both from X-rays \citep{Owen2012} and EUV photons \citep{Alexander2007}. 
One can see a tight correlation between the mass accretion rate, $\dot{M}_{\rm acc}$, measured at the inner boundary of  $r_{\rm in}=$\,0.01\,au of the calculation, and the  radially integrated mass loss rate, $\dot{M}_{\rm MCW}$, with $\dot{M}_{\rm MCW}/\dot{M}_{\rm acc} ({\rm r_{in}})\approx 10$ throughout the evolution. 
Note that in this model the large $\dot{M}_{\rm MCW}$ is driven by a small lever arm and a large ratio of $r_{\rm out}/r_{\rm in}$ (eq.~\ref{eq:Mwind_Macc_lambda}).
 The wind mass loss rate is also overestimated since, inside of the MRI-active region ($\le 1$\,au), a significant fraction of the gas will accrete onto the star \citep[][\S~\ref{sect:massloading} and \ref{sect:innerdisk}]{Takasao2018,2021A&A...647A.192J}.

Importantly, the tight correlation between $\dot{M}_{\rm MCW}$ and $\dot{M}_{\rm acc}$  results from the mutual coupling between the accretion and the wind; the removal of the angular momentum via the MCW induces mass accretion, and the portion of the accretional energy that is converted into kinetic energy drives the MCW wind.
Apart from its obvious dependence on the amount of magnetic flux entrained in the disk, $\dot{M}_{\rm MCW}$ generally has a positive correlation with the disk surface density.  Since $\dot{M}_{\rm acc}$ is directly linked to the surface density, $\dot{M}_{\rm MCW}$ and $\dot{M}_{\rm acc}$ show similar declining trends with time, which can be directly compared to the observed trends in Table \ref{tab:jetwindprop} of \S~\ref{sect:obs}.
On the other hand, the mass loss rate of the PE winds included in this model is determined by the XEUV luminosity of the central star 
(see \S~\ref{sect:photoevaporation} for details) which is kept constant with time. 
At early times when the disk surface density is high, $\dot{M}_{\rm MCW}$ dominates PE mass loss rates while, beyond 2\,Myr, $\dot{M}_{\rm X}$ exceeds $\dot{M}_{\rm MCW}$. An inner cavity is formed after 3\,Myr so that the contribution from EUV PE becomes also gradually important. The gas in the outer region is finally dispersed by  X-ray and EUV photoevaporation at 6\,Myr.  

In contrast, mass ejection in \citet{Carrera2017} is purely from PE  and take into account all the radiation components (X-ray, EUV, and FUV) and dust evolution with a detailed grain size distribution, based on earlier models by \citet{Gorti2015}. Dust coagulation/fragmentation, wind entrainment, radial drift and criteria for planetesimal formation via the streaming instability are all considered, but they do not include MHD winds. They instead consider a layered accretion model where the midplane has a low $\alpha=10^{-4}$ while the disk still accretes onto the star with a relatively large $\alpha=10^{-2}$.  The curves presented in Fig.~\ref{fig:t-Mdot} are from their standard model and show that $\dot{M}_{\rm acc}$  declines in time, in proportion to the declining surface density as the disk evolves. 
At early times there is very little penetration of XEFUV photons through an assumed inner MHD wind (see Fig.~\ref{fig:colden}, and \citealt{Gorti2009}) and minimal mass loss  occurs, while beyond $\sim 1$\,Myr $\dot{M}_{\rm PE}$ becomes comparable to $\dot{M}_{\rm acc}$ and they decline together. This is because mass loss is mainly driven by the FUV component that is, in turn, determined by the accretion rate. An important distinction to note is that while $\dot{M}_{\rm MCW}$ determines $\dot{M}_{\rm acc}$ in the MHD evolutionary model, here the situation is reversed---$\dot{M}_{\rm acc}$ determines the FUV accretion luminosity in these $\alpha-$disk models which subsequently determines $\dot{M}_{\rm PE}$. Therefore, the disk lifetime and surface density evolution are quite sensitive to the choice of $\alpha$ (see also \citealt{Gorti2015}). In the model of \citet{Carrera2017}, viscous accretion spreads material to the outer disk where it is removed by FUV photoevaporation (see also \S~\ref{sect:GasDiskRadius} in relation to the evolution of the outer radius). We note that disk dispersal is not contingent on gap opening as in the case of X-ray and EUV driven photoevaporative winds.

\section{\textbf{\uppercase{Comparison of Wind Models to Observations}}}\label{sect:link}

Here, we will interpret the observations presented in \S~\ref{sect:obs} in the context of the disk wind models  summarized in \S~\ref{sect:theory}. 
The central questions to be addressed are whether MHD disk winds are present and strong enough to be the prime enabler of disk accretion via the extraction of angular momentum, whether MHD disk winds are radially extended and can affect the spatial distribution of mass and its evolution in the disk, and whether PE winds are present and are the primary agent of disk dissipation. We further seek to understand the relative balance between these two genres of winds as disks evolve.

\subsection{\textbf{Spatially Resolved Disk Wind Diagnostics Towards the Highest Accretors}} \label{sect:SpatiallyResolvedWindDiagnostics}
As we discuss in detail below, we find that radially extended disk winds provide the most likely origin for the spatially resolved wind diagnostics outlined in \S~\ref{sect:obs}. While photoevaporative flows are sometimes invoked in the literature (e.g., the CO emission from DG~Tau~A, \citealt{Gudel2018}), there are  drawbacks in extending such an interpretation to all spatially resolved winds.  First, PE winds are predicted to turn mostly atomic as soon as they leave the disk surface, due to their high temperature and irradiation 
\citep[e.g.,][]{WangGoodman2017}, while outflowing molecular cones are found to extend up to several thousands of au. The second issue is screening by an inner jet+wind (see Fig.~\ref{fig:colden}): XEUV photons from Class~0 and I sources, which have  jet mass loss rates in the range $\sim 10^{-6} - 10^{-7}$\,M$_\odot$/yr (Table~\ref{tab:jetwindprop}), will not reach the disk surface.   A vertically offset lamppost-like  illumination such as that provided from X-rays produced in jets has been suggested to reduce
this inner screening  \citep{Weber2020}. However, at the time of writing, there are only two sources which are known to have stationary shocks producing X-ray emission offset from the central star
(L1551, Class~I: \citealt{Schneider2011A&A...530A.123S} and DG~Tau~A, Class~II: \citealt{Gudel2008A&A...478..797G}) and for the less obscured one, DG~Tau~A, the jet X-ray luminosity is an order of magnitude lower than the stellar X-ray luminosity. 
Even if XEFUV photons would make it to the disk, predicted PE mass-loss rates are much lower  than observationally inferred mass fluxes from spatially resolved winds in Class~0 and I sources (compare values in Table~\ref{tab:jetwindprop} and Fig.~\ref{fig:PE-r2-Sigma-dot}). Therefore, it appears that PE winds can play a role only in the latest Class~II evolutionary stage.

Radially extended MHD disk winds do not have these limitations and, instead, can reproduce most of the  properties of the spatially resolved outflows: nested velocity structure and associated opening angles, 
persistence of molecular gas over large vertical extents, consistent rotation and  geometrical radii, and large mass fluxes. We address each of these in turn below. 

\subsubsection{Velocity structure and molecular content}
The resolved outflows discussed in \S~\ref{sect:Class0I} and \ref{sect:ClassII} clearly  show a large velocity gradient  with polar angle: from $> 200$\,\kms\ at angles of 4$^\circ$ down to a few \kms\ at 30$^\circ$ from the jet axis (see also Fig.~\ref{fig:dgtaua}).
As discussed in \S~\ref{sect:innerdisk}, this ``onion-like" kinematic structure, in combination with the large range of speeds, is a  characteristic feature of radially extended MCWs launched from the co-rotation radius outward  (see eq.~\ref{eq:vp-lambda}).

MHD winds can remain molecular to large distances due to a combination of factors: gas and dust grains carried in the wind partly screen stellar high-energy photons,  and ion-neutral drift heating increases only gradually with height \citep[see][and \S~\ref{sect:thermal}]{Panoglou2012,Wang2019, Gressel2020}.  As the wind mass-flux drops with time, self-screening and radiative cooling are less efficient, the wind is globally hotter, and the CO-rich wind region moves to larger launch radii, from $\leq$ 0.2 au in Class 0 to a few au in Class II sources \citep[e.g.,][]{Panoglou2012, Wang2019}. In addition, within a given wind, outer streamlines are cooler than inner ones, hence they will be highlighted in low-excitation CO lines. 
The predicted ``onion-like" gradient in chemistry and temperature will naturally produce a hollow appearance for slow CO molecular winds resolved with ALMA, with a rotating atomic flow at  intermediate velocities nested inside the molecular ``cavity", as observed in DG Tau A \citep{AgraAmboage2014}.
Ambipolar diffusion heating also predicts a ``warm" irradiated wind chemistry with enhanced abundances of SO \& SO$_2$ \citep[][in agreement with the extended  wind seen in Fig.~\ref{fig:hh212} for HH~212]{Panoglou2012}, H$_2$O \citep[][in agreement with \textit{Herschel} line profiles]{Yvart2016}, and HCN \citep{Gressel2020}.

\subsubsection{Consistent rotation and launch radii}\label{sect:rotation_launch}
Observations provide two independent constraints on the launch radius of an MHD disk wind candidate.  In a first, model-independent approach, one can extrapolate the observed cones at constant opening angle down to the disk midplane to obtain a ``geometrical" radius $r_{\rm geom}$ \citep{Bjerkeli2016}. Since the opening angle of an MCW should decrease with height (at least initially) due to magnetic recollimation (see \S~\ref{sect:MHD}), $r_{\rm geom}$ sets an upper limit to the true MCW launch radius. 

In a second, independent approach, the launch radius $r_{\rm MCW}$ is derived from the observed  wind rotation in the ``cold" MHD approximation of negligible enthalpy and pressure gradients (\S~\ref{sect:MHD}). Assuming the disk wind is axisymmetric and steady, $r_{\rm MCW}$ is inferred by solving for $\Omega_0$ the following ideal-MHD invariant \citep{Anderson2003ApJ...590L.107A}:
\begin{equation}
J =  \frac{V_p^2 + V_\phi^2}{2} + \Phi_g - rV_\phi \Omega_0 = - \frac{3}{2} (GM_\star \Omega_0)^{2/3}, 
\label{eq:anderson}
\end{equation}
where $V_p$ and $V_\phi$ are the (observed) wind poloidal and rotational velocity, 
$j = rV_\phi$ the specific angular momentum in the form of matter rotation, 
$\Phi_g$ the local gravitational potential, and $\Omega_0$ is the Keplerian angular frequency at the footpoint $r_{\rm MCW}$ that one wishes to determine. When $V_p \gg V_{\rm kep}(r_{\rm MCW})$, the right-hand side term can be neglected, yielding a simpler expression \citep[see eq.~5 in][]{Anderson2003ApJ...590L.107A}. However, at the low speeds $V_p \le 10$ \kms\ of small-scale rotating molecular flows detected e.g. with ALMA, this approximation does not apply and the exact form of eq.~\ref{eq:anderson} has to be used. For example, in CB26 the inferred $r_{\rm MCW}$ is 32\,au with the approximate expression, and\,12 au when the right-hand side term of eq.~\ref{eq:anderson} is taken into account.

When the wind is under-resolved (e.g., DG~Tau~A) or edge-on (e.g., HH~212), $j$ and $r_{\rm MCW}$ can be significantly underestimated  \citep{Tabone2020A&A...640A..82T}. Therefore, synthetic observations are necessary to properly account for observational biases \citep[e.g.,][ for the rotating atomic and $H_2$ flow in DG~Tau~A with inferred launch radii $\sim 3-10$\,au]{Pesenti2004,AgraAmboage2014}. For HH~212 a proper model of different flow surfaces along the line of sight gives an outermost launch radius of $\simeq$\,40\,au instead of 1\,au (when neglecting flow stratification) for the rotating SO wind and 0.2\,au instead of 0.05\,au for the SiO jet \citep{Tabone2017,Tabone2020A&A...640A..82T}. Modeling of Position-Velocity (PV) cuts at several heights suggests $\lambda \simeq 3.5$ \citep{LeeTabone2021ApJ...907L..41L}.

Table~\ref{tab:jetwindprop} gives the full range of geometric $r_{\rm geom}$ and inferred launch radii $r_{\rm MCW}$ in the cold MHD approximation among current disk wind candidates. In each individual source, one verifies that $r_{\rm MCW} \le r_{\rm geom}$. This is also verified streamline by streamline in two Class~I sources, TMC1A and DG Tau B (\citealt{Bjerkeli2016}, de Valon et al. 2022 in press). Therefore the observed geometrical radii and specific angular momenta are both consistent with an origin in an MCW. 

Interestingly, the inferred outermost wind launch radii are substantially smaller than the full disk extent in Class~I and Class~II sources. 
he apparent limited radial extent of these winds may result from an observational bias (e.g., filtering out of extended faint emission by ALMA, see the case of DG~Tau~B from de Valon et al. 2022, in press.) or partial dissociation of the outer wind by the ambient UV field. Complementary tracers should be explored to determine the full radial extent of rotating outflows from Class I/II disks.

\subsubsection{Wind mass fluxes}
An important question is whether the large mass fluxes estimated in spatially resolved MHD disk wind candidates could extract enough angular momentum to drive  accretion. 
As discussed in \S~\ref{sect:massloading}, this condition is fulfilled if the wind mass-flux  verifies eq.~\ref{eq:Mwind_Macc_lambda} with $f_J$=1 at disk radii with a dead-zone interior.  

The inner ($r_{\rm in}$) and outer ($r_{\rm out}$) launch radii of the emitting wind zone under study can be estimated from observations using cold MHD theory (see previous section). The wind magnetic lever arm parameter $\lambda$ can  be also inferred, once r$_{\rm MCW}$ is calculated from eq.~\ref{eq:anderson}, as $\lambda = j / \sqrt{GM_\star r_{\rm MCW}}$, assuming $j$ has reached its maximum asymptotic value (i.e., all  angular momentum extracted in the form of magnetic torsion has been converted into matter rotation). In most molecular wind candidates (e.g., HH30, DG Tau B, \citealt{Louvet2018}, de Valon et al. 2022 in press), the inferred $\lambda \simeq 1.6-1.8$ is close to the minimum value of 1.5 for magnetic acceleration, consistent with current models of winds from the dead zone in weakly ionized disks \citep[e.g., ][]{Wang2019, Gressel2020, Lesur2021A&A...650A..35L}.
Note that  in eq.~\ref{eq:Mwind_Macc_lambda} $\dot{M}_{\rm acc}(r_{\rm in})$ is the mass accretion rate at the inner launch radius, which is approximated by the mass accretion rate onto the star (or by ten times the jet accretion rate if the star is too embedded).

In all cases where values for $\lambda$, $r_{\rm out}/r_{\rm in}$, $\dot{M}_{\rm wind}$, and $\dot{M}_{\rm acc}$ are sufficiently well determined to carry out the above test (DG Tau A, HH~30, HH~212, DG~Tau~B), the test is conclusive. The estimated angular momentum flux carried by the rotating wind appears sufficient (within a factor 2-3) to drive steady accretion across the wind launching zone at the requested rate \citep{Bacciotti2002,AgraAmboage2011,2014A&A...565A.110M,Louvet2018,Tabone2020A&A...636A..60T,deValon2020}.

\subsubsection{\textbf{Disk wind-jet connection}}
\label{sec:wind-jet}
Previous Protostars and Planets chapters on high-velocity jets \citep{Ray2007,Frank2014} considered the tentative rotation signatures identified across a few atomic jets to infer HVC launching radii $r_{\rm MCW} \simeq$ 0.5-1\,au, thus preferring an extended MHD wind to the other jet launching mechanisms discussed in \S~\ref{sect:innerdisk}. However, these estimates were based on transverse velocity shifts in atomic jets, which can be contaminated by asymmetric shocks or jet wiggles \citep[e.g.,][]{Coffey2015,Louvet2016A&A...596A..88L,Erkal2021A&A...650A..46E} 
so that the previously inferred jet launching radii are only upper limits. Much smaller values are inferred from 
recent rotation measurements at higher angular and spectral resolution 
with ALMA across the SiO jet in HH212: 
$r_{\rm MCW} \simeq $ 0.05 au for a single emitting ring \citep{Lee2017NatAs...1E.152L} and 0.05-0.3 au for a radially extended MCW \citep{Tabone2017}. 
A jet origin within the dust sublimation radius would be consistent with the short dynamical age of SiO knots in HH212, and the sharp increase of SiO detection rate with jet mass-flux (\S~\ref{sect:Class0I}), both predicted by dust-poor chemistry \citep{Tabone2020A&A...636A..60T,Lee2020ARAA}. Thus, while it remains unclear whether HVC jets are the manifestation of one or more of the MHD ejection processes discussed in \S~\ref{sect:innerdisk}, they most certainly are launched within $<$0.3\,au.

An open question is then whether the same wind responsible for  the slow cones of H$_2$ \& CO and the atomic/ionic emission at medium- and low-velocity 
extends continuously inward to produce the HVC jets, or whether there are ``gaps" of low mass-flux due to e.g., magnetic instabilities gathering the flux in some disk regions (cf. chapter by Lesur et al.). This opens the possibility of a scenario where the inner time variable jet could be disconnected from the outer disk wind, but interact with it through bowshocks. 

The interaction of a time-variable jet with an outer disk wind was predicted and first modeled in \citet{Tabone2018A&A...614A.119T}. The jet drives large bow shocks that sweep up a thin conical shell 
inside the disk wind, except at small heights where the disk wind has time to partly ``refill" the cavity.
Recent ALMA observations of HH~212 \citep{LeeTabone2021ApJ...907L..41L} compare favorably with the key predictions of the model. Above $50$\,au, the flow morphology and kinematics at different heights  can be globally reproduced assuming that MHD disk wind streamlines launched from $\sim 0.2-4$\,au were swept-out by a large jet bowshock (seen in SiO) into a thin radially expanding ``shell", while outer MHD disk wind streamlines, launched from 8 to 40\,au, remain unaffected.  
Such a jet-disk wind interaction might contribute to
the hollow geometry and constant opening angles (up to a thousand times the launch radius) of resolved  rotating CO cones in Class I/II sources.
Below 40\,au, the signs of perturbation are more ambiguous, with PV cuts consistent with replenished, unperturbed MHD disk wind down to ALMA resolution \citep{Tabone2020A&A...636A..60T}. 
Numerical simulations of the jet interaction with a more realistic MHD disk wind would be important to further test this scenario and compare with observations of both resolved CO flows and unresolved LVC components, which would trace replenished wind regions near the disk surface at radii $\sim 0.2-4$\,au.

\subsubsection{\textbf{An alternative scenario of entrained material}} \label{sect:alternative}

Although we have interpreted the small scale ($<$500 au) nested atomic and molecular cones of outflowing gas that surround the jet as evidence for winds emerging from the disk over radii out to $\sim$\,50\,au,  an alternate interpretation merits discussion: whether they might instead trace the base of the entrainment of ambient material swept up by a fast wide angle  X-wind (see \S~\ref{sect:innerdisk}).

This possibility has been invoked in the literature as the source of some slow $\le 500$ au scale resolved H$_2$ and CO flows observed in Class 0/I/II sources \citep[e.g.,][]{Pyo2003ApJ590340P,Takami2006ApJ...641..357T, Beck2008,Zapata2015,Lee2018ApJ...863...94L,Bjerkeli2019A&A...631A..64B}. However, on such small scales the ambient environment will no longer be a static self-gravitating $1/r^2$ core, as for large scale entrained flows \citep[][and references therein]{Shang2007prpl.conf..261S,Shang2020ApJ...905..116S}, but rather an infalling rotating envelope, where centrifugal deflection and gravity of the central star become dominant. Wind-swept shells in such small scale environments have been modeled both analytically for the ballistic infall solution of \citet{Ulrich1976ApJ...210..377U}, see \citet{Lopez2019,Lopez2020ApJ...904..158L,Liang2020ApJ...900...15L}, and numerically for the infalling sheet solution of \citet{Hartmann1994ApJ...430L..49H}, see
\citet{Delamarter2000ApJ...530..923D,Cunningham2005ApJ...631.1010C}. When the ambient material being impacted by a fast wide angle wind is an infalling rotating envelope, several observed  characteristics of small-scale molecular cones  (\S~\ref{sect:obs}) appear difficult to explain. We discuss the main ones below: 

{\bf – Base radius}: The predicted radius at the base of the swept-up shell is close to the disk centrifugal radius, as a consequence of the high equatorial speed of the X-wind 
(e.g., Fig.~1 in \citealt{Cunningham2005ApJ...631.1010C}, Fig.~9 in \citealt{Lopez2019}, Fig.2 in \citealt{Liang2020ApJ...900...15L}). In contrast, resolved molecular cones in Class~I/II  appear anchored at radii $r_{\rm geom} \simeq 5-40$ au generally much smaller than the gas disk radius (see \S~\ref{sect:obs}). Having smaller shell base radii would require a strong obstacle to the X-wind e.g., a vertical magnetic field in the disk. However, a magnetic field able to stall an X-wind at 10\,au would inevitably drive a powerful MHD wind which would induce accretion at 300 times the X-wind mass flux and largely dominate over any entrained material (compare eq.~8 in \citealt{Cabrit2007LNP...723...21C} with eq.~\ref{eq:bfield-macc-two}).

{\bf – Kinematics and rotation}:
Semi-analytical models of expanding shells driven by a fast wide-angle wind into the rotating infalling envelope of \citet{Ulrich1976ApJ...210..377U} could reproduce the observed mass and poloidal speeds in the low-mass Class I rotating flow of CB26, but the rotation speeds are too low by an order of magnitude from the observed
ones \citep{Lopez2019}. In the case of
DG~Tau~B, observed at $\sim$\,15\,au resolution, there has been a detailed comparison of the observed CO conical shell with models from \citet{Lee2001ApJ...557..429L} and \citet{Lopez2019}, showing that the predicted radial expansion
does not reproduce the kinematics, and the specific
angular momentum is again smaller than observed (de Valon et al. 2022, in press).

{\bf – Fast replenishment}:  the inferred age of a wind-driven shell, $\simeq R/V$, is much shorter than the source age in many cases: e.g. 15\,yr in DG Tau A, 45\,yr in HH212 and DG Tau B, 1,000\,yr in CB26 \citep{AgraAmboage2014,Lee2018ApJ...863...94L,Lopez2019}. 
This implies that new shells need to be periodically generated by wind outbursts, with a typical recurrence timescale of 20-1,000\,yrs. Once the infalling envelope is cleared out by previous shells, however, replenishing  on such short timescales will be challenging, and eventually no ambient material would remain to be entrained on small scales. Replenishment ``from below'' by a PE wind can be ruled out, as the rate of mass to be entrained in the CO shells (a few $10^{-6}$ \ms/yr in the Class 0 HH212 and a few $10^{-7}$ \ms/yr in Class I/II) largely exceeds the expected mass-outflow rate in a PE wind for realistic irradiation conditions (\S~\ref{sect:photoevaporation} and~\ref{sect:SpatiallyResolvedWindDiagnostics}).
The envelope replenishment issue could be avoided if the swept-up shell is stationary instead of expanding, and infalling material is entrained in a mixing-layer parallel to the shell walls \citep{Liang2020ApJ...900...15L}. However, the predicted shell base radius is still close to the disk radius, hence much too large to explain 
Class I/II observations (see above).

We conclude that the MHD radial disk wind scenario is better able to reproduce  all of the observed properties of slow rotating flows on scales $< 500$\,au, while the scenario of entrainment by a fast wide-angle X-wind has several drawbacks.

\subsection{\textbf{Spatially Unresolved Wind Diagnostics Through to the Lowest Accretors}} \label{sect:spatiallyunresolved}

Since PPVI, a growing number of atomic and molecular transitions with line profiles indicative of slow winds have been identified (\S~\ref{sect:Class0I} and \ref{sect:ClassII}). They include: the LVC of optical forbidden lines like the \OIa ; near-infrared CO and H$_2$ transitions; 
infrared forbidden lines like the \neii{} at 12.8\,\micron ; and  CO  rotational lines at millimeter wavelengths. At the same time, both MHD and PE wind models have incorporated rather detailed heating and cooling mechanisms (\S~\ref{sect:GlobalDiskWindSimulations} and \ref{sect:photoevaporation}) and have published line luminosities and profiles that can be confronted with observations. Our goal here is to summarize such comparisons, assess which transitions are most likely to trace MHD or PE winds, and discuss new theoretical predictions that can be tested with further observations.

The LVCs of oxygen forbidden lines in Class II sources have recently seen a renewed interest as disk wind tracers as they are bright, blueshifted, and require high temperature for excitation (see also Fig.~\ref{fig:n-t-xe-lines}). They also allow us to probe the relative contributions of both MHD and photoevaporative winds, as many Class II sources have low enough column densities through the wind that high energy radiation may penetrate to the disk beyond the critical radius for thermal ejection.  Indeed, \citet{ErcolanoOwen2010MNRAS.406.1553E} were among the first to predict line emission signatures due to X-ray photoevaporation as X-rays can heat neutral gas to high temperatures and naturally excite the forbidden transitions.  

The behavior of the LVC is sufficently complex, however, with both broad and narrow kinematic components, that neither MHD nor PE winds have yet been able to fully account for their structure. The sources with composite BC+NC profiles are of particular interest. Photoevaporative winds cannot be invoked for the BC, as the large FWHMs and shifts trace gas within the gravitational potential well of the star, requiring an MHD rather than a thermal flow \citep[e.g.,][]{Simon2016,Picogna2019}.  The empirical correlations between the BC and NC described in \S~\ref{sect:ClassII}, 
led \citet{Banzatti2019} to argue for a common origin of both components in different parts of the same MHD wind. While this argument is compelling there is as yet no explanation for why distinct BC and NC profiles would arise in a radially continuous MHD wind. \cite{Weber2020} suggest a combination of MHD and PE winds to explain the composite BC+NC profiles, combining an analytic MHD inner wind model with the \citet{Picogna2019} photoevaporative model where an inner MHD wind produces the BC while a PE wind produces the NC. The correlations reported in \citet{Banzatti2019} are accounted for in this dual wind scenario  by a single common correlation between line luminosity and accretion luminosity which is modelled as an extra EUV component heating the line-emitting region in both winds. 

One example is now documented where the NC cannot be attributed to a PE wind.  In a spectroastrometric study of the high accretion rate Class II source RU~Lup,   \citet{Whelan2021ApJ...913...43W}  measure the centroid of the NC emission out to a de-projected distance of $\sim$8\,au in \OIa{} and $\sim$40\,au in \SIIc{} and show a smoothly decreasing velocity with increasing vertical offset, as expected in a disk wind (see their Fig.~8). The observed NC blueshifts of $\sim$30\,km/s  at $\sim$2\,au from the disk midplane are too fast and too deep within the potential gravitational well of the star for thermal winds \citep[e.g.,][]{Ballabio2020}. A similar spectro-astrometric signal was detected in the CO fundamental for this source \citep{Pontoppidan2011}, where the molecular gas traces the slower portion of the wind closer to the disk surface (see Fig.~9 in \citealt{Whelan2021ApJ...913...43W}).

Furthermore, LVC with composite BC+NC features are most common among the higher accretion rate Class II sources, but as the accretion rate falls, the LVC profiles simplify to single Gaussians, with a range of FWHM (100 to 10 km/s) that encompass the widths attributed to BC and NC. There is as yet no explanation for why some sources would have BC, and others NC, characteristics. Importantly, however, the NC becomes the most common feature as the accretion rate continues dropping and the inner disk is dissipating (\S~\ref{sect:ClassII}). This is where the \neii{} 12.8\,\micron{} LVC, with small blueshifts (by a few up to several km/s) and FWHMs ($\sim 20\kms$), appears. 
The new VLT/VISIR2 survey discussed in \S~\ref{sect:ClassII} has shown that \neii{} LVC luminosities display an opposite behaviour to the \OIa{} LVC in that they increase in systems with depleted inner dust and typically lower mass accretion rates  \citep{Pascucci2020}. The \neii{} LVC  is well reproduced by photoevaporative models \citep[e.g.,][]{Alexander2014,Ballabio2020} and the persistent blueshifts in their centroids point to a wind beyond the dust cavity
while the weaker \oi{} emission, which is often centered at the stellar velocity in these evolved sources, arises within. These new findings suggest an evolution in disk winds:  MHD winds appear to dominate in  high-accreting, radially continuous disks while PE winds begin to emerge in disks with an inner dust cavity, lower accretion, and weaker inner wind (see Fig.~9 in \citealt{Pascucci2020}). MIRI MRS observations with JWST should be able to spatially resolve the \neii{} emission from the disks with the largest dust cavities and thus test whether this line truly traces a PE wind.

Moving to longer wavelengths, \citet{Yvart2016} have carried out a detailed comparison between predicted and spectrally resolved H$_2$O profiles observed with {\it Herschel}/HIFI toward  Class~0 and I sources. The authors adopted an analytic MHD disk wind model used to explain jet rotation in T~Tauri stars and upgraded the thermo-chemistry developed in \cite{Panoglou2012} along dusty streamlines. By varying only four parameters (stellar mass, mass accretion rate, the outermost launch radius, and viewing angle), they show that this typical dusty MHD disk wind can reproduce the variety of broad wing  emission profiles beyond $\sim 5\kms$ from systemic velocity. An interesting prediction of these models is that the launch radius beyond which most of the molecular (H$_2$) content survives moves radially outward with evolutionary stage. This is because as the wind density drops going from Class~0 through to Class~II, stellar X-rays and FUV photons penetrate deeper in the wind dissociating more molecular hydrogen \citep[e.g., Figs.~6 and 7 in][]{Panoglou2012}. Observations with JWST/NIRSpec that could spatially resolve winds in H$_2$ ro-vibrational transitions might be able to test this prediction.

Photoevaporative winds also have a layer in between the cold (T$< 100$\,K) midplane and the mostly ionized hot ($T\ge 10,000$\,K) surface where H$_2$ and other molecules survive and could leave the system at moderate radial velocities ($\sim 5-10\kms$), see e.g. Fig.~1 in \citet{WangGoodman2017}. \citet{Gressel2020} have recently computed synthetic line profiles (and moment 1 maps) of selected atomic and molecular transitions, e.g., \oi{} 63.2\,\micron{} and HCN (3-2) through to (8-7), to distinguish a PE from an MHD disk wind. While the far-infrared lines are too faint to be reliably detected with SOFIA, the millimeter transitions are detectable with ALMA. Because the main difference between their PE and MHD wind models is the presence of a more collimated closer-in outflow in the latter case, predicted line profiles are significantly more asymmetric and blueshifted in the MHD than in the PE wind at most disk inclinations (e.g., their Fig.~16), an asymmetry that is even more evident in moment 1 maps (e.g., their Fig.~17).

\subsection{\textbf{Winds in Transition Disks}}\label{sect:TransitionDisks}
Disks with dust inner cavities inferred from their spectral energy distributions, also known as transition disks \citep{Strom1989}, have been commonly assumed to experience a global (dust and gas) inside-out clearing and their properties and frequencies have been used to constrain disk dispersal mechanisms \citep[e.g.,][]{Alexander2014}. Since PPVI \citep{Espaillat2014}, an unbiased search and characterization of these disks \citep{vanderMarel2016A&A...592A.126V}, together with improved and expanded $\dot{M}_{\rm acc}$ measurements \citep[e.g.,][]{Manara2014A&A...568A..18M}, has revealed that transition disks are in fact a diverse class of objects. 

PE models that assume solar metallicity disks can only explain half of the transition disk population, i.e. those with $  \dot{M}_{\rm acc} \lesssim 10^{-9}$\,M$_\odot$/yr and cavities up to $\sim 30$\,au \citep{ErcolanoPascucci2017,Picogna2019}. As a cavity is only opened when the disk accretion rate falls below the PE mass loss rate, a way to explain higher accreting TDs is to further enhance photoevaporation which can be obtained in carbon-depleted disks \citep[][and \S~\ref{sect:photoevaporation}]{Ercolano2018,Wolfer2019}. Alternatively, \citet{WangGoodman2017transitions} have suggested that magnetized winds are driving inner disk accretion at transonic radial speeds, which are much larger than those obtained via typical viscous torques \citep[see also][]{Combet2010A&A...519A.108C}. When the ambipolar diffusion parameter is in the right range for neutral midplane gas to be moderately coupled to the magnetic field, the gas surface density inside the cavity can be relatively low ($\Sigma \sim 7 \times 10^{-3}$\,g\,cm$^{-2}$) but still result in significant accretion ($ \dot{M}_{\rm acc} \sim 4 \times 10^{-9}$\,M$_\odot\,{\rm yr}^{-1}$). 

Gas in the cavity of transition disks has been detected via several diagnostics, including H$_2$ fluorescent \citep[e.g.,][]{Hoadley2015ApJ...812...41H}, \OIa{} \citep[e.g.,][]{Fang2018}, and CO  fundamental emission \citep[e.g.,][]{Pontoppidan2011}. 
The \OIa{} profile, which is a good tracer of inner winds in accreting T~Tauri stars with full disks (\S~\ref{sect:ClassII}), here lacks an HVC, is typically fit by just one Gaussian NC centered on the stellar velocity, and is consistent with bound gas inside the dust cavity \citep[][]{Simon2016,Pascucci2020}. However, the \OIa{} could still form in an inner wind, as both approaching and receding material contribute to the profile \citep[e.g.,][]{Ercolano2016}. In a few cases, a wind is certainly present but detected  beyond the cavity via blueshifts in the \neii{} line centroids \citep[][]{Pascucci2020}. Unfortunately, these UV to IR tracers are beset with uncertain abundances and excitation conditions, hence are not ideal to constrain the gas surface density within the cavity. Significant improvements could be achieved with ALMA by selecting nearly optically thin lines and spatially resolving the cavities. By simultaneously modeling  the source SED, continuum visibilities, and
$^{13}$CO spectra and visibilities, \citet{vanderMarel2016A&A...585A..58V,vanderMarel2018ApJ...854..177V} and \citet{Dong2017ApJ...836..201D}  find that the gas surface density within the
dust cavity is depleted by more than two orders of magnitude than outside (except in RY Lup,
where modeling could be complicated by a likely warped
inner disk). These upper limits are encouraging but, in most cases, not stringent enough to test Wang \& Goodman's accretion scenario. As such, dynamical clearing by giant planets remains the favored mechanism for explaining high accretors with large dust cavities \citep[e.g.,][]{vanderMarel2018ApJ...854..177V}. At least in the case of PDS~70, two accreting giant planets have been detected \citep{Keppler2018A&A...617A..44K,Wagner2018ApJ...863L...8W,Haffert2019NatAs...3..749H,Wang2021AJ....161..148W} within the large ($\sim 50$\,au) millimeter dust gap \citep{Keppler2019A&A...625A.118K}. Although PDS~70 accretes at low levels ($\sim 10^{-10}$\,M$_\odot$/yr, \citealt{Thanathibodee2020ApJ...892...81T}), the inferred depletion of gas surface density at the location of one the planets \citep{Keppler2019A&A...625A.118K} lends strong support to dynamical clearing.

A more direct way to test the scenario proposed by Wang \& Goodman would be to detect gas accreting at transonic speeds within the large dust cavities of high accreting stars. \citet{Rosenfeld2014ApJ...782...62R} have shown that fast radial inflow should be detectable in channel maps of optically thick lines via isovelocity contours that are rotated with respect to the disk axes and twisted isophotes in maps covering velocities close to systemic. However, they also caution that similar observational diagnostics are produced by a warped inner disk as e.g. in HD~142527 \citep{Casassus2015ApJ...811...92C}. For more details on kinematic signatures from fast accreting gas inside dust cavities we refer the reader to the chapter by Pinte et al. in this volume.

\subsection{\textbf{Tentative and Indirect Disk Wind Diagnostics}}\label{sect:effect}
In addition to direct evidence for disk winds, there are several indirect ways to establish their presence or absence. Demographic studies of global gas and dust disk properties are one such approach and covered in the chapter by Manara et al. in this volume. Here, we focus on a few complementary indirect disk wind diagnostics that have been further investigated since PPVI.   

\subsubsection{\textit{\textbf{Free-free Emission}}}\label{sect:Freefree}
A fully or partially ionized disk wind should produce free-free thermal emission as well as H recombination lines at radio wavelengths \citep[][]{Pascucci2012,Owen2013}. While the line emission is probably too weak to be detected even with the ngVLA \citep{Pascucci2018ASPC..517..155P}, the number of candidate disk wind sources identified from cm emission in excess of the dust thermal emission has grown since PPVI \citep[e.g.,][]{Pascucci2014,Coutens2019A&A...631A..58C}. When available, multi-wavelength cm observations  have been used to exclude other possible sources of excess emission, such as gyro-synchrotron non-thermal radiation and emission from very large cm-size or very small nanometer-size grains, but it is clear that higher sensitivity and spatial resolution are needed to turn this diagnostic into a direct tracer of disk winds. At the time of writing there exist only two tentative morphological detections of wide angle cm emission compatible with disk winds: DG~Tau~A \citep[eMERLIN,][]{Ainsworth2013MNRAS.436L..64A} and GM~Aur \citep[VLA,][]{Macias2016ApJ...829....1M}. Recently, \citet{Ricci2021ApJ...913..122R} used the X-ray PE model by \citet{Picogna2019} and the analytic MHD disk model from \citet{Weber2020} to simulate ngVLA observations at milliarcsecond angular resolution for the nearby disk of TW~Hya. They showed that such observations can spatially resolve the wind and further distinguish between the two models as the PE disk would present two spatially distinct emission regions, with only the one further away tracing unbound gas, while the MHD disk would have a smoother more compact free-free emission. 

Along with possibly tracing a wind, sensitive observations at cm wavelengths have been also used to set useful upper limits on the EUV photon luminosity ($\Phi_{\rm EUV}$) impinging on the disk surface. This is because the free-free luminosity from an ionized disk surface is linearly proportional to $\Phi_{\rm EUV}$ \citep{Pascucci2012}, yet other sources of cm emission discussed above cannot be disentangled.
 This simple approach led to the identification of a significant number ($> 20$) of disks receiving $\le 10^{41}$ EUV photons per second \citep{Pascucci2014,GalvanMadrid2014A&A...570L...9G,Macias2016ApJ...829....1M,Coutens2019A&A...631A..58C}, a $\Phi_{\rm EUV}$ too low for EUV-driven photoevaporation alone to disperse protoplanetary disks. In addition, for the subset of disks having a \neii{} 12.8\,\micron{} detection, the inferred $\Phi_{\rm EUV}$ are too low to efficiently ionize Ne atoms and reproduce the observed line luminosities \citep{Pascucci2014}. This points to  Ne atoms being primarily ionized by  1\,keV X-ray photons. Importantly, because these \neii{} profiles are also blueshifted, the outflowing gas is located deeper in the disk
than what predicted by EUV-only PE models,  implying higher mass loss rates.

\subsubsection{\textit{\textbf{Dust Entrained in Winds}}} \label{sect:EntrainedDust}

Aerodynamically well coupled dust grains can be stirred upward by the vertical upflows accompanying  MHD or PE disk winds \citep[e.g., ][see also \S\ref{sect:implications}]{Owen2011MNRAS.411.1104O,Miyake2016}.
Unless turbulence is very strong in the disk ($\alpha \ge 0.01$), the mechanism through which dust is delivered to the wind is vertical advection rather than diffusion and results in $\sim 0.01-1$\,\micron{} grains efficiently  removed from the disk \citep{BoothClarke2021}. 

These uplifted  grains produce near-infrared emission, and hence  can increase the excess emission above the stellar photosphere and affect the vertical height of disks measured in scattered light \citep[e.g.,][]{Bans2012,Riols2018,Franz2020}. In addition, \citet{McJunkin2014} suggest that they could contribute to extinguish stellar photons and thus explain the discrepancy between the higher $A_V$ obtained from optical and IR observations versus the lower interstellar extinction calculated from Ly-$\alpha$ absorption. In the case of MHD winds, the large variability of magnetic activity within $\sim 1$\,au is expected to drive time-variable upflows \citep{Flock2017}. Dust grains  dragged upward in a time-dependent manner would form local dust concentrations which, depending on the viewing angle, could temporally obscure the central star and lead to time-dependent optical fading and infrared brightening \citep[e.g.,][]{Ellerbroek2014,Fernandes2018ApJ...856..103F}. This phenomenon might also explain some, but not all, of the so-called ``dippers", young stars that show  day-long dimmings up to several tens of percent \citep[e.g.,][]{Ansdell2020}. The deep irregular dimming of RW~Aur~A from the U through to the M band have been also interpreted as obscuration from a dusty wind launched beyond $\sim 0.1$\,au \citep{Dodin2019MNRAS.482.5524D}.

Finally, crystalline grains detected in the outer disk, beyond $\sim 10-30$\,au  \citep[e.g.,][]{Sturm2013}, may also serve as an indirect diagnostic of disk winds. These grains are thought to have undergone thermal annealing at temperatures $> 600-1,000$\,K \citep{Yamamoto2018,Hallenbeck2000}, much larger than those in the outer disk. \citet{Giacalone2019} recently adopted a self-similar solution for a cold MHD disk wind and showed that grains beyond the dust sublimation radius can be efficiently uplifted from the disk surface, thermally annealed by stellar irradiation in the wind, and transported outward;  intermediate size ($\sim 0.1\!-\!1$\,\micron ) grains would thus re-enter and settle in the outer disk (see \S~\ref{sect:implications}). For their classic T~Tauri disk model, the bulk of the re-entering mass would be crystalline if launched within $\sim 0.5$\,au, that is, well inside the potential well of the star.  
The efficiency of the global circulation induced by the MHD disk wind may thus be measured by the observed ratio of crystalline to amorphous grains. 

Synthetic observations of dusty winds are necessary to test theoretical models and establish how efficiently MHD and PE winds entrain dust grains. As pointed out in \citet{Franz2020}, the different mass loss profiles of these winds should result in different size distributions and densities for the entrained dust versus radius (see also \citealt{RD2022arXiv220101478R}) -- with the resulting dust emission likely being more concentrated toward the star for MHD than for PE winds. Scattered light images at near-infrared wavelengths should be most sensitive to the sub-micron- and micron-size grains expected to be entrained in the wind \citep[e.g.,][]{Owen2011MNRAS.411.1104O}. However, recent synthetic observations of XEUV winds show that the disk height is only marginally affected by the wind and its unambiguous detection will be challenging even with JWST \citep{Franz2022A&A...657A..69F}.

\subsubsection{\textit{\textbf{Local Gas and Dust Substructures}}} \label{sect:substructures}
 The formation of local substructures can also be an indirect diagnostic for the removal of angular momentum and disk mass via MCWs through relatively strong poloidal magnetic field with $\beta_{\rm midpl}\sim 10^2 - 10^4$.  An initial small perturbation of gas density leads to the radial advection of vertical magnetic flux. The excited perturbation of the magnetic field strength is anti-correlated with the initial density perturbation
as a result of the viscous diffusion of magnetic flux from denser regions.
If stronger mass loss is driven from the less dense stronger-field region and/or if faster accretion is induced by the more efficient removal of angular momentum there, the density of this gap region further decreases to reinforce the initial density perturbation \citep{Lubow1994,Riols2019} -- as a result, rings and gaps spontaneously form in disks with MHD winds. \citet{Suriano2017,Suriano2018,Suriano2019} demonstrated that rings and gaps are created by this instability in both 2D and 3D global non-ideal MHD simulations.  
In their 2D simulations, \citet{Riols2020} found a typical separation between rings of 2.5 and 5 times the local scale  height at $R \approx 20$ au for 
$\beta_{\rm midpl}=10^4$ and $10^3$, respectively. Thus, the separation between rings is found to be wider for stronger vertical fields ($\beta_{\rm midpl}=10^3$) but the dependence is weak. While in principle such ring-gap features can be long lived, potential secondary non-axisymmetric instabilities, which may not be resolved by current simulations, might disrupt them. Dense rings are obviously a favorable site for the concentration and growth of dust particles, which will be discussed in \S~\ref{sect:implications}.

\subsubsection{\textit{\textbf{Evolution of the Gas Disk Outer Radius}}}\label{sect:GasDiskRadius}
When mass accretion is predominantly driven by the removal of angular momentum through MCWs, disk evolution is different from that governed by viscous accretion. First, the reduced turbulent dissipation suppresses viscous heating, resulting in lower temperatures in the disk midplane \citep{Mori2019}. Second, the disk outer radius does not expand much because of reduced viscous spreading \citep[e.g.,][]{Shadmehri2019, chambers2019}, which is important for planet formation models (see \S~\ref{sect:implications}) and can be tested observationally \citep[e.g.,][]{Najita2018}.    

Inferring gas disk sizes from observations requires proper treatments that account for the optical depth, freeze-out, and photo-dissociation of gas-phase tracers such as CO \citep[e.g.,][]{Facchini2017A&A...605A..16F,Trapman2020}. Recently, \citet{Trapman2020} generated 
synthetic emission maps based on viscously evolving disk models and found that most observed CO gas disk sizes in Lupus can be reconciled with low levels of turbulence (i.e., $\alpha = 10^{-4}$ to $10^{-3}$). These low values compare well with those obtained in recent global simulations of weak MRI turbulence as a result of ambipolar diffusion \citep{CuiBai2021}. 
A potential caveat for the viscous disk scenario is that disks would have to start out small (radii $\simeq 10\,$au) which requires efficient removal of angular momentum during disk formation \citep[e.g.,][]{Machida2011}. The gas disk radii measured  so far for the earliest Class~0 evolutionary stage are significantly larger than this and often comparable to those inferred in the later Class~I and II stages (see Table~\ref{tab:jetwindprop}) suggesting, if anything, only moderate viscous spreading over millions of years. 

The radial extent of a viscously evolving disk  is further affected by FUV photoevaporation \citep{Gorti2009, Gorti2015}. Since the latter is more effective at larger radii, gas removal here halts the expansion of the outer gas disk, further complicating interpretations of observed disk radii in the context of viscous spreading. In fact, FUV PE models indicate a modest shrinking in the disk size over the final $\sim$ Myr even in disks with high $\alpha$ of $10^{-2}$ (see Fig.~8 in \citealt{Gorti2015}). 

\section{\textbf{\uppercase{Implications for planet formation and migration}}}\label{sect:implications}
Disk winds also affect the evolution of the solid component of a protoplanetary disk and the subsequent formation, evolution, and migration of planets. 

Both PE and MHD disk winds are primarily outflowing gas from the disk surface but small dust particles can be also entrained and uplifted
(see \S~\ref{sect:EntrainedDust}). There are no conclusive results on the amount of entrained dust. However,  \citet{Hutchison2016} point out that it is the dust scale height, which is determined by turbulence, that controls the amount of grains brought into the launch regions. As such, the amount of entrained dust is likely linked to the wind launching region. 
\citet{Hutchison2021} find that EUV photoevaporation creates a vertical gas flow within the disk that assists turbulence in supplying dust to the ionization front and predict a maximum size $a_{\rm max}\approx 4$\,\micron\ for the uplifted grains beyond $\sim 40$\,au. The denser XEUV wind models by  \citet{Franz2020} find instead that dust grains of $a_{\rm max}\approx 10$ $\mu$m  
can be coupled  to the gas and lifted  from 15-30\,au. Models of dust entrained in winds driven by all high-energy components, including FUV photons, are not available but efficient dust entrainment is also expected beyond $\sim 10$\,au.
In contrast to PE models, MHD disk winds can be stronger and denser closer in and dust entrainment is expected to be more efficient at smaller radii (see \citealt{Hansen2014MNRAS.440.3545H} for an early work on this topic). A simple radial scaling based on the MMSN model suggests $a_{\rm max}\approx 25$ $\mu$m$\left(\frac{r}{1{\rm au}}\right)^{-3/2}$ \citep{Miyake2016}.
Dust grains with size $\approx a_{\rm max}(r)$ are expected to float at $\approx$ several scale heights from the midplane, which can be used as an indirect signature of MHD disk winds (\S~\ref{sect:EntrainedDust}). 
Recent 2D simulations further show the importance of radiation pressure in pushing out dust grains, particularly those within 1\,au \citep{Vinkovic2021}, thus assisting in the outward transport and reentering in the disk outer regions \citep[see e.g.,][and \S~\ref{sect:EntrainedDust}]{Giacalone2019}. This large-scale transport is not  seen in PE models \citep[e.g.,][]{Hutchison2021}. 

\subsection{Winds and planetesimal formation}
A direct outcome of the preferential removal of gas by PE or MHD disk winds is the increase of the disk dust-to-gas ratio  \citep[e.g.,][]{Suzuki2010, Gorti2015}, which promotes the growth of solid particles.
\citet{Carrera2017} adopted the PE wind model from \citet{Gorti2015} and showed that the  enhanced dust-to-gas ratio favors the formation of planetesimals by the streaming instability, whereas the amount of the planetesimals sensitively depends on the detailed treatment of the solid component. While earlier X-ray PE models did not succeed in finding such enhancements \citep{Ercolano2017}, recent work on dust dynamics in XEUV photoevaporative winds shows very rapid increase in the dust-to-gas ratio reaching unity in $\sim 10^5$ years \citep{Franz2022A&A...657A..69F}.

PE and MHD disk winds also affect the radial drift of solid particles.
In a standard protoplanetary disk with a MMSN surface density, boulder-sized ($\approx 1$\,m) objects suffer severe inward drift to the central star \citep[e.g.,][]{Adachi1976}, the so-called radial drift barrier against the formation of planets. 
However, the radial profile of the gas component is largely altered in the presence of  disk winds.
An inner cavity is often formed by the effect of PE at late stages of evolution.
The radial drift of solid particles slows at the outer edge of the cavity, which enhances the local dust-to-gas ratio. 
Combined with the preferential removal of gas, the inner cavity is a suitable site for the formation of planetesimals \citep{Carrera2017}.
 MHD disk winds generally disperse the gas in an inside-out manner \citep{Suzuki2010}.
As a result, the radial slope, $p$, of the surface density ($\Sigma_{\rm g}\propto r^{-p}$) is reduced which slows down the inward drift of solid particles from small dust grains to planetesimals.
\citet{Taki2021} used the wind-driven accretion models of \citet{Suzuki2016} and found that drifting dust grains pile up to form planetesimals in a region around $1-2$\,au  when $\Sigma_{\rm d}/\Sigma_{\rm g} \gtrsim  \lvert \eta \rvert$,
where $\eta$ is the dimensionless radial-pressure gradient force.
If the inside-out dispersal is more drastic \citep{Takahashi2018}, $\eta$ could be negative so that dust particles drift outward, which could provide an alternative explanation for crystalline grains in the outer disk regions \citep{Arakawa2021}. 
However, we should note that $\eta$ is  uncertain as it depends on the properties of advected poloidal magnetic field and detailed microphysical processes \citep{Bai2016}. Once planetesimals are concentrated within  $\sim 1$\,au, they can form rocky planets with properties similar to those in our Solar System \citep{Ogihara2018a}.    

Local substructures such as rings and gaps created by the MHD disk winds (\S\ref{sect:substructures}) also play a role in the growth of dust grains because dense gaseous rings accumulate solid particles \citep{Riols2018}.
3D global MHD simulations 
by \citet{Riols2020} demonstrate that dust grains concentrate into  pressure maxima of axisymmetric gaseous rings to eventually form dusty rings.

\subsection{Winds and planetary migration}
In a typical protoplanetary disk, low-mass planets migrate inward due to the gravitational interaction with the background gas disk \citep{Ward1997,Tanaka2002,Paardekooper2011}. Importantly, the altered gaseous radial profile of wind-driven accretion disks also affects the migration of (proto)planets in two ways. 
First, the decrease of the inner disk surface density by MHD disk winds slows down, halts, or can even reverse the inward migration of planetary embryos with $\approx 0.1M_{\oplus}$ and Earth-size planets \citep{Ogihara2015}.
As a result, low-mass planets can survive in the inner region and their  radial distribution can reproduce the observed distribution of close-in super-Earths in exoplanetary systems \citep{Ogihara2018b}. 
Second, when accretion is triggered by the removal of angular momentum in MHD disk winds \citep{Bai2017,Gressel2020}, the accretion flow is laminar, in contrast with the turbulent diffusivity of viscous accretion models.  
While in a viscous accretion flow gas diffuses between the corotation region and the surrounding regions at both the inner and outer edges, in a laminar case the librating flow in the corotation region is more asymmetric and induces a positive corotation torque on the planet \citep{McNally2017,Ogihara2017}.
Therefore, the inward migration of low-mass planets tends again to be suppressed or even reversed in a wind-driven accreting disk, whereas 
the velocity of the migration also depend on other detailed conditions of  
the vertical profile of the radial velocity of the gas \citep{Ogihara2017} and the initial relative motion between the gas and the planet \citep{McNally2018}. 

Finally, wind-driven accretion also affects the migration of more massive planets that open a gap in the disk.
Similarly to lower-mass planets, mass supply to the co-orbital gap region is largely altered from viscous accretion disks.
2D simulations by \citet{Kimmig2020} show that in cases of strong wind-driven accretion Jupiter- and Saturn-size planets rapidly migrate outward thanks to the positive corotation torque excited by the asymmetric librating flow structure. 3D simulations should be carried out to test the robustness of this phenomenon.

\section{\textbf{\uppercase{Conclusions and outlook}}}\label{sect:outlook}
Since PPVI, significant progress has been made in detecting and characterizing mass ejection from low-mass stars and in building more realistic models of disk evolution and dispersal. Through this review we find that, at each evolutionary stage, there is evidence of powerful collimated fast ($\ge 100$\,\kms ) jets being surrounded by  radially extended slow outflows with morphologies and velocities consistent with MHD disk winds. These winds arise from a large range of disk radii, from within the dust sublimation radius out to $\sim$100\,au, with outflow velocity decreasing with radius.  While still preliminary, best estimates are that the wind mass loss exceeds that in the jet and is comparable to the accretion rate onto the star. Such high mass loss rates are consistent with MHD winds as a primary driver of accretion in the planet formation region ($\sim 1-30$\,au). 
Unresolved outflow signatures in Class~II sources further show that, 
as the accretion rate falls, disk winds persist after the jet is no longer detectable, and as the accretion rate continues to decline further, the  wind appears to evolve, from MHD to PE dominated. 

Alongside with observations, global disk simulations that couple gas dynamics with magnetic field, thermal energy transport, and chemistry find that radially extended disk winds are a common outcome and can drive accretion at the observed levels. More realistic photoevaporative wind models coupling hydrodynamics and chemistry have clarified that FUV photons play an important role in driving thermal flows and that mass loss rates may range from  a few $10^{-9}$\,M$_\odot$/yr up to $10^{-8}$\,M$_\odot$/yr for typical low-mass stars, a factor of a few lower than  those from MHD models. Thus, both observations and theory suggest that MHD disk winds drive the early evolution of protoplanetary disks while photoevaporative winds  become critical at later stages.

There remain a number of open questions and controversies in this broad brush evolutionary picture which set future priorities for observations and theoretical investigations. For instance,  it remains unclear whether jets and winds are the manifestation of the same physical process, as expected in MHD wind models, or rather constitute physically distinct mechanisms. In addition, the interpretation of observed, outer molecular cones as disk winds is still debated for the Class~0/I stage and the alternative scenario  of entrained gas from fast inner winds is not completely ruled out. Further progress in these areas can be achieved with  more sensitive high-resolution ALMA images testing e.g. whether jets are typically hollow and if warmer MHD disk wind streamlines are present at radial distances closer to the star than  the cooler molecular cones.

In the Class~II stage, where wind diagnostics remain mostly spatially unresolved, we do not know how different forbidden LVC components could arise from a radially continuous MHD wind, wind mass loss rates bear large uncertainties, and the transition from MHD to PE winds remains elusive. As discussed in  this chapter, spectro-astrometry in the LVC holds the promise to constrain the vertical extent of the BC and NC and thus help to pin down wind mass loss rates. Sensitive IFU spectro-imaging, from the ground and soon from space with JWST/NIRSpec and MIRI, will expand the number of Class~II sources with spatially resolved jets and winds. This sample will enable further testing of the ``onion-like" kinematic structure expected from radially extended MHD disk winds and verify whether more mass is typically lost in the wind than in the jet as suggested by the few systems available so far. Additionally, when targeting transition disks with large dust cavities, IFU spectro-imaging could test if their winds arise just exterior to the cavity as expected in the photoevaporative  scenario. While several ALMA surveys of Class~II disks have been carried out so far, they are either too shallow or limited in spatial coverage due to high resolution to adequately resolve  cool molecular winds. Sensitive multi-configuration ALMA observations such as those for DG~Tau~B and HH~30 are necessary to expand the sample of  spatially resolved  molecular winds and establish if the mass ejection is continuous or  if more mass is removed in the outer vs the inner disk.

From the theoretical side, global MHD disk simulations have become reasonably realistic but are still confined to snapshots in time and remain mostly relevant to typical Class~II accretors. It would be valuable to expand the set of simulations and explore different evolutionary stages that could be  applicable to the earlier Class0/I stage as well as the later Class~II stages when PE winds might start to dominate.  As yet it remains unclear how the effective $\alpha$ evolves radially and in time, whether it smoothly connects to MRI-active regions, and whether  there are specific radial distances where mass is preferentially removed, which could drive gaps and structures in disks.  One of the grand challenges is to conduct a long-time global simulation (including MRI-active and non MRI-active regions) from the Class-0 stage to the Class-II stage with appropriate microphysics. 
Along with tracking basic disk parameters,  synthetic line observations at all evolutionary stages would be crucial to further test disk wind models as well as to constrain some of the critical input parameters such as the amount and evolution of the magnetic flux threading the disk. Current theoretical investigations have shown that wind-driven accretion significantly impacts the evolution of the solids in a disk as well as planet formation, evolution, and migration.  Confronting disk wind observations and theory is  crucial to building a comprehensive theory of disk evolution and dispersal and thus clarifying  the role of disk winds in the assembly and final architecture of planetary systems.
%
%
%

\bigskip
\noindent\textbf{Acknowledgments}  
We thank C. F. Lee and A. deValon for creating the composite images of HH~212 and DG~Tau~B respectively, and K. Stapelfeldt and D. Padgett for providing the HST data tracing the DG~Tau~B jet. We are grateful to
M. G\"udel for providing the ALMA $^{12}$CO data of DG~Tau~A, to J. Tobin for inputs on the VANDAM survey, to T. Beck for comments on the H$_2$ near-infrared resolved wind morphologies, and to J. Ferreira for insightful comments on the theory of MHD disk winds. I.P., U.G., and S.E. acknowledge support from a Collaborative NSF Astronomy \& Astrophysics Research grant (ID: 1715022,
ID:1713780, and ID:1714229). S.C. acknowledges support by the Programme National “Physique et Chimie du Milieu Interstellaire” (PCMI) of CNRS/INSU with INC/INP co- funded by CEA and CNES, and by the Conseil Scientifique of Observatoire de Paris. T.K.S. is supported by Grants-in-Aid for Scientific Research from the MEXT of Japan, 17H01105 and 21H00033. This material is based on work
supported by the National Aeronautics and Space Administration
under agreement No. NNX15AD94G for the program
{\it Earths in Other Solar Systems}. The results reported herein
benefited from collaborations and/or information exchange
within NASA’s Nexus for Exoplanet System Science 
research coordination network sponsored by NASA’s Science
Mission Directorate.

\bigskip
\bibliographystyle{pp7}
\bibliography{diskwinds.bib}
\end{document}